\documentclass[12pt]{article}
\catcode`\@=11

\@addtoreset{equation}{section}
\def\theequation{\arabic{equation}}
\def\theequation{\thesection\arabic{equation}}

\newcommand{\mc}{\mathcal}

\def\vth{\vartheta}
\def\tld{\tilde}

\def\t{\tau}

\def\g{\gamma}

\def\vth{\vartheta}

\def\L{\Lambda}
\def\e{\epsilon}

\def\part{\partial}
\def\tld{\tilde}

\newcommand{\ra}{\rightarrow}

\def\@normalsize{\@setsize\normalsize{15pt}\xiipt\@xiipt
\abovedisplayskip 14pt plus3pt minus3pt%
\belowdisplayskip \abovedisplayskip
\abovedisplayshortskip  \z@ plus3pt%
\belowdisplayshortskip  7pt plus3.5pt minus0pt}
\def\small{\@setsize\small{13.6pt}\xipt\@xipt
\abovedisplayskip 13pt plus3pt minus3pt%
\belowdisplayskip \abovedisplayskip
\abovedisplayshortskip  \z@ plus3pt%
\belowdisplayshortskip  7pt plus3.5pt minus0pt
\def\@listi{\parsep 4.5pt plus 2pt minus 1pt
            \itemsep \parsep
            \topsep 9pt plus 3pt minus 3pt}}

\def\underline#1{\relax\ifmmode\@@underline#1\else
        $\@@underline{\hbox{#1}}$\relax\fi}
\@twosidetrue
\relax

\catcode`@=12

\catcode`\@=11

\def\section{\@startsection{section}{1}{\z@}{3.5ex plus 1ex minus
   .2ex}{2.3ex plus .2ex}{\large\bf}}
\def\subsection{\@startsection{subsection}{1}{\z@}{3.5ex plus 1ex minus
   .2ex}{2.3ex plus .2ex}{\it}}
\def\subsubsection{\@startsection{subsubsection}{1}{\z@}{3.5ex plus 1ex minus
   .2ex}{2.3ex plus .2ex}{\it}}
\def\thesection{\arabic{section}.}
\def\thesubsection{\arabic{section}.\arabic{subsection}}

\def\ps@headings{\def\@oddfoot{}\def\@evenfoot{}
\def\@oddhead{\hbox{}\hfill
        \makebox[.5\textwidth]{\raggedright\ignorespaces --\thepage{}--
        \hfill }}
\def\@evenhead{\@oddhead}
\def\subsectionmark##1{\markboth{##1}{}} }
\ps@headings

\catcode`\@=12

\relax

\def\figcap{\section*{Figure Captions\markboth
        {FIGURECAPTIONS}{FIGURECAPTIONS}}\list
        {Fig. \arabic{enumi}:\hfill}{\settowidth\labelwidth{Fig. 999:}
        \leftmargin\labelwidth
        \advance\leftmargin\labelsep\usecounter{enumi}}}
 \relax
\def\tablecap{\section*{Table Captions\markboth
        {TABLECAPTIONS}{TABLECAPTIONS}}\list
        {Table \arabic{enumi}:\hfill}{\settowidth\labelwidth{Table 999:}
        \leftmargin\labelwidth
        \advance\leftmargin\labelsep\usecounter{enumi}}}
 \relax
\def\reflist{\section*{References\markboth
        {REFLIST}{REFLIST}}\list
        {[\arabic{enumi}]\hfill}{\settowidth\labelwidth{[999]}
        \leftmargin\labelwidth
        \advance\leftmargin\labelsep\usecounter{enumi}}}
 \relax

\catcode`\@=11

\def\marginnote#1{}
\newcount\hour
\newcount\minute
\newtoks\amorpm
\hour=\time\divide\hour by60
\minute=\time{\multiply\hour by60 \global\advance\minute by-
\hour}
\edef\standardtime{{\ifnum\hour<12 \global\amorpm={am}%
    \else\global\amorpm={pm}\advance\hour by-12 \fi
    \ifnum\hour=0 \hour=12 \fi
    \number\hour:\ifnum\minute<100\fi\number\minute\the\amorpm}}
\edef\militarytime{\number\hour:\ifnum\minute<100\fi\number\minute}
\def\draftlabel#1{{\@bsphack\if@filesw {\let\thepage\relax
  \xdef\@gtempa{\write\@auxout{\string
    \newlabel{#1}{{\@currentlabel}{\thepage}}}}}\@gtempa
    \if@nobreak \ifvmode\nobreak\fi\fi\fi\@esphack}
     \gdef\@eqnlabel{#1}}
\def\@eqnlabel{}
\def\@vacuum{}
\def\draftmarginnote#1{\marginpar{\raggedright\scriptsize\tt#1}}
\def\draft{\oddsidemargin -.5truein
        \def\@oddfoot{\sl preliminary draft \hfil
        \rm\thepage\hfil\sl\today\quad\militarytime}
        \let\@evenfoot\@oddfoot \overfullrule 3pt
        \let\label=\draftlabel
        \let\marginnote=\draftmarginnote
   
\def\@eqnnum{(\theequation)\rlap{\kern\marginparsep\tt\@eqnlabel}%
\global\let\@eqnlabel\@vacuum}  }
\def\preprint{\twocolumn\sloppy\flushbottom\parindent 1em
        \leftmargini 2em\leftmarginv .5em\leftmarginvi .5em
        \oddsidemargin -.5in    \evensidemargin -.5in
        \columnsep 15mm \footheight 0pt
        \textwidth 250mmin      \topmargin  -.4in
        \headheight 12pt \topskip .4in
        \textheight 175mm
        \footskip 0pt
        
\def\@oddhead{\thepage\hfil\addtocounter{page}{1}\thepage}
        \let\@evenhead\@oddhead \def\@oddfoot{} \def\@evenfoot{}  }
\def\titlepage{\@restonecolfalse\if@twocolumn\@restonecoltrue\onecolumn
     \else \newpage \fi \thispagestyle{empty}\c@page\z@
        \def\thefootnote{\fnsymbol{footnote}} }
\def\endtitlepage{\if@restonecol\twocolumn \else  \fi
        \def\thefootnote{\arabic{footnote}}
        \setcounter{footnote}{0}}  
\catcode`@=12
\relax

\def\ps@headings{\def\@oddfoot{}\def\@evenfoot{}
\def\@oddhead{\hbox{}\hfill
        \makebox[.5\textwidth]{\raggedright\ignorespaces --\thepage{}--
        \hfill }}
\def\@evenhead{\@oddhead}
\def\subsectionmark##1{\markboth{##1}{}} }
\ps@headings
\relax
\def\firstpage#1#2#3#4#5#6{
\begin{document}
\begin{titlepage}
\nopagebreak
\title{\begin{flushright}
        \vspace*{-1.8in}
        {\normalsize ROM2F-03/07}\\[-9mm]
        {\normalsize hep-th/0305224}\\[4mm]
\end{flushright}
\vfill {#3}}
\author{\large #4 \\[1.0cm] #5}
\maketitle
\vskip -7mm     
\nopagebreak 
\begin{abstract} {\noindent #6}
\end{abstract}
\vfill
\begin{flushleft}
\rule{16.1cm}{0.2mm}\\[-3mm]
May 2003
\end{flushleft}
\thispagestyle{empty}
\end{titlepage}}
\def\t{\tau}
\def\simlt{\stackrel{<}{{}_\sim}}
\def\simgt{\stackrel{>}{{}_\sim}}
\newcommand{\dal}{\raisebox{0.085cm} {\fbox{\rule{0cm}{0.07cm}\,}}}
\newcommand{\dt}{\partial_{\langle T\rangle}}
\newcommand{\dtbar}{\partial_{\langle\overline{T}\rangle}}
\newcommand{\al}{\alpha^{\prime}}
\newcommand{\mst}{M_{\scriptscriptstyle \!S}}
\newcommand{\mpl}{M_{\scriptscriptstyle \!P}}
\newcommand{\dv}{\int{\rm d}^4x\sqrt{g}}
\newcommand{\lv}{\left\langle}
\newcommand{\rv}{\right\rangle}
\newcommand{\ph}{\varphi}
\newcommand{\abar}{\overline{a}}
\newcommand{\sbar}{\,\overline{\! S}}
\newcommand{\xbar}{\,\overline{\! X}}
\newcommand{\fbar}{\,\overline{\! F}}
\newcommand{\zbar}{\overline{z}}
\newcommand{\dbar}{\,\overline{\!\partial}}
\newcommand{\tbar}{\overline{T}}
\newcommand{\taubar}{\overline{\tau}}
\newcommand{\ubar}{\overline{U}}
\newcommand{\ybar}{\overline{Y}}
\newcommand{\phb}{\overline{\varphi}}
\newcommand{\cm}{Commun.\ Math.\ Phys.~}
\newcommand{\prl}{Phys.\ Rev.\ Lett.~}
\newcommand{\pr}{Phys.\ Rev.\ D~}
\newcommand{\pl}{Phys.\ Lett.\ B~}
\newcommand{\ibar}{\overline{\imath}}
\newcommand{\jbar}{\overline{\jmath}}
\newcommand{\np}{Nucl.\ Phys.\ B~}
\newcommand{\F}{{\cal F}}
\renewcommand{\L}{{\cal L}}
\newcommand{\A}{{\cal A}}
\newcommand{\beq}{\begin{eqnarray}}
\newcommand{\eeq}{\end{eqnarray}}
\newcommand{\be}{\begin{equation}}
\newcommand{\ee}{\end{equation}}
\newcommand{\bay}{\begin{array}}
\newcommand{\eay}{\end{array}}
\newcommand{\ba}{\begin{eqnarray}}
\newcommand{\bea}{\begin{eqnarray}}
\newcommand{\ea}{\end{eqnarray}}
\newcommand{\eea}{\end{eqnarray}}
\newcommand{\dslash}{{\not\!\partial}}
\newcommand{\gsi}{\,\raisebox{-0.13cm}
{$\stackrel{\textstyle >}{\textstyle\sim}$}\,}
\newcommand{\lsi}{\,\raisebox{-0.13cm}
{$\stackrel{\textstyle <}{\textstyle\sim}$}\,}
\date{}
\firstpage{3118}{IC/95/34} {\large\bf Magnetized
Four-Dimensional $Z_2 \times Z_2$ Orientifolds}  {M. \ Larosa \ 
and \ G.\ Pradisi}
{\normalsize Dipartimento di Fisica, Universit\`a di Roma 
``Tor Vergata''\\ [-3mm] \normalsize INFN,
Sezione di Roma ``Tor Vergata''\\[-3mm]\normalsize\sl 
Via della Ricerca Scientifica 1, I-00133 Roma,
Italy}
{We study deformations of $Z_2 \times Z_2$ (shift-)orientifolds in four 
dimensions in the presence of both uniform Abelian internal 
magnetic fields and  quantized NS-NS 
$B_{ab}$ backgrounds, that are shown to be equivalent to asymmetric 
shift-orbifold projections.  These models are related by $T$-duality to
orientifolds with $D$-branes intersecting at angles.  As 
in corresponding six-dimensional examples,
 $D9$-branes magnetized along two internal directions acquire a 
charge with respect to the R-R six 
form, contributing to the tadpole of the orthogonal 
$D5$-branes (``brane transmutation'').  The resulting models exhibit 
rank reduction of the gauge group 
and multiple matter families, due both to the quantized 
$B_{ab}$ and to the 
background magnetic fields.
Moreover, the low-energy spectra are chiral and anomaly 
free if additional $D5$-branes longitudinal to the magnetized directions
are present, and if there are no Ramond-Ramond tadpoles in the 
corresponding twisted sectors 
of the undeformed models.}
\tableofcontents

\section{Introduction}

Perturbative type I vacua \cite{cargese,zori1,ps} (for reviews see 
\cite{review,dudas}) are a small corner 
in the moduli space of the
unified underlying and yet poorly understood 
eleven dimensional M-theory \cite{mtheory}.
Nowadays, however, they are the most promising perturbative models
to test possible ``stringy'' effects at the next generation of accelerators 
\cite{aahdd}.
Indeed, while in the usual heterotic SUSY-GUT scenarios \cite{die1} 
the string scale
is directly tied to the Planck scale, making it hard to conceive probes
of low-energy effects, the type I string scale 
is basically an independent parameter, that can be lowered down 
to a few 
TeV's \cite{laq}.  In this setting, and more generally in the 
context of ``brane-world
scenarios'' \cite{bw,aahdd}, the gauge degrees of freedom are confined to some 
(stacks of) branes
while the gravitational interactions invade the whole higher dimensional 
spacetime.  In order
to respect the experimental limits on gravitational interactions, 
the extra dimensions orthogonal to the branes 
could be up to sub-millimeter size \cite{exper}, while the extra
dimensions longitudinal to the branes should be quite tiny
(at least of TeV scale) but
still testable in future experiments.  In this context, however, 
several aspects 
of the conventional Standard Model picture like, 
for instance, the problem of scale hierarchies and the unification
of the running coupling constants at a scale of order $10^{16} GeV$ in
the MSSM desert hypothesis, have to be reconsidered \cite{aahdd,dudas,anto}.  
Some other issues, like supersymmetry breaking,
find instead new possibilities in type I perturbative vacua.  

There are 
basically four known ways to break supersymmetry within String Theory.  
The first is to combine left and right moving modes 
in a non-supersymmetric fashion, like for instance in the type $0$ models 
\cite{type0}
and in the corresponding lower dimensional 
compactifications and orientifolds \cite{zori1,zori2,zori3,branes0}.  
Some type I-like instances, the type $0^{\prime}$B and its 
compactifications \cite{zori2}, 
are also free of the tachyons 
that typically plague this kind of models.  
The second is the generalization to String Theory of the 
Scherk-Schwarz mechanism \cite{scsc},
available also in the heterotic case, in which the breaking is 
due to a generalized Kaluza-Klein compactification that 
involves different periodicities for bosons and fermions, thereby
not respecting supersymmetry \cite{sshet}.  In this framework, the 
scale of supersymmetry breaking is inversely proportional to the 
volume of the internal manifold and some residual global 
supersymmetries may be left at tree level on some branes 
(``brane supersymmetry'') \cite{ssopen,shift}.  
The third possibility is related to 
models with supersymmetry non linearly realized 
on some branes, as a result 
of the simultaneous presence of branes and antibranes
of the same \cite{bantib,aadds,abg} or of different types \cite{bsb,aadds,bmp},
or of the introduction of ``exotic'' orientifold planes \cite{sugi}.  These 
are referred to as ``brane supersymmetry breaking'' models, and   
the corresponding scale 
of supersymmetry breaking can be tied generically to the string scale.
Finally, supersymmetry can be broken \cite{magwit,bachas,penta} 
by the introduction of internal magnetic fields \cite{ft} in the open sectors 
\cite{bachas,aads,magnbsb,ted}.  In 
refs. \cite{aads,magnbsb}, six-dimensional 
$(T^2 \times T^2) / Z_2$ orientifold models with a pair of 
internal magnetic fields turned on inside 
the internal tori have been discussed 
in detail.  

In this paper we extend the construction of \cite{aads,magnbsb}
to four dimensional models obtained as magnetic deformations of 
$Z_2 \times Z_2$ orientifolds, possibly combined with momentum or 
winding shifts along some of the internal directions \cite{shift,aadds}.  
Some preliminary results have already appeared in refs. \cite{noi}.  
As in the six-dimensional models of \cite{aads,magnbsb}, for generic 
values of the magnetic fields Nielsen-Olesen instabilities 
\cite{nielsole} manifest
themselves in the presence of tachyonic excitations, and supersymmetry
is broken due to the unpairing of states of different spins.
However, the compactness of the internal space allows
self-dual or antiself-dual Abelian field configurations with 
non-vanishing instanton number, that can compensate the R-R charge excess, 
eliminating the tachyons and retrieving supersymmetry if the BPS bound
is saturated, or giving rise to ``brane supersymmetry breaking'' models 
if the magnetized $D9$-branes transmute into anti $D5$-branes 
\cite{aads,magnbsb}.
The resulting models exhibit several interesting features, to wit 
Chan-Paton gauge groups of reduced rank and several families of matter
multiplets, linked in a natural way to the degeneracy of the Landau levels.
Moreover, in the presence of $D$-branes longitudinal to the directions 
along which the magnetic fields are turned on, the four dimensional models can
also be chiral.  It should be stressed that these models 
with internal background (open) magnetic fields are connected by T-duality 
to orientifolds
with $D$-branes intersecting at angles 
\cite{intbranes,phen1,phen2,csu,phen3,phen4,phen5,phen6,fluxes}, 
that have received much attention
in the last few years in attempts to recover 
(extensions of) the Standard
Model as low-energy limits of String Theory or M-theory.  A byproduct of this 
analysis is  a precise link between a quantized NS-NS $B_{ab}$ and
shift-orbifolds.

This paper is organized as follows: in Section 2, after a brief discussion
of the relation between shifts and the quantized NS-NS two form $B_{ab}$, 
we shortly review the
basic facts characterizing $Z_2 \times Z_2$ (shift-)orientifolds in its 
presence.  
In Section 3 we discuss general aspects of the inclusion of uniform
internal  Abelian magnetic fields in the open-string sector and 
shortly review the models in \cite{aads}.  In Section 4 we describe 
four dimensional examples based on magnetic deformations of the 
$Z_2 \times Z_2$ (shift-)orientifolds previously introduced.  
Section 5 is devoted to one ``brane supersymmetry breaking'' 
example and finally Section 6 contains our 
Conclusions.  Notations, conventions and Tables are 
displayed in the Appendices.  In particular, 
Appendix A collects the relevant lattice sums that enter the
one-loop partition functions.  The 
$Z_2 \times Z_2$ characters are defined in Appendix B, while Appendix C
is a collection of Tables that summarize massless spectra, 
gauge groups and tadpole cancellation conditions for {\bf all}  
 the models considered in this paper.

\section{$Z_2 \times Z_2$ Orientifolds}

In this Section we review some four dimensional type I vacua obtained as 
orientifolds of 
$Z_2 \times Z_2$ orbifolds, or of 
freely acting $Z_2 \times Z_2$ shift-orbifolds.
The starting point is an orbifold of the type IIB superstring compactified
on an internal six-torus that, without any loss of generality for our 
purposes, can be chosen 
to be a product of three two-tori $T^{45}$, $T^{67}$ and $T^{89}$ 
along the three complex directions $Z^1=X^4 + i X^5$, 
$Z^2=X^6 + i X^7$ and $Z^3=X^8 + i X^9$.   Each two-torus can be equipped 
with a NS-NS background two-form $B_i$ of rank $r_i$ (with $r_i=0$ or $2$) 
that, if the orientifold projection is induced by the  
world-sheet parity operator
$\Omega$, is a discrete modulus and may thus 
take only quantized values \cite{toroidal,carlo}. 
The orbifold group will be taken 
to be the combination of the  $Z_2 \times Z_2$ generated by the elements
\be 
g: (+,-,-) \qquad {\rm and} \qquad h: (-,-,+) \quad ,
\label{gens}
\ee
where the minus signs indicate the two-dimensional $Z_2$ 
inversion of the corresponding coordinates $(Z^i \rightarrow - Z^i)$, 
with momenta and/or winding shifts along the real part of (some of) 
the three complex directions.

The conventional $Z_2 \times Z_2$ orbifolds, allowing for the introduction 
of a ``discrete torsion''\cite{disto}, 
give rise to supersymmetric orientifolds \cite{erice,bl,csu} as 
well as to orientifolds with ``brane supersymmetry breaking'' \cite{aadds}.  
Moreover, the freely acting $Z_2 \times Z_2$ (shift-)orbifolds produce 
ten classes of orientifolds with different amounts of supersymmetry 
(models with ``brane supersymmetry'' \cite{ssopen,shift}), 
together with a huge number of variants with 
brane-antibrane pairs and ``brane supersymmetry breaking''\cite{aadds}.

\subsection{NS-NS $B_{ab}$ and Shifts}

Before moving to the detailed analysis of the models, it is worth 
clarifying the relation between a non-vanishing discretized NS-NS two form 
$B_{ab}$ on orientifolds \cite{toroidal,carlo} 
and momentum and winding shifts.  In particular,
we want to show that the presence of a quantized 
NS-NS two-form field $B_{ab}$ on a two-torus 
is exactly equivalent to an asymmetric shift-orbifold in which a momentum 
shift along the first direction of the torus
is accompanied by 
a winding shift along the other direction\footnote{This argument was 
partly developed in collaboration with E. Dudas and J. Mourad.  See also 
\cite{kaku} and, for 
related observations in the context of duality chains, \cite{sethi}.}. 
For simplicity, let us take for the two-torus a product of circles both of 
radius $R$.  Parametrizing the discretized two form as 
\cite{toroidal} 
\ba
B \ = \ \frac{\alpha^{\prime}}{2} \left( \begin{array}{rrrr}
0&1\\
-1&0\end{array} \right) \quad , 
\label{bdiscrete}
\ea
the generalized momenta of eqs. (\ref{pleft}, \ref{pright}) 
reduce to the expressions
\ba
p_{{\rm (L,R)} \, a} \ = \ m_a^{\prime} \ \pm \ \frac{1}{\alpha'} \ g_{ab} 
\ n^b \qquad , 
\label{pwithqb} 
\ea 
where 
$m_a^{\prime} \, = \, m_a \, - \, \frac{1}{2} \, \epsilon_{ab} \, n^{b}$. 
The presence 
of $B_{ab}$ makes the $m_a^{\prime}$'s integers 
or half-integers depending on the 
oddness or evenness of the integer $n^a$'s.  As a result, omitting
the prime on the dummy $m$-variables for the rest of this Section,  
the torus partition function of eq. (\ref{2dtorus}) 
can be decomposed in the form
\ba
 \!\!\! \Lambda(B) \!\!\! &=& \!\!\! \Lambda(m_1, m_2, 2n_1, 2n_2) +  
\Lambda(m_1+1/2, m_2, 2n_1, 2n_2+1) \label{torob} \\
\!\!\! &+& \!\!\! \Lambda(m_1, m_2+1/2, 2n_1+1, 2n_2) +  
\Lambda(m_1+1/2, m_2+1/2, 2n_1+1, 2n_2+1) \, , \nonumber 
\ea
where $\Lambda(m_1, m_2, n_1, n_2)$ denotes the 
two-dimensional lattice sum over momenta $m_a / R$
and windings $n^a R$.  

The same 
partition function can be obtained as an asymmetric 
shift-orbifold.  
Indeed, projecting the conventional ($B=0$) $\Lambda(m_1, m_2, n_1, n_2)$ 
under the action of a $p_1 w_2$ shift and completing the modular invariant 
with the addition of twisted sectors, the resulting partition function is
\ba
\Lambda(p_1 w_2) \!\!\! &=& \!\!\!  \frac{1}{2} \ \bigl[ \, 
\Lambda(m_1, m_2, n_1, n_2) + (-1)^{m_1+n_2} \, 
\Lambda(m_1, m_2, n_1, n_2) \\ 
\!\!\!\!\!\!\! &+& \!\!\! \Lambda(m_1, m_2+1/2, n_1+1/2, n_2) + (-1)^{m_1+n_2} \,
\Lambda(m_1, m_2+1/2, n_1+1/2, n_2)\, \bigr] \ . \nonumber 
\label{sheqb}
\ea
This expression is exactly the one in eq. (\ref{torob}) after 
doubling the radius $(R \rightarrow 2 R)$ along the first direction. 
Thus, it should not come as a surprise that 
in some cases the effect of the shifts
can be compensated by the presence of a quantized $B_{ab}$.  However,
all the models in this paper display a reduction of the rank of the
Chan-Paton group when a non-vanishing quantized $B_{ab}$ is turned on.
The reason is that the shifts we consider 
affect only one {\it real} coordinate of
the tori, rather than two as in the previous asymmetric shift-orbifold
construction.  Nonetheless, in some cases, the shifts make the multiplicities
of the matter multiplets independent of the rank of $B_{ab}$. 
By $T$-duality, 
other allowed discrete moduli \cite{ab,bgk,z3} like, 
for instance, 
the off-diagonal components of the metric in orientifolds 
of the type IIA superstring, 
can also be related to suitable shift-orbifolds.
 
A nice geometric interpretation of the rank reduction of the Chan-Paton 
groups can also be given resorting to the 
asymmetric shift-orbifold description of 
$B_{ab}$.  As we shall extensively see in the next Sections, a momentum 
shift orthogonal to $D$-branes splits them into multiple images. 
After a $T$-duality along
the second direction of the two-torus, the $p_1 w_2$ 
becomes a $p_1 p_2$ shift-orbifold, that 
admits orientifold projections containing $D1$-branes 
parallel to $p_1$ and orthogonal to $p_2$. The corresponding annulus amplitude
can be written  
\ba
{\cal A} \ = \ \frac{1}{2} \ N^2 \ \big( \, P_1 + P_1^{1/2} \, \big) \ 
\big( \, W_2 + W_2^{1/2} \, \big) \ , 
\label{anelp1p2}
\ea
where $P_i$ and $W_i$ are the usual one-dimensional momentum and 
winding lattice sums \cite{toroidal}, respectively, 
while $P_i^{1/2}$ and $W_i^{1/2}$ are
the corresponding shifted ones, and the consistent M\"obius amplitude, 
describing the unoriented projection, is
\ba 
{\cal M} \ = \ - \frac{1}{2} \ N \ \big( \, \hat{P}_1 \, \hat{W}_2 \, 
+ \, \hat{P}_1^{1/2} \, 
\hat{W}_2^{1/2} \, \big) \ .
\label{moebp1p2}
\ea
Eqs. (\ref{anelp1p2}, \ref{moebp1p2}) neatly display the expected doublet 
structure of the $D1$-brane configuration, and the analysis of the
tadpole cancellation conditions reveals the related rank reduction
of the Chan-Paton group.
For instance, an equivalent eight-dimensional 
$p_1 p_2$ shift-orientifold compactification of the 
type IIB superstring, would yield type I models with an $SO(16)$ 
gauge group, thus providing a rank reduction by a factor of two.

\subsection{$Z_2 \times Z_2$ Models and Discrete Torsion}

Let us now turn to the analysis of the {$Z_2 \times Z_2$ models.  Aside 
from the identity, the  $Z_2 \times Z_2$ elements can be  grouped
together in the matrix 
\ba
\sigma_0 \ = \ \left( \begin{array}{rrrr}
+&-&-\\
-&+&-\\
-&-&+\end{array} \right) \quad , 
\label{sigmazero}
\ea
whose rows represent the action of $g$, $f = g \circ h$ and $h$  
on the three internal torus coordinates $Z^i$.  
The one-loop closed partition function
can be obtained supplementing the $Z_2 \times Z_2$ projections of 
the toroidal amplitude with the 
inclusion of three twisted sectors, located at the three fixed
tori, to complete the modular invariant.
There are actually two options, related to the freedom of introducing
a discrete torsion \cite{disto}, {\it i.e.} a relative sign between two
disconnected orbits of the modular group.  The result is
\ba
{\cal T}\!\!\!&=&\!\!{1 \over 4} \Biggl\{ 
|T_{oo}|^2 \Lambda_1(B_1) \Lambda_2(B_2) \Lambda_3(B_3)
+ \big[ \, |T_{og}|^2  \Lambda_1(B_1) + |T_{of}|^2 \Lambda_2(B_2) 
+ |T_{oh}|^2 \Lambda_3(B_3) \, \big] \ 
\left|{4\eta^2 \over \vartheta_2^2}\right|^2 \nonumber \\
&+& \big[ \ |T_{go}|^2 \Lambda_1(B_1) + |T_{fo}|^2  \Lambda_2(B_2) 
+ |T_{ho}|^2 \Lambda_3(B_3) \ \big] \ 
\left|{4\eta^2 \over \vartheta_4^2}\right|^2 \nonumber \\
&+&  \big[ \ |T_{gg}|^2 \Lambda_1(B_1) + |T_{ff}|^2 \Lambda_2(B_2)
+ |T_{hh}|^2 \Lambda_3(B_3) \ \big] \ 
\left|{4\eta^2 \over \vartheta_3^2}\right|^2  \nonumber \\
&+& \omega \left( |T_{gh}|^2 + |T_{gf}|^2 + |T_{fg}|^2 + |T_{fh}|^2 
+ |T_{hg}|^2
+ |T_{hf}|^2 \right) \left|{8\eta^3 \over \vartheta_2 \vartheta_3 
\vartheta_4} \right|^2 
\Biggr\} \quad  , 
\label{a1} 
\ea
where the $\Lambda_i$'s are the two-dimensional Narain 
lattice sums for the three internal tori (see Appendix A), that 
depend on the two-dimensional blocks $(B_i)$ of the NS-NS 
two-form $B_{ab}$, 
and $\omega=\pm 1$ is the sign associated to the discrete
torsion. We have 
expressed the torus amplitude in terms of the 16 quantities $(i=o,g,h,f)$
\ba
T_{io} &=&  \tau_{io} +  \tau_{ig} + \tau_{ih} + \tau_{if} \quad , \qquad
T_{ig} =  \tau_{io} +  \tau_{ig} - \tau_{ih} - \tau_{if} \quad , \nonumber \\
T_{ih} &=&  \tau_{io} -  \tau_{ig} + \tau_{ih} - \tau_{if} \quad , \qquad
T_{if} =  \tau_{io} -  \tau_{ig} - \tau_{ih} + \tau_{if} \quad ,
\label{tiab}
\ea
where the $16$ $Z_2 \times Z_2$ characters $\tau_{il}$ \cite{erice}, 
combinations of products of level-one SO(2) characters, are displayed
in Appendix B.  The geometric model, related to the 
``charge conjugation'' modular invariant, corresponds to the choice
$\omega = -1$, as can be deduced from the massless spectra reported in 
Table \ref{clorz2z2}.  
It is a compactification on (a singular limit of) a Calabi-Yau
threefold with Hodge numbers $(h_{11}=51, h_{21}=3)$, while the $\omega = 1$ 
choice, linked in this context to the T-dual compactification, 
leads to (a singular limit of) the 
mirror symmetric Calabi-Yau threefold, with  $h_{11}=3, h_{21}=51$.

The starting point for the orientifold construction are 
the Klein-bottle amplitudes 
\ba
{\cal K} \!\! &=& \!\! \frac{1}{8} \biggl\{ ( P_1 P_2 P_3 + 2^{-4} 
P_1 W_2(B_2) W_3(B_3) +  2^{-4}  W_1(B_1) P_2
W_3(B_3) +   2^{-4} W_1(B_1) W_2(B_2) P_3 ) T_{oo} \nonumber \\ &+ & \!\! 
2 \times 16  \bigl[2^{-\frac{r_2}{2}-\frac{r_3}{2}} \omega_1 
(P_1 + \omega \, 2^{-2} W_1(B_1) ) T_{go} 
+  2^{-\frac{r_1}{2}-\frac{r_3}{2}} \omega_2 (P_2 + \omega \, 2^{-2} 
W_2(B_2) ) T_{fo} \nonumber \\
 &+ & \!\!  2^{-\frac{r_1}{2}-\frac{r_2}{2}} \omega_3 
(P_3 + \omega \, 2^{-2} W_3(B_3) ) T_{ho} \bigr] 
\left( \frac{\eta}{\vartheta_4} \right)^2 \biggr\} \ ,
\label{a5}
\ea
that project the oriented closed spectra into unoriented ones.  The signs
$\omega_i$ are linked to the discrete torsion through 
the ``crosscap constraint''\cite{croco} by the relation \cite{aadds}
\be
\omega_1 \ \omega_2 \ \omega_3 \ = \omega \quad ,
\label{s4}
\ee
The transverse channel amplitude, obtained performing an  
$S$ modular transformation, is
\ba
\tilde{\cal K}\!\!&=&\!\! \frac{2^5}{8} \biggl\{ 
\bigl( v_1v_2v_3 W^e_1 W^e_2 W^e_3
+  2^{-4} \frac{v_1}{v_2 v_3} W_1^e P_2^e(B_2) P_3^e(B_3) \nonumber \\
\!\!\!\!\! &+&\!\!\!\! 2^{-4} \frac{v_2}{v_1 v_3} P_1^e W_2^e(B_2) 
P_3^e(B_3) + 2^{-4} \frac{v_3}{v_1 v_2} P_1^e(B_1) P_2^e(B_2) 
W_3^e \bigr) \ T_{oo} \nonumber \\
\!\!\!\!\! &+&\!\!\!\! 2 \ \bigl[ \ 2^{-\frac{r_2}{2}-\frac{r_3}{2}}
\omega_1 ( \ v_1 W_1^e + \omega \, 2^{-2} 
\frac{P_1^e(B_1)}{v_1} \ ) \ T_{og}  \nonumber \\
\!\!\!\!\! &+&\!\!\!\!  2^{-\frac{r_1}{2}-\frac{r_3}{2}} \, 
\omega_2 \,  ( \ v_2 W_2^e + \omega \, 2^{-2} \frac{P_2^e(B_2)}{v_2} \ ) \ T_{of}
\nonumber \\
\!\!\!\!\! &+&\!\!\!\!
2^{-\frac{r_1}{2}-\frac{r_2}{2}} \, \omega_3 \, ( \ v_3 W_3^e + \omega \, 2^{-2} 
\frac{P_3^e(B_3)}{v_3} \ ) \ T_{oh} \, \bigr] 
\left( \frac{2
\eta}{\theta_2} \right)^2 \biggr\}  \quad ,
\label{ktraz2z2}
\ea
where the superscript $e$ denotes the usual restriction of the sums
to  even subsets  and the $v_i$ denote the volumes of the
three internal tori. At the origin of the lattices, the reflection 
coefficients are perfect squares,
\ba
\!\!\!\!\! \tilde{\cal K}_0 \!\!&=&\!\! \frac{2^5}{8} \Biggl\{ \, \left( \sqrt{v_1v_2v_3} +
 2^{-\frac{r_2}{2}-\frac{r_3}{2}} \omega_1 \sqrt{\frac{v_1}{v_2 v_3}} + 
 2^{-\frac{r_1}{2}-\frac{r_3}{2}} \omega_2 \sqrt{\frac{v_2}{v_1 v_3}}  
+ 2^{-\frac{r_1}{2}-\frac{r_2}{2}} \omega_3  
\sqrt{\frac{v_3}{v_1 v_2}} \right)^2 \tau_{oo} 
\nonumber \\ 
\!\!\! &+&\!\!\!\! \left( \sqrt{v_1v_2v_3} +
  2^{-\frac{r_2}{2}-\frac{r_3}{2}} \omega_1 \sqrt{\frac{v_1}{v_2 v_3}} - 
2^{-\frac{r_1}{2}-\frac{r_3}{2}} \omega_2  \sqrt{\frac{v_2}{v_1 v_3}}  - 
2^{-\frac{r_1}{2}-\frac{r_2}{2}} \omega_3 
\sqrt{\frac{v_3}{v_1 v_2}} \right)^2 \tau_{og} 
\nonumber \\
\!\!\! &+&\!\!\!\! \left( \sqrt{v_1v_2v_3} -
  2^{-\frac{r_2}{2}-\frac{r_3}{2}} \omega_1 \sqrt{\frac{v_1}{v_2 v_3}} + 
2^{-\frac{r_1}{2}-\frac{r_3}{2}} \omega_2  \sqrt{\frac{v_2}{v_1 v_3}}  - 
2^{-\frac{r_1}{2}-\frac{r_2}{2}} \omega_3 
\sqrt{\frac{v_3}{v_1 v_2}} \right)^2 \tau_{of} 
\nonumber \\
\!\!\! &+&\!\!\!\! \left( \sqrt{v_1v_2v_3} -
  2^{-\frac{r_2}{2}-\frac{r_3}{2}} \omega_1 \sqrt{\frac{v_1}{v_2 v_3}} - 
2^{-\frac{r_1}{2}-\frac{r_3}{2}} \omega_2  \sqrt{\frac{v_2}{v_1 v_3}}  + 
2^{-\frac{r_1}{2}-\frac{r_2}{2}} \omega_3 
\sqrt{\frac{v_3}{v_1 v_2}} \right)^2 \tau_{oh} 
 \, \Biggr\} ,
\label{k0z2z2}
\ea
and encode the presence, together with the conventional Orientifold $9$-planes 
($O9_{+}$-planes from now on), of three kinds of $O5$-planes, 
that we shall denote $O5_{1\alpha}$, $O5_{2\alpha}$ 
and $O5_{3\alpha}$.  These (non-dynamical) planes are fixed under 
the combined action of $\Omega$ and the inversion along the 
directions orthogonal to them, namely $g$ for the 
$O5_{1\alpha}$, $f$ for the $O5_{2\alpha}$ and $h$ for the $O5_{3\alpha}$,  
and the index $\alpha$ reflects their R-R charge. 
We shall use the $+$ sign to indicate $O$-planes 
with tension and R-R charge opposite to the corresponding
quantities for the $D$-branes, and the $-$ sign for the ``exotic''
Orientifold planes with reverted tension and R-R charge.  
As is evident from eq. (\ref{k0z2z2}), the $\omega_i$ 
are proportional to the R-R charges of the $O5_i$.  While
manifestly compatible with the usual positivity requirements, the eight 
different choices reported, for the case with 
$B_{ab} = 0$, in Table \ref{clunz2z2},
affect the tadpole conditions.  In particular, the presence of
``exotic'' $O5_i$  
requires the introduction of antibranes in order to globally neutralize 
the R-R charge of the vacuum configuration.  In this respect, 
according to \cite{bsb}, $\omega=-1$ implies the
reversal of at least one of the $O5$-plane charges, 
producing type I vacua with 
``brane supersymmetry breaking'' \cite{aadds}.  
Moreover, the presence of the NS-NS two
form blocks $B_i$ affects the reflection coefficients in front of a
crosscap by the familiar powers of two, responsible for the rank 
reduction of the Chan-Paton gauge groups \cite{toroidal,carlo}.
 
In this Section, we shall limit ourselves to the discussion of 
the orientifolds of the unique supersymmetric model with $\omega_i=+1$, 
leaving to Section 5 some examples with ``brane supersymmetry breaking''. 
The unoriented closed spectra are reported in 
Table \ref{clunomeg1}, while the annulus amplitude can be written as
\ba
{\cal A} &=& \frac{1}{8} \biggl\{ \bigl( N^2 \ 2^{r - 6} \ P_1(B_1) P_2(B_2) 
P_3(B_3) + 
D_1^2 \ 2^{r_1 - 2} \ P_1(B_1) W_2 W_3 \nonumber \\
 &+& D_2^2 \ 2^{r_2 - 2} 
\ W_1 P_2(B_2) W_3 + D_3^2 \ 2^{r_3 - 2} \ W_1 W_2 P_3(B_3) \bigr) \ T_{oo} 
\nonumber \\
&+& \bigg[ \ 2^{\frac{r_2}{2}+\frac{r_3}{2}} \ 
\bigl( 2 N D_1 \ 2^{r_1 - 2} \ P_1(B_1) + 2 D_2 D_3 \ W_1 \bigr) \ T_{go} 
\nonumber\\
&+& \ 2^{\frac{r_1}{2}+\frac{r_3}{2}} \ 
\bigl( 2 N D_2 \ 2^{r_2 - 2} \ P_2(B_2) + 2 D_1 D_3 \ W_2 \bigr) \ T_{fo} 
\nonumber\\ &+& \ 2^{\frac{r_1}{2}+\frac{r_2}{2}} \   
\bigl( 2 N D_3 \ 2^{r_3 - 2} \ P_3(B_3) + 2 D_1 D_2 \ W_3 \bigr) 
\ T_{ho} \bigg]
\ \left( \frac{\eta}{\theta_4} \right)^2  \biggl\}
\quad ,
\label{az2z2conb}
\ea
where $r=r_1+r_2+r_3$ is the total $B$-rank.  Aside 
from the standard NN open-strings, there are the three
types of open-strings with Dirichlet boundary conditions along two of the 
three internal directions, as well as mixed ND open strings.
The corresponding vacuum-channel amplitude 
displays four independent squared reflection coefficients, related to
the ubiquitous D9-branes on which the NN strings end, 
and to three types of $D5$-branes.  In particular, we call  
$D5_i$-branes those with world-volume that invades the four-dimensional 
space-time and the 
internal $Z^i$ coordinate.  Again, the presence of the NS-NS two form reflects
itself in the generic appearance of additional matter multiplets whose 
multiplicities depend on the rank of the $B_i$-blocks along the directions
orthogonal to the fixed tori.  $N$ and $D_i$ in eq. (\ref{az2z2conb}) 
indicate the traces of the Chan-Paton matrices, or Chan-Paton multiplicities, 
corresponding to the D9 and D5 branes, respectively.

Standard methods \cite{cargese,zori1,ps,review} 
determine the direct-channel M\"obius
amplitude
\ba
{\cal M} &=& - \frac{1}{8} \, \biggl\{ \,
\bigl[\, 2^{\frac{r - 6}{2}} \, N \, P_1(B_1,\g_{\e_1}) \, 
P_2(B_2,\g_{\e_2}) \, P_3(B_3,\g_{\e_3}) \nonumber \\
 &+& 2^{\frac{r_1 - 6}{2}} \, D_1 \, 
P_1(B_1,\g_{\e_1}) W_2(B_2,\tld\g_{\e_2}) W_3(B_3,\tld\g_{\e_3}) \nonumber \\
&+& 2^{\frac{r_2 - 6}{2}} \, D_2 \, 
W_1(B_1,\tld\g_{\e_1}) P_2(B_2,\g_{\e_2}) W_3(B_3,\tld\g_{\e_3}) \nonumber \\
&+& 2^{\frac{r_3 - 6}{2}} \, D_3 \,
W_1(B_1,\tld\g_{\e_1}) W_2(B_2,\tld\g_{\e_2}) P_3(B_3,\g_{\e_3}) 
\, \bigr] \, \hat{T}_{oo}  \nonumber \\
&-& \! \bigl[ \,  2^{\frac{r_1 - 2}{2}} \, (N \!+\! D_1 ) 
P_1(B_1,\g_{\e_1}) \!+\!  2^{-1} \, (D_2 \!+\! D_3) W_1(B_1,\tld\g_{\e_1})
\bigr] \, \hat{T}_{og} 
\left( \frac{2 \hat{\eta}}{\hat{\theta_2}}\right)^2 \nonumber \\
&-& \! \bigl[ \, 2^{\frac{r_2 - 2}{2}} \, (N \!+\! D_2 ) 
P_2(B_2,\g_{\e_2}) \!+\!  2^{-1} \, (D_1 \!+\! D_3) W_2(B_2,\tld\g_{\e_2})
\bigr] \, \hat{T}_{of} 
\left( \frac{2 \hat{\eta}}{\hat{\theta_2}}\right)^2 \nonumber \\
&-& \! \bigl[ \, 2^{\frac{r_3 - 2}{2}} \, (N \!+\! D_3 ) 
P_3(B_3,\g_{\e_3}) \!+\!  2^{-1} \, (D_1 \!+\! D_2) W_3(B_3,\tld\g_{\e_3})
\bigr] \, \hat{T}_{oh} 
\left( \frac{2 \hat{\eta}}{\hat{\theta_2}}\right)^2 \, \biggr\} \ ,
\label{s13}
\ea
where the hatted version of the blocks in eq. (\ref{tiab}) is linked,
as usual, to the choice of a {\it real} basis of characters. 
A proper particle interpretation of the annulus and M\"obius strip 
amplitudes requires  a rescaling of the charges in such a way that 
$N=2n$ and $D_i=2 d_i$.
The (untwisted) tadpole conditions reported in Table \ref{cpz2z2}
emphasize the usual rank reduction due to the presence of 
quantized values of $B_{ab}$ and demand that the signs 
$\gamma_\epsilon$ and $\tilde{\gamma}_\epsilon$ 
satisfy the conditions
\be
\sum_{\epsilon_i=0,1} \gamma_{\epsilon_i} = 2 \, \quad , \qquad \qquad \sum_
{\epsilon_i = 0,1 \, \in {\rm Ker}(B)} 
\tilde{\gamma}_{\epsilon_i} = 2^{(2-r_i)/2} \, .
\label{transversegamma}
\ee  
There are several solutions for the allowed gauge groups, that depend on the
additional signs $\xi_i$ and $\eta_i$ defined by 
\be
\sum_{\epsilon_i=0,1 } 
\tilde{\gamma}_{\epsilon_i} =  2 \, \xi_i \, \quad , \qquad \qquad  
\sum_{\epsilon_i = 0,1 \, \in {\rm Ker}(B)} 
{\gamma}_{\epsilon_i} = 2^{(2-r_i)/2} \, \eta_i 
   \, . \label{directgamma}
\ee
As shown in Table \ref{cpz2z2}, they are products of four factors, 
chosen to be $USp$ or $SO$ depending on the values of $\xi_i$ and $\eta_i$.
The massless unoriented open spectra, encoded in the annulus and
M\"obius  amplitudes at the lattice origin, 
\ba 
{\cal A}_0 &=& {\frac{1}{2}} \ [ \, ( \, n^2 + d_1^2 +d_2^2 + d_3^2 
\, ) \, ( \tau_{oo} +
\tau_{og} + \tau_{oh} + \tau_{of} ) \nonumber \\ 
&+&
2^{\frac{r_2}{2}+\frac{r_3}{2}} \ 
(2 n d_1 + 2 d_2 d_3 ) ( \tau_{go} +
\tau_{gg} + \tau_{gh} + \tau_{gf} )
\nonumber \\  
&+& 
2^{\frac{r_1}{2}+\frac{r_3}{2}} \ (2 n d_2 + 2 d_1 d_3 ) ( \tau_{fo} +
\tau_{fg} + \tau_{fh} + \tau_{ff} ) \nonumber \\ 
&+& 
2^{\frac{r_1}{2}+\frac{r_2}{2}} \ 
(2 n d_3 + 2 d_1 d_2 ) ( \tau_{ho} +
\tau_{hg} + \tau_{hh} + \tau_{hf} )\, ]
\label{a0z2z2}
\ea
and
\ba
{\cal M}_0   &=&  - \frac{1}{2} \ \biggl[ \  \tau_{oo} \  [ \frac{n}{2} \,  
(\eta_1 \eta_2 \eta_3 -  \eta_1 - \eta_2 -  \eta_3) + \frac{d_1}{2} ( \,
\eta_1 \xi_2 \xi_3 -  \eta_1 - \xi_2 -  \xi_3) \nonumber \\
&+& \frac{d_2}{2} \, 
(\xi_1 \eta_2 \xi_3 -  \xi_1 - \eta_2 -  \xi_3) +  \frac{d_3}{2} ( \,
\xi_1 \xi_2 \eta_3 -  \xi_1 - \xi_2 -  \eta_3)] \nonumber \\
&+& \tau_{og} \ [ \frac{n}{2} \, 
(\eta_1 \eta_2 \eta_3 -  \eta_1 + \eta_2 +  \eta_3) + \frac{d_1}{2} ( \,
\eta_1 \xi_2 \xi_3 -  \eta_1 + \xi_2 +  \xi_3) \nonumber \\
&+& \frac{d_2}{2} \,
(\xi_1 \eta_2 \xi_3 -  \xi_1 + \eta_2 +  \xi_3) +  \frac{d_3}{2} ( \,
\xi_1 \xi_2 \eta_3 -  \xi_1 + \xi_2 + \eta_3)] \nonumber \\
&+& \tau_{of} \ [ \frac{n}{2} \,
(\eta_1 \eta_2 \eta_3 +  \eta_1 - \eta_2 +  \eta_3) + \frac{d_1}{2} ( \,
\eta_1 \xi_2 \xi_3 +  \eta_1 - \xi_2 +  \xi_3) \nonumber \\
&+& \frac{d_2}{2} \,
(\xi_1 \eta_2 \xi_3 +  \xi_1 - \eta_2 +  \xi_3) +  \frac{d_3}{2} ( \,
\xi_1 \xi_2 \eta_3 +  \xi_1 - \xi_2 + \eta_3)] \nonumber \\
&+& \tau_{oh} [ \frac{n}{2} \,
(\eta_1 \eta_2 \eta_3 +  \eta_1 + \eta_2 - \eta_3) + \frac{d_1}{2} ( \,
\eta_1 \xi_2 \xi_3 +  \eta_1 + \xi_2 - \xi_3) \nonumber \\
&+& \frac{d_2}{2} \,
(\xi_1 \eta_2 \xi_3 + \xi_1 + \eta_2 - \xi_3) +  \frac{d_3}{2} ( \,
\xi_1 \xi_2 \eta_3 +  \xi_1 + \xi_2 - \eta_3)] \biggr] \, ,
\label{m0z2z2}
\ea
are reported in Table \ref{ouz2z2}. Being non chiral, these models 
are clearly free of anomalies.

\subsection{$Z_2 \times Z_2$ Shift-orientifolds and Brane Supersymmetry}

In this Section we review how $(\delta_L,\delta_R)$ shifts 
can be combined with $Z_2 \times Z_2$ orbifold 
operations in the open descendants of type IIB compactifications. 
As in \cite{shift,aadds}, we
shall distinguish between symmetric {\it momentum} shifts 
$(p)=(\delta,\delta)$
and antisymmetric {\it winding} shifts 
$(w)=(\delta,-\delta)$, since the two  
have very different effects on the resulting spectra. 
These orbifolds correspond to singular limits of Calabi-Yau manifolds
with Hodge numbers $(19,19)$, $(11,11)$ and $(3,3)$ in the cases of
one, two and three shifts, respectively, as shown in Table \ref{clorz2z2sh}.
Let us begin by introducing a convenient notation
to specify the orbifold action $Z_i \to \sigma(Z_i)$ on 
the complex coordinates of the three internal tori. 
There are several ways to combine the three 
operations $g$, $f$ and $h$ of the matrix (\ref{sigmazero})
with shifts consistently with the $Z_2 \times Z_2$ group
structure. However, up to T-dualities and corresponding redefinitions
of the $\Omega$ projection, all non-trivial possibilities are captured 
by \cite{shift}
\ba
\sigma_1(\delta_1,\delta_2,\delta_3) = \left( \begin{array}{rrrr}
\delta_1 & -\delta_2 & -1 \\
-1 & \delta_2 & -\delta_3 \\ -\delta_1 & -1 & \delta_3 \end{array}
\right) \quad , \qquad
\sigma_2(\delta_1,\delta_2,\delta_3) = \left( \begin{array}{rrrr} 
\delta_1 & -1 & -1 \\
-1 & \delta_2 & -\delta_3 \\ -\delta_1 & -\delta_2 & \delta_3 \end{array} 
\right)
\ ,
\label{sigmas}
\ea
where the three lines refer to the new operations,
that we shall continue to denote by $g$, $f$ and $h$, and where
$-\delta_i$ indicates the combination of a shift in the
real part of the $i$-th coordinate 
with the orbifold inversion.  Notice that when a
line of the table contains $p$ or $-w$, the corresponding $D5$-branes are 
eliminated.  
One thus obtains the ten different classes of 
models reported in Table \ref{oldmodels}, with 
the $w_2p_3$ model now linked to the $\sigma_1$ action, 
correcting a misprint in ref. \cite{shift}.  
In listing these models, we have followed the 
choices of axes 
made in ref. \cite{shift}, so that when a single set of $D5$ branes 
is present, this is always
the first, $D5_1$, and when two sets are present, these are always
$D5_1$ and $D5_2$. All these freely acting orientifolds 
have $N=1$ supersymmetry in the closed sector, but exhibit interesting
instances of ``brane supersymmetry'' in the open part:  
additional 
supersymmetries are present 
for their massless modes, that in some cases extend also to the
massive ones \cite{ssopen} confined to some branes.  
\begin{table}[ht]
\begin{center}
\begin{tabular}{||c|c|c|c|c||}\hline\hline
{\rm models}&{\rm shift}&{$D9$ susy}&{$D5_{1}$ susy}&{$D5_{2}$ susy} 
\\ \hline\hline
{$p_3$}&{$\sigma_1$}&{N=1}&{N=2} &{N=2}\\
{$w_2p_3$}&{$\sigma_1$}&{N=2}&{N=2} &{N=4}\\ 
{$w_1w_2p_3$}&{$\sigma_2$}&{N=4}&{N=4}&{N=4} \\  \hline
{$p_2 p_3$}&{$\sigma_2$}&{N=1}&{N=2}&{--} \\
{$w_1p_2$}&{$\sigma_2$}&{N=2}&{N=4}&{--} \\
{$w_1p_2p_3$}&{$\sigma_2$}&{N=2}&{N=4}&{--} \\ 
{$w_1p_2w_3$}&{$\sigma_1$}&{N=4}&{N=4}&{--} \\ \hline
{$p_1 p_2 p_3$}&{$\sigma_1$}&{N=1}&{--}&{--} \\  
{$p_1w_2w_3$}&{$\sigma_2$}&{N=2}&{--} &{--}\\ 
{$w_1w_2w_3$}&{$\sigma_1$}&{N=4}&{--}&{--}
\\ \hline\hline
\end{tabular} 
\caption{Shifts and ``brane supersymmetry'' for the various models.}
\label{oldmodels}
\end{center}
\end{table}
Table \ref{oldmodels} 
also collects the number of 
supersymmetries of the massless modes for the various branes present 
in each model. 
The unoriented truncations and the open spectra are generically affected
by the shifts,  that lift in mass some tree-level closed string 
terms eliminating the corresponding tadpoles, and determine the
brane content of the models, related to the presence of the projectors
\ba
\Pi_1 &\sim& 1 + (-1)^{\delta_1+ \delta_2} + (-1)^{\delta_2+ \delta_3} + 
(-1)^{\delta_1+ \delta_3} \quad ,\\
\Pi_2 &\sim& 1 + (-1)^{\delta_1} + (-1)^{\delta_2+\delta_3} + 
(-1)^{\delta_1+ \delta_2+ \delta_3} \quad ,
\label{g8}
\ea
for the $\sigma_1$ and $\sigma_2$ tables respectively, along the tube.  
   
The open-string spectra of the models in Table \ref{oldmodels} 
are shown in Table \ref{opunundef}.
They correspond to peculiar and interesting brane configurations, 
related to the fact that some projections are absent in the
NN or $D9-D9$ string contributions, as well as in the DD or $D5-D5$ string
contributions.  These features 
admit a nice geometrical interpretation: they are linked 
to the presence of multiplets of branes, associated with 
multiplets of tori fixed by
some $Z_2 \times Z_2$ elements and interchanged by the action of the 
remaining ones.  
Only the  
projections introduced by the former elements are thus present, since in
these sectors the physical states are combinations of multiplets 
localized on the image branes.  
If one attempts to insert all branes 
at a fixed point, the other operations inevitably move them, giving rise to 
multiple images.  Equivalently, as already discussed 
in Section 2.1, brane multiplets may be traced to 
the presence of {\it momentum} shifts orthogonal to 
the branes \cite{ssopen}.  
\begin{figure}[ht]
\begin{center}
\input epsf \centerline{ \epsfbox{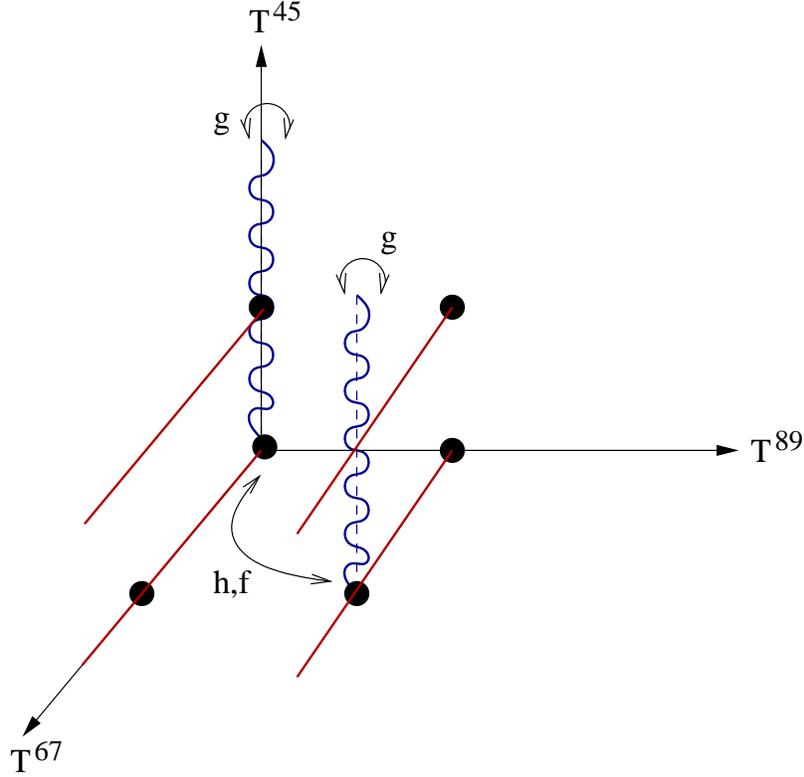}}
\caption{$D5_1$ and $D5_2$ brane configurations for the $w_2 p_3$ model}
\label{doppietti}
\end{center}
\end{figure}
As a 
consequence, the massless modes (or 
at times the full spectrum) exhibit enhanced supersymmetry.
Figure \ref{doppietti} displays the brane
configuration of the $w_2p_3$ model. 
The three axes refer to the three 
two-tori $T^{45}$, $T^{67}$ and $T^{89}$,
and the $D5_1$ branes, drawn as wavy blue lines, occupy a pair of fixed tori, 
while the $D5_2$ branes, the solid red lines, occupy
all the four fixed tori along the $Z^2$ direction.  
The $D5_1-D5_1$ configurations correspond 
to doublets of branes, associated to the pair of tori fixed by
$g$ and interchanged by $f$ and $h$.  As expected and as shown in Table 
\ref{oldmodels}, there is an 
$N=2$ supersymmetry associated to the $D5_1$ brane doublets together with an
$N=4$ supersymmetry associated to the quartets of $D5_2$ branes.

\section{Magnetic Deformations and Brane Transmutation}

A uniform background magnetic field $H_i$ on the $i$-th 
torus is a monopole field, 
a $U(1)$ 
bundle with non-trivial transition functions gluing two local charts, 
whose consistency with particle dynamics requires a  
\emph{Dirac quantization condition}
\be
2 \pi \alpha^{\prime} q_i H_i = \frac{k_i}{v_i} \qquad ,
\label{dirac}
\ee 
where $q_i$ is the  $U(1)$-charge, $v_i$ is the volume of the torus and
$k_i$ is an integer defining the (quantized) number of elementary 
fluxons or, equivalently, the Landau-level degeneracy.  
The background magnetic field couples solely to the open string ends 
through a boundary term in the action \cite{ft}. 
Thus, denoting with $q_L$ and $q_R$ the 
$U(1)$-charges at the two ends of the open strings and defining 
the total charge $Q = q_L + q_R$, one has to distinguish 
between ``neutral'', $Q=0$, sectors and ``charged'', $Q \ne 0$, ones.
Charged open strings with $Q = \pm 1$ or $Q = \pm 2$ 
are characterized by oscillators whose frequencies are shifted 
according to 
\be
z_i = \frac{1}{\pi} [\ arctg (2 \pi  \alpha^{\prime} q_L H_i) + 
arctg (2 \pi  \alpha^{\prime} q_R H_i) \ ] \ .
\ee
Moreover, there is a Landau-level degeneracy that exactly 
amounts to the $k_i$ factor 
in eq. (\ref{dirac}) and reflects the fact that 
the zero-modes along the 
two directions of the torus do not commute.  Moreover, on
a $Z_N$-orbifold 
the unit cell has a volume reduced by a factor  $N$ with respect to the 
torus.  As a result, the degeneracies of Landau levels should be multiples
of $N$.  For instance, $k_i$ are typically even on the $Z_2$-orbifold.  
However, a non-vanishing quantized NS-NS $B_{ab}$ could make 
some orbifold projections uneffective, thus allowing more choices 
for the $k_i$  \cite{magnbsb}.
On the other hand, neutral ``dipole'' strings
with $q_L = - q_R$ ($Q = 0$), have integer-mode frequencies, 
but the structure of their zero-modes is rather peculiar.  Indeed, 
both momenta and windings are now allowed, but the effect of 
the  magnetic field on this sector is simply to introduce rescalings 
in the momentum and winding lattices entering the one-loop 
annulus partition function.  
An analysis of the resulting projector along the tube shows that 
in the open one-loop channel the lattice sums over momenta and 
windings are indeed subjected to the ``complex boosts''   
\be
m_i \ra \frac{m_i}{\sqrt{1 + ( 2 \pi  \alpha^{\prime} q_i H_i)^2}}\ , \quad 
n_i \ra n_i \sqrt{1 + ( 2 \pi  \alpha^{\prime} q_i H_i)^2} \ .  
\ee
Let us stress that  a uniform Abelian magnetic field can
be given a dual interpretation in terms of rotated branes \cite{intbranes}.  
In particular, 
after a T-duality along one of the two directions of one torus, 
the modified boundary conditions can be rephrased in terms of rotations 
of the corresponding $D$-branes by  angles $\theta_i$ 
such that  
$$
\tan{\theta_i} = 2 \pi  \alpha^{\prime} q_i H_i \quad.
$$
The Dirac quantization condition in eq. (\ref{dirac}) can then 
be interpreted as the 
requirement that the D-branes wrap exactly $k_i$-times the
tori. Notice that we always normalize the electric charge of the 
open-strings in such a way that it corresponds to the elementary quantum.
This is the reason why we can describe all spectra using just one
integer $k_i$ for each two-torus or, in other words, 
setting to one the electric winding number of the corresponding $D$-branes.

Let us briefly review how these configurations manifest themselves in the 
orientifold of type IIB on a magnetized 
$(T^2 \times T^2) / Z_2$ \cite{aads}.  This is a deformation of 
the six-dimensional $(T^2 \times T^2) / Z_2$ \cite{zori1,usedici} 
model, whose massless 
oriented closed spectrum is reported in Table \ref{coz2d6} and 
comprises $N=(2,0)$ supergravity coupled to $21$ tensor multiplets, as 
expected for a singular limit of a $K_3$ compactification.
The unoriented spectra, not affected by the constant magnetic backgrounds,  
are reported in table \ref{cuz2d6}
and exhibit at zero mass $N=(1,0)$ supergravity modes coupled
to hypermultiplets and tensor multiplets whose numbers depend upon 
the total rank $r$ of $B_{ab}$.  
In the open sector, the two magnetic fields $H_1$ and 
$H_2$ turned on inside 
the two $T^2$'s are aligned along the same $U(1)$ subgroup 
of the Chan-Paton gauge group. 
Generic fluxes of  $H_1$ and 
$H_2$ produce the breaking of supersymmetry and the presence of  
Nielsen-Olesen instabilities, signaled by the appearance of 
tachyonic states.  Absence of tachyons and supersymmetry can be 
attained choosing self-dual field configurations, 
that at the same time allow to compensate 
the non-vanishing instanton density with additional 
lower dimensional branes.  
As a result, the magnetized D9-branes mimic the behaviour of the D5-branes 
and couple to the R-R six-form, contributing to the tadpole cancellation 
conditions.  This is the ``brane transmutation'' phenomenon, first described,
in this context, in ref. \cite{aads} and linked there to the peculiar
Wess-Zumino couplings in the D-brane actions \cite{bdl}.
  There are several solutions,
reported in Tables \ref{cpz2d6}, \ref{ouz2d6}, \ref{cpz2d6r}, \ref{ouz2d6r} 
(see Appendix C for notations and conventions used in the Tables), 
depending on the $O$-plane configurations, 
or equivalently on the signs, 
analogous to the ones in Eq. (\ref{directgamma}), 
contained in the M\"obius-strip amplitudes.  In particular
 the R-R tadpole cancellation 
conditions and the resulting 
Chan-Paton gauge groups are shown in Table \ref{cpz2d6} 
for the models with complex charges and 
in Table \ref{cpz2d6r} for the models with real charges.  
Moreover, the untwisted 
NS-NS tadpoles are related to the derivatives of the Born-Infeld action
with respect to the untwisted moduli, while the twisted ones are associated 
with corresponding couplings in the effective Lagrangian, as in 
\cite{aads,magnbsb}.  
The resulting
open spectra, reported in Table \ref{ouz2d6} for complex charges and
in Table \ref{ouz2d6r} for real charges, can be easily recognized 
as deformations of the models without background magnetic fields 
\cite{zori1,usedici}.  
Several interesting facts, however, emerge from the analysis of 
 Tables \ref{ouz2d6} and \ref{ouz2d6r}.  First, there is an unusual rank
reduction of the Chan-Paton gauge group.  Second, some 
matter multiplets appear in multiple families, because of the degeneracy
introduced by the Landau levels.  Third, as already stressed 
the magnetized D9-branes 
behave exactly like D5-branes.  This is particularly evident for
the models {\it without} D5-branes, related for instance to the choice
$r=0$, $k_2 = k_3 = 2$, 
$d = 0$, $n=12$ and $m=4$ in Table \ref{ouz2d6}, and 
 corresponds \cite{smallins} to the fact that the D5-branes can be 
interpreted as instantons of vanishing size.  So, in the presence of 
self-dual configurations of the internal magnetic fields, a stack of 
$2^{\frac{r}{2}}|k_2 k_3|$ D5-branes is replaced
by a ``fat'' instanton that invades the whole ten-dimensional space-time, in a
transition related by $T$-duality \cite{csu} to the inverse 
small-instanton transition discussed in \cite{smallins}.
Finally, introducing antiself-dual configurations for the magnetic field,
the magnetized $D9$-branes mimic anti-$D5$-branes \cite{magnbsb}.  
Tachyons are 
again eliminated, but supersymmetry is broken at tree level in the 
open sector at the string scale or, better, is 
non-linearly realized in the anti $D5$-brane sector, as 
discussed in refs. \cite{dudmo,prari}.

\section{Magnetized Four-Dimensional Models}

The four-dimensional models developed in refs. \cite{shift,aadds} display 
$D9$ and $D5$ branes in their perturbative spectra, and are thus a
natural arena to build consistent magnetized models 
sharing the qualitative features of the six-dimensional 
$T^4 / Z_2$ model.  As we shall see, their spectra exhibit rank reductions 
of the gauge groups and multiple matter families.  In addition, 
as in the six-dimensional examples, there is the option of introducing 
pairs of magnetic fields aligned along the same $U(1)$ subgroup.  
This is allowed only if the undeformed model contains corresponding 
$O5$-planes orthogonal
to the two magnetized directions, or, equivalently, if there are sources 
that add to the magnetized $D9$-branes 
in such a way as to compensate the R-R charge excess.  
We shall always 
introduce uniform Abelian background magnetic fields $(H_2, H_3)$ 
along the $(Z^2,Z^3)$ 
directions, thus requiring just the presence of $O5_1$-planes to 
balance the
R-R charge excess of magnetized $D9$-branes and $D5_1$-branes.

If $D5$-branes whose world-volume invades
coordinates longitudinal to the magnetized directions are 
also present, 
one obtains, as a bonus, chiral fermions.  
Chirality is connected on the one hand
to the intersection of two sets of orthogonal $D5$-branes, and 
on the other 
to the chiral asymmetry in the ``pure magnetic'' sector.
Moreover, the phenomenon of brane transmutation acquires in this setting 
its full-fledged form.  Indeed, as stressed in Section 2, shift-orientifolds
are characterized by the presence of multiplets of defocalized D5-branes.
This implies, as mentioned before, 
that some of the D5-branes cannot be put on the same
fixed tori but have to be distributed among the images interchanged  
by the action of some orbifold group elements.  The magnetized
D9-branes ``remember'' the localized distribution of the D5-branes they 
are mimicking. Indeed, although they
invade the whole internal space, the centers of the  corresponding 
 classical Landau orbits
organize themselves in multiplets that reflect 
the structure of the D5-branes.

\subsection{Magnetized $Z_2 \times Z_2$ Orientifolds}

Let us begin the discussion of magnetic deformations by 
considering the 
``plain'' $Z_2 \times Z_2$ model with $\omega_i = 1$ (for related examples 
in the language of intersecting $D$-brane models, see \cite{csu}).
As stated in Section 2, the remaining models in Table 
\ref{clunz2z2} give rise to 
orientifolds with ``brane supersymmetry breaking'', that, 
for brevity, will not be explicitly discussed
here.  However, in Section 5 we shall describe 
one class of orientifolds with ``brane supersymmetry 
breaking'' related to a class of $w_2 p_3$ shift-orbifold
models. 

The closed string amplitudes, not affected by the introduction of 
 constant background magnetic fields $H_2$ and $H_3$ 
along the $Z^2$ and $Z^3$ directions, as well as the $O$-plane
content, are described in Section 2.2. 
In addition, the closed unoriented spectra 
are collected in Table \ref{clunz2z2}.  
The annulus amplitude can be 
obtained using the techniques of \cite{aads}, reviewed 
in Section 3, as a deformation of the annulus 
amplitudes of eq. (\ref{az2z2conb}). The result can be
cast into the sum of the following  
three contributions (for notations and conventions see the Appendices)
\bea
\mc{A}_{(Q=0)} & = & \frac{1}{8} \biggl\{ \bigl( N^2 \ 2^{r - 6} \ 
P_1(B_1) P_2(B_2) P_3(B_3) + 2 m \bar{m}  \ 2^{r - 6} \ P_1(B_1) 
\tld{P}_2(B_2) \tld{P}_3(B_3) \nonumber \\
&+& D_1^2 \ 2^{r_1 - 2} \ P_1(B_1) W_2 W_3 + 
D_2^2 \ 2^{r_2 - 2} \ W_1 P_2(B_2) W_3 \nonumber \\ &+& D_3^2 \ 2^{r_3 - 2} 
\ W_1 W_2 P_3(B_3) \bigr) 
T_{oo}(0;0;0) \nonumber \\
&+&  2^{\frac{r_2}{2}+\frac{r_3}{2}} \ T_{go}(0;0;0)  \bigg[ 
2 N D_1 \ 2^{r_1 - 2} \ P_1(B_1) + 2 D_2 D_3 \ W_1 \bigg] \, 
\Big( \frac{ \eta}{\vth_4(0)} \Big)^2  \nonumber \\
&+&  2^{\frac{r_1}{2}+\frac{r_3}{2}} \ T_{fo}(0;0;0)  \bigg[ 
2 N D_2 \ 2^{r_2 - 2} \ P_2(B_2) + 2 D_1 D_3 \ W_2 \bigg] \, 
\Big( \frac{ \eta}{\vth_4(0)} \Big)^2  \nonumber \\
&+&  2^{\frac{r_1}{2}+\frac{r_2}{2}} \ T_{ho}(0;0;0)  \bigg[ 
2 N D_3 \ 2^{r_3 - 2} \ P_3(B_3) + 2 D_1 D_2 \ W_3 \bigg] \, 
\Big( \frac{ \eta}{\vth_4(0)} \Big)^2  \biggl\} 
\eea
for the zero-charge sectors, 
\bea
\mc{A}_{(Q=1)} & = & \frac{1}{8} \biggl\{ - 2^{r - 2} 2 m N 
T_{oo}(0;z_2 \t;z_3 \t) P_1(B_1) \  
\frac{k_2 \eta}{\vth_1(z_2 \t)} \ \frac{k_3 \eta}{\vth_1(z_3 \t)} 
\nonumber \\
&-& 2^{r - 2} 2 \bar{m} N T_{oo}(0;- z_2 \t;- z_3 \t) P_1(B_1) \  
\frac{k_2 \eta}{\vth_1(- z_2 \t)} \ \frac{k_3 \eta}{\vth_1(- z_3 \t)} 
\nonumber \\
&+&  2^{\frac{r_2}{2}+\frac{r_3}{2}} \ T_{go}(0;z_2 \t;z_3 \t) \bigg[ 
2 m D_1 \ 2^{r_1 - 2} \ P_1(B_1) \bigg] \, 
 \frac{ \eta}{\vth_4(z_2 \t)} \  \frac{ \eta}{\vth_4(z_3 \t)} \nonumber \\ 
&+&  2^{\frac{r_2}{2}+\frac{r_3}{2}} \ T_{go}(0;-z_2 \t;-z_3 \t) \bigg[ 
2 \bar{m} D_1 \ 2^{r_1 - 2} \ P_1(B_1) \bigg] \, 
 \frac{ \eta}{\vth_4(-z_2 \t)} \  \frac{ \eta}{\vth_4(-z_3 \t)} \nonumber \\ 
&+&  2^{\frac{r + r_2}{2}} \ T_{fo}(0;z_2 \t;z_3 \t) \bigg[ 
- 2 i m D_2 \bigg] \, \frac{ \eta}{\vth_4(0)} \ 
 \frac{ k_2 \eta}{\vth_1(z_2 \t)} \  \frac{ \eta}{\vth_4(z_3 \t)} \nonumber \\ 
&+&  2^{\frac{r + r_2}{2}} \ T_{fo}(0;-z_2 \t;-z_3 \t) \bigg[ 
2 i \bar{m} D_2 \bigg] \, \frac{ \eta}{\vth_4(0)} \ 
 \frac{ k_2 \eta}{\vth_1(-z_2 \t)} \ \frac{ \eta}{\vth_4(-z_3 \t)} 
\nonumber \\ 
&+&  2^{\frac{r + r_3}{2}} \ T_{ho}(0;z_2 \t;z_3 \t) \bigg[ 
- 2 i m D_3 \bigg] \, \frac{ \eta}{\vth_4(0)} \ 
 \frac{\eta}{\vth_4(z_2 \t)} \  \frac{k_3 \eta}{\vth_1(z_3 \t)} \nonumber \\ 
&+&  2^{\frac{r + r_3}{2}} \ T_{ho}(0;-z_2 \t;-z_3 \t) \bigg[ 
2 i \bar{m} D_3 \bigg] \, \frac{ \eta}{\vth_4(0)} \ 
 \frac{\eta}{\vth_4(-z_2 \t)} \ \frac{k_3 \eta}{\vth_1(-z_3 \t)} \biggl\}  
\eea
for the charge-one sectors, and 
\bea
\mc{A}_{(Q=2)} & = & \frac{1}{8} \biggl\{- 2^{r - 2} m^2 
T_{oo}(0;2 z_2 \t;2 z_3 \t) P_1(B_1) \ 
\frac{2 k_2 \eta}{\vth_1(2 z_2 \t)} \ \frac{2 k_3 \eta}{\vth_1(z_3 \t)} 
\nonumber \\
&-& 2^{r - 2} \bar{m}^2  T_{oo}(0;- 2 z_2 \t;- 2 z_3 \t) P_1(B_1) \  
\frac{2 k_2 \eta}{\vth_1(- 2 z_2 \t)} \ \frac{2 k_3 \eta}{\vth_1(- 2 z_3 \t)} 
\biggl\} 
\eea
for the total charge-two sectors. 
The M\"obius amplitude can be obtained 
in a similar way from the undeformed case, distinguishing only the 
uncharged ($Q=0$) sector from the charged ($Q=2$) one, since the 
$Q=1$ sector is absent in $\mc{M}$ because of the oriented nature of the
corresponding open-strings. Thus
\bea 
\mc{M}_{(Q=0)} \!\!&=&\!\!\! - \frac{1}{8} \biggl\{ \,
\bigl[\, 2^{\frac{r - 6}{2}} \, N \, P_1(B_1,\g_{\e_1}) \, 
P_2(B_2,\g_{\e_2}) \, P_3(B_3,\g_{\e_3}) \nonumber \\
   &+& 2^{\frac{r_1 - 6}{2}} \, D_1 \, 
P_1(B_1,\g_{\e_1}) W_2(B_2,\tld\g_{\e_2}) W_3(B_3,\tld\g_{\e_3}) \nonumber \\
   &+& 2^{\frac{r_2 - 6}{2}} \, D_2 \, 
W_1(B_1,\tld\g_{\e_1}) P_2(B_2,\g_{\e_2}) W_3(B_3,\tld\g_{\e_3}) \nonumber \\
   &+& 2^{\frac{r_3 - 6}{2}} \, D_3 \,
W_1(B_1,\tld\g_{\e_1}) W_2(B_2,\tld\g_{\e_2}) P_3(B_3,\g_{\e_3}) 
\, \bigr] \, \hat{T}_{oo}(0;0;0)\\
&-& \! \bigl[ \,  2^{\frac{r_1 - 2}{2}} \, (N \!+\! D_1 ) 
P_1(B_1,\g_{\e_1}) \!+\!  2^{-1} \, (D_2 \!+\! D_3) W_1(B_1,\tld\g_{\e_1})
\bigr] \, \hat{T}_{og}(0;0;0) 
\left( \frac{2 \hat{\eta}}{\hat{\theta_2}}\right)^2 \nonumber \\ 
&-& \! \bigl[ \, 2^{\frac{r_2 - 2}{2}} \, (N \!+\! D_2 ) 
P_2(B_2,\g_{\e_2}) \!+\!  2^{-1} \, (D_1 \!+\! D_3) W_2(B_2,\tld\g_{\e_2})
\bigr] \, \hat{T}_{of}(0;0;0) 
\left( \frac{2 \hat{\eta}}{\hat{\theta_2}}\right)^2 \nonumber \\ 
&-& \! \bigl[ \, 2^{\frac{r_3 - 2}{2}} \, (N \!+\! D_3 ) 
P_3(B_3,\g_{\e_3}) \!+\!  2^{-1} \, (D_1 \!+\! D_2) W_3(B_3,\tld\g_{\e_3})
\bigr] \, \hat{T}_{oh}(0;0;0) \left( \frac{2 \hat{\eta}}{\hat{\theta_2}}\right)^2 \biggr\} \quad , \nonumber
\eea
and
\bea 
\mc{M}_{(Q=2)} \!\!&=&\!\!\ - \frac{1}{8} \biggl\{ \,
- 2^{\frac{r - 2}{2}} m \hat{T}_{oo}(0;2 z_2 \t;2 z_3 \t) \, 
P_1(B_1,\g_{\e_1}) \, \frac{2 k_2 \hat{\eta}}{\hat{\vth}_1(2 z_2 \t)} 
\frac{2 k_3 \hat{\eta}}{\hat{\vth}_1(2 z_3 \t)}  \nonumber \\
&-& 2^{\frac{r - 2}{2}} \bar{m} \hat{T}_{oo}(0;-2 z_2 \t;-2 z_3 \t) \, 
P_1(B_1,\g_{\e_1}) \, \frac{2 k_2 \hat{\eta}}{\hat{\vth}_1(-2 z_2 \t)} 
\frac{2 k_3 \hat{\eta}}{\hat{\vth}_1(-2 z_3 \t)}  \nonumber \\
&-& 2^{\frac{r_1 - 2}{2}} m \hat{T}_{og}(0;2 z_2 \t;2 z_3 \t) \, 
P_1(B_1,\g_{\e_1}) \, \frac{2 \hat{\eta}}{\hat{\vth}_2(2 z_2 \t)} 
\frac{2 \hat{\eta}}{\hat{\vth}_2(2 z_3 \t)}  \nonumber \\
&-& 2^{\frac{r_1 - 2}{2}} \bar{m} \hat{T}_{og}(0;-2 z_2 \t;-2 z_3 \t) \, 
P_1(B_1,\g_{\e_1}) \, \frac{2 \hat{\eta}}{\hat{\vth}_2(-2 z_2 \t)} 
\frac{2 \hat{\eta}}{\hat{\vth}_2(-2 z_3 \t)}  \nonumber \\
&+& i \, m \, 2^{\frac{r_2}{2}} \, \hat{T}_{of}(0;2 z_2 \t;2 z_3 \t) \, 
\frac{2 \hat{\eta}}{\hat{\vth}_2(0)}\, 
\frac{2 k_2 \hat{\eta}}{\hat{\vth}_1(2 z_2 \t)} 
\frac{2 \hat{\eta}}{\hat{\vth}_2(2 z_3 \t)}  \nonumber \\
&-& i \, \bar{m} \, 
2^{\frac{r_2}{2}} \, \hat{T}_{of}(0;-2 z_2 \t;-2 z_3 \t) \, 
\frac{2 \hat{\eta}}{\hat{\vth}_2(0)}\, 
\frac{2 k_2 \hat{\eta}}{\hat{\vth}_1(-2 z_2 \t)} 
\frac{2 \hat{\eta}}{\hat{\vth}_2(-2 z_3 \t)}  \nonumber \\
&+& i \, m \, 2^{\frac{r_3}{2}} \, \hat{T}_{oh}(0;2 z_2 \t;2 z_3 \t) \, 
\frac{2 \hat{\eta}}{\hat{\vth}_2(0)}\, 
\frac{2 \hat{\eta}}{\hat{\vth}_2(2 z_2 \t)} 
\frac{2 k_3 \hat{\eta}}{\hat{\vth}_1(2 z_3 \t)}  \nonumber \\
&-& i \, \bar{m} \, 
2^{\frac{r_3}{2}} \, \hat{T}_{oh}(0;-2 z_2 \t;-2 z_3 \t) \, 
\frac{2 \hat{\eta}}{\hat{\vth}_2(0)}\, 
\frac{2 \hat{\eta}}{\hat{\vth}_2(-2 z_2 \t)} 
\frac{2 k_3 \hat{\eta}}{\hat{\vth}_1(-2 z_3 \t)} \biggr\} \quad .
\eea
These amplitudes describe the couplings of conventional $D9$ 
and $D5_i$ branes with an additional set of (magnetized) $D9$-branes.
This natural interpretation is encoded into the four $o$, $g$, $f$ 
and $h$ projections, 
present in the ($Q=2$) 
sector of the M\"obius 
amplitudes.
The tadpole cancellation conditions may be extracted combining 
the transverse (tree)
channel Klein-bottle amplitude of eq. (\ref{ktraz2z2}) with the
transverse annulus amplitudes
\bea
\tld{\cal{A}} & = & \frac{2^{-5}}{8} \ \biggl\{ \ [ 2^{r-6} v_1 v_2 v_3 
N^2 W_1(B_1) W_2(B_2) W_3(B_3) \ + \ \frac{v_1}{v_2 v_3} \ 2^{r_1-2} \ 
D_1^2 W_1(B_1) P_2 P_3 \nonumber \\ 
&+& \frac{v_2}{v_1 v_3}  \ 2^{r_2-2} \ D_2^2 P_1 W_2(B_2) P_3 \ + \ 
\frac{v_3}{v_1 v_2}  \ 2^{r_3-2} \ D_3^2 P_1 P_2 W_3(B_3) \nonumber \\
&+& 2^{r-6} v_1 v_2 v_3 \  2 m \bar{m} 
(1+q^2 H_2^2) (1+q^2 H_3^2) W_1(B_1) \tld{W_2}(B_2) \tld{W_3}(B_3)  \ ] \  
T_{oo}(0;0;0) \nonumber \\
            & + &  2^{r - 2} \ 8 \ N \ 
[ m T_{oo}(0;z_2;z_3)  +   \bar{m} 
T_{oo}(0;-z_2;-z_3) ] v_1 W_1(B_1)\  
\frac{k_2 \eta}{\vth_1(z_2)} \frac{k_3 \eta}{\vth_1(z_3)} \nonumber \\
            & + & 2^{r - 2} \ 4 \ [  m^2 T_{oo}(0;2z_2;2z_3) + \bar{m}^2 
T_{oo}(0;-2z_2;-2z_3) ] v_1 W_1(B_1) \ \frac{2 k_2 \eta}{\vth_1(2 z_2)} 
\frac{2 k_3 \eta}{\vth_1(2 z_3)} \nonumber \\
            &+&  \ 2^{\frac{r_2}{2}+\frac{r_3}{2}} \  
T_{og}(0;0;0) \Big[ 2^{r_1-2} \ 2 N D_1 v_1 W_1(B_1) + 2 D_2 D_3 
\frac{P_1}{v_1} \Big] \  \frac{2 \eta}{\vth_2(0)} \  
\frac{2 \eta}{\vth_2(0)} \nonumber \\
          &+&  \ 2^{\frac{r_1}{2}+\frac{r_3}{2}} \  
T_{of}(0;0;0) \Big[ 2^{r_2-2} \ 2 N D_2 v_2 W_2(B_2) + 2 D_1 D_3 
\frac{P_2}{v_2} \Big] \  \frac{2 \eta}{\vth_2(0)} \  
\frac{2 \eta}{\vth_2(0)} \nonumber \\
          &+&  \ 2^{\frac{r_1}{2}+\frac{r_2}{2}} \  
T_{oh}(0;0;0) \Big[ 2^{r_3-2} \ 2 N D_3 v_3 W_3(B_3) + 2 D_1 D_2 
\frac{P_3}{v_3} \Big] \  \frac{2 \eta}{\vth_2(0)} \  
\frac{2 \eta}{\vth_2(0)} \nonumber \\
           &+&  \ 2^{\frac{r_2}{2}+\frac{r_3}{2}} \ 
T_{og}(0;z_2;z_3) \Big[ 2^{r_1-2} \ 2 m D_1 v_1 W_1(B_1) \Big] \ 
\frac{2 \eta}{\vth_2 (z_2)} \ \frac{2 \eta}{\vth_2 (z_3)} \nonumber \\ 
            &+&  \ 2^{\frac{r_2}{2}+\frac{r_3}{2}} \ 
T_{og}(0;-z_2;-z_3) \Big[ 2^{r_1-2} \ 2 \bar{m} D_1 v_1 W_1(B_1) \Big] \ 
\frac{2 \eta}{\vth_2 (-z_2)} \ \frac{2 \eta}{\vth_2 (-z_3)} \nonumber \\ 
             &+&  \ 2^{\frac{r_1}{2}+\frac{r_3}{2}} \ 
T_{of}(0;z_2;z_3) \ 2 m D_2 \ \frac{2 \eta}{\vth_2 (0)} \  
\frac{2 k_2 \eta}{\vth_1 (z_2)} \ \frac{2 \eta}{\vth_2 (z_3)} \nonumber \\ 
              &+&  \ 2^{\frac{r_1}{2}+\frac{r_3}{2}} \ 
T_{of}(0;-z_2;-z_3) \ 2 \bar{m} D_2 \ \frac{2 \eta}{\vth_2 (0)} \  
\frac{2 k_2 \eta}{\vth_1 (-z_2)} \ \frac{2 \eta}{\vth_2 (-z_3)} \nonumber \\ 
             &+&  \ 2^{\frac{r_1}{2}+\frac{r_2}{2}} \ 
T_{oh}(0;z_2;z_3) \ 2 m D_3 \ \frac{2 \eta}{\vth_2 (0)} \  
\frac{2 \eta}{\vth_2 (z_2)} \ \frac{2 k_3 \eta}{\vth_1 (z_3)} \nonumber \\ 
              &+&  \ 2^{\frac{r_1}{2}+\frac{r_2}{2}} \ 
T_{oh}(0;-z_2;-z_3) \ 2 \bar{m} D_3 \ \frac{2 \eta}{\vth_2 (0)} \  
\frac{2 \eta}{\vth_2 (-z_2)} \ \frac{2 k_3 \eta}{\vth_1 (-z_3)} \ \biggr\} 
\quad ,
\eea
and with the transverse M\"obius amplitudes
\bea
\tld{\cal{M}} & = & - \frac{2}{8} \ \biggl\{  \ [ \ 2^{\frac{r-6}{2}} \ 
N \ v_1 v_2 v_3 \ W_1(B_1,\g_{\e_1}) \ 
 W_2(B_2,\g_{\e_2}) \  W_3(B_3,\g_{\e_3})] \ \hat{T}_{oo}(0;0;0) \nonumber \\
&+& [ \ 2^{\frac{r_1-6}{2}} \ \frac{v_1}{v_2 v_3} \ D_1 \ 
W_1(B_1,\g_{\e_1}) \  
P_2(B_2,\tld\g_{\e_2}) \ P_3(B_3,\tld\g_{\e_3}) \ ] \ \hat{T}_{oo}(0;0;0)  
\nonumber \\
&+& [ \ 2^{\frac{r_2-6}{2}} \ \frac{v_2}{v_1 v_3} \ D_2 \ 
P_1(B_1,\tld\g_{\e_1}) \ W_2(B_2,\g_{\e_2}) \  
 P_3(B_3,\tld\g_{\e_3}) \ ] \ \hat{T}_{oo}(0;0;0)  
\nonumber \\
&+& [ \ 2^{\frac{r_3-6}{2}} \ \frac{v_3}{v_1 v_2} \ D_3 \ 
P_1(B_1,\tld\g_{\e_1}) \  
P_2(B_2,\tld\g_{\e_2}) \ W_3(B_3,\g_{\e_3}) \ ] \ \hat{T}_{oo}(0;0;0)  
\nonumber \\
&+&  2^{\frac{r_1}{2} + r_2+r_3-1} \ 4 \ 
[ \ m \hat{T}_{oo}(0;z_2;z_3) +  \bar{m} 
\hat{T}_{oo}(0;-z_2;-z_3) \ ] \ v_1 W_1(B_1,\g_{\e_1}) 
\frac{k_2 \hat{\eta}} {\hat{\vth}_1(z_2)} 
\frac{k_3 \hat{\eta}}{\hat{\vth}_1(z_3)} \nonumber \\
          &+&   \hat{T}_{og}(0;0;0) 
\ [  2^{\frac{r_1}{2} -1} (N + D_1) v_1 W_1(B_1,\g_{\e_1}) \ + \  
\frac{2^{-1}}{v_1} (D_2 + D_3) P_1(B_1,\tld\g_{\e_1}) ] \ \frac{2 \hat{\eta}}
{\hat{\vth}_2(0)} \frac{2 \hat{\eta}}{\hat{\vth}_2(0)}\nonumber \\
          &+&   \hat{T}_{of}(0;0;0) 
\ [  2^{\frac{r_2}{2} -1} (N + D_2) v_2 W_2(B_2,\g_{\e_2}) \ + \  
\frac{2^{-1}}{v_2} (D_1 + D_3) P_2(B_2,\tld\g_{\e_2}) ] \ \frac{2 \hat{\eta}}
{\hat{\vth}_2(0)} \frac{2 \hat{\eta}}{\hat{\vth}_2(0)}\nonumber \\
          &+&   \hat{T}_{oh}(0;0;0) 
\ [  2^{\frac{r_3}{2} -1} (N + D_3) v_3 W_3(B_3,\g_{\e_3}) \ + \  
\frac{2^{-1}}{v_3} (D_1 + D_2) P_2(B_2,\tld\g_{\e_2}) ] \ \frac{2 \hat{\eta}}
{\hat{\vth}_2(0)} \frac{2 \hat{\eta}}{\hat{\vth}_2(0)}\nonumber \\
&+&  2^{\frac{r_1}{2}-1} [ \ m \hat{T}_{og}(0;z_2;z_3) +  \bar{m} 
\hat{T}_{og}(0;-z_2;-z_3) \ ] \ v_1 W_1(B_1,\g_{\e_1}) \ 
\frac{2 \hat{\eta}}{\hat{\vth}_2(z_2)} \ 
\frac{2 \hat{\eta}}{\hat{\vth}_2(z_3)} \nonumber \\
&+&   [ \ m \hat{T}_{of}(0;z_2;z_3) -  \bar{m} 
\hat{T}_{of}(0;-z_2;-z_3) \ ] \ \frac{2 \hat{\eta}}{\hat{\vth}_2(0)}
\ \frac{2 k_2 \hat{\eta}}{\hat{\vth}_1(z_2)} \ \frac{2 \hat{\eta}}
{\hat{\vth}_2(z_3)} \nonumber \\
          & + &  [ \ m \hat{T}_{oh}(0;z_2;z_3) -  \bar{m} 
\hat{T}_{oh}(0;-z_2;-z_3) \ ] \ \frac{2 \hat{\eta}}{\hat{\vth}_2(0)}
 \ \frac{2 \hat{\eta}} {\hat{\vth}_2(z_2)} \ \frac{2 k_3 \hat{\eta}}
{\hat{\vth}_1(z_3)} \ \biggr\} \quad  . 
\label{mtshift}
\eea
Apart from the $m=\bar{m}$ condition, automatic for the unitary
gauge group selected by the 
magnetic background, all R-R tadpole cancellation 
conditions directly tied to four-dimensional
non Abelian anomalies  
arise from the untwisted sector.  After the charges are 
rescaled and parametrized 
in such a way that $N=2n$, $D_i=2 d_i$ and $m \rightarrow 2m$, the 
resulting R-R conditions are as in Table \ref{cpz2z2m}, 
provided the signs $\gamma_\epsilon$ and $\tilde{\gamma}_\epsilon$ 
satisfy the same identities (\ref{transversegamma}) 
of the undeformed case.  
The NS-NS tadpoles, canceled only at the
supersymmetric $H_2 = - H_3$ point, 
are related to the derivatives of the Born-Infeld action
with respect to the moduli, exactly as in the six-dimensional case 
\cite{aads}.
Several choices of gauge group are again allowed by the
additional signs $\xi_i$ and $\eta_i$, in eq.(\ref{directgamma}). 
Introducing the combinations
\bea
2 \rho_{\alpha,o} \ = \ a_1 a_2 a_3 - a_1 - a_2 - a_3 \ , \nonumber \\
2 \rho_{\alpha,g} \ = \ a_1 a_2 a_3 - a_1 + a_2 + a_3 \ , \nonumber \\
2 \rho_{\alpha,f} \ = \ a_1 a_2 a_3 + a_1 - a_2 + a_3 \ , \nonumber \\
2 \rho_{\alpha,h} \ = \ a_1 a_2 a_3 + a_1 + a_2 - a_3 \ , 
\eea
where $a_i=\eta_i$ if $\alpha=n$ while $a_i=\eta_i$ but $a_k=\xi_k, k\ne i$
if $\alpha=d_i$, the massless spectra are encoded in
\bea
&&{\mc{A}}_0 + {\mc{M}}_0 \, = \, 
\t_{oo}(0) \  \Big[ \frac{n (n-\rho_{no})}{2} + 
\frac{d_1 (d_1 - \rho_{d_1 o})}{2} + 
\frac{d_2 (d_2 - \rho_{d_2 o})}{2} + 
\frac{d_3 (d_3 - \rho_{d_3 o})}{2} + m \bar{m} \Big] \nonumber \\
   & + &  \t_{og}(0) \Big[ \frac{n (n-\rho_{ng})}{2} + 
\frac{d_1 (d_1 - \rho_{d_1 g})}{2} + 
\frac{d_2 (d_2 - \rho_{d_2 g})}{2} + 
\frac{d_3 (d_3 - \rho_{d_3 g})}{2} + m \bar{m} \Big] \nonumber \\
& + & \t_{oh}(0) \Big[ \frac{n (n-\rho_{nh})}{2} + 
\frac{d_1 (d_1 - \rho_{d_1 h})}{2} + 
\frac{d_2 (d_2 - \rho_{d_2 h})}{2} + 
\frac{d_3 (d_3 - \rho_{d_3 h})}{2} + m \bar{m} \Big] \nonumber \\
& + & \t_{of}(0) \Big[ \frac{n (n-\rho_{nf})}{2} + 
\frac{d_1 (d_1 - \rho_{d_1 f})}{2} + 
\frac{d_2 (d_2 - \rho_{d_2 f})}{2} + 
\frac{d_3 (d_3 - \rho_{d_3 f})}{2} + m \bar{m} \Big] \nonumber \\ 
   & + &  \Big[ \ \t_{gh}(0) \ + \ \t_{gf}(0) \ \Big] 
 \ 2^{\frac{r_2}{2}+\frac{r_3}{2}} \ ( n d_1 + d_2 d_3 ) \nonumber \\
   & + & \Big[ \ \t_{fg}(0) \ + \ \t_{fh}(0) \ \Big] 
\ 2^{\frac{r_1}{2}+\frac{r_3}{2}} \ ( n d_2 + d_1 d_3 ) \nonumber \\
  & + & \Big[ \ \t_{hg}(0) \ + \ \t_{hf}(0) \ \Big] 
\ 2^{\frac{r_1}{2}+\frac{r_2}{2}} \ ( n d_3 + d_1 d_2 ) \nonumber \\
   & + &  \Big[ \ \t_{oh}(+) \ + \ \t_{of}(+) \ \Big]
 \ 2^{r_2 + r_3} \ | k_2 k_3 | \ n m \ + \  
\Big[ \ \t_{oh}(-) \ + \ \t_{of}(-) \ \Big]
 \  2^{r_2 + r_3} \ | k_2 k_3 | \ n \bar{m} \nonumber \\
   & + &  \Big[ \ \t_{gh}(+) \ + \ \t_{gf}(+) \ \Big]
 \ 2^{\frac{r_2}{2}+\frac{r_3}{2}} \ m d_1 \ + \  
 \Big[ \ \t_{gh}(-) \ + \ \t_{gf}(-) \ \Big]
 \ 2^{\frac{r_2}{2}+\frac{r_3}{2}} \ \bar{m} d_1 \nonumber \\
& + &  \Big[ \ \t_{fh}(+) \ \Big]  \ 
2^{\frac{r_1}{2}+\frac{r_3}{2}} \ m d_2 |k_2| \ + \ \Big[ \ 
\t_{fg}(-) \ \Big] 
 2^{\frac{r_1}{2}+\frac{r_3}{2}} \bar{m} d_2 |k_2| \nonumber \\
   & + &  \Big[ \ \t_{hg}(+) \ \Big] \ 
 2^{\frac{r_1}{2}+\frac{r_2}{2}} \ m d_3 |k_3| \ + \ \Big[ \ 
\t_{hf}(-) \ \Big] 
 2^{\frac{r_1}{2}+\frac{r_2}{2}} \ \bar{m} d_3 |k_3|  \nonumber \\
   & + &  \Big[ \ \t_{oh}(2+) \ \Big] \ 
\frac{m (m-1)}{2} \  \Big[ 2^{r_2+r_3} 2 |k_2 \, k_3| + 2^{\frac{r_2+r_3}{2}}
\eta_1 |k_2 \, k_3| + \eta_1 +  2^{\frac{r_2}{2}} |k_2| - 
2^{\frac{r_3}{2}} |k_3|  \Big]  \nonumber \\
  & + &  \Big[ \ \t_{oh}(2+) \ \Big] \ 
\frac{m (m+1)}{2} \  \Big[ 2^{r_2+r_3} 2 |k_2 \, k_3| - 2^{\frac{r_2+r_3}{2}}
\eta_1 |k_2 \, k_3| - \eta_1 -  2^{\frac{r_2}{2}} |k_2| +  
2^{\frac{r_3}{2}} |k_3|  \Big]  \nonumber \\
  & + &  \Big[ \ \t_{of}(2+) \ \Big] \ 
\frac{m (m-1)}{2} \  \Big[ 2^{r_2+r_3} 2 |k_2 \, k_3| + 2^{\frac{r_2+r_3}{2}}
\eta_1 |k_2 \, k_3| + \eta_1 -  2^{\frac{r_2}{2}} |k_2| + 
2^{\frac{r_3}{2}} |k_3|  \Big]  \nonumber \\
   & + &  \Big[ \ \t_{of}(2+) \ \Big] \ 
\frac{m (m+1)}{2} \  \Big[ 2^{r_2+r_3} 2 |k_2 \, k_3| - 2^{\frac{r_2+r_3}{2}}
\eta_1 |k_2 \, k_3| - \eta_1 +  2^{\frac{r_2}{2}} |k_2| - 
2^{\frac{r_3}{2}} |k_3|  \Big]  \nonumber \\
  & + &  \Big[ \ \t_{of}(2-) \ \Big] \ 
\frac{\bar{m} (\bar{m}-1)}{2} \  \Big[ 2^{r_2+r_3} 2 |k_2 \, k_3| + 
2^{\frac{r_2+r_3}{2}}
\eta_1 |k_2 \, k_3| + \eta_1 +  2^{\frac{r_2}{2}} |k_2| - 
2^{\frac{r_3}{2}} |k_3|  \Big]  \nonumber \\
  & + &  \Big[ \ \t_{of}(2-) \ \Big] \ 
\frac{\bar{m} (\bar{m}+1)}{2} \  
\Big[ 2^{r_2+r_3} 2 |k_2 \, k_3| - 2^{\frac{r_2+r_3}{2}}
\eta_1 |k_2 \, k_3| - \eta_1 -  2^{\frac{r_2}{2}} |k_2| + 
2^{\frac{r_3}{2}} |k_3|  \Big]  \nonumber \\
  & + &  \Big[ \ \t_{oh}(2-) \ \Big] \ 
\frac{\bar{m} (\bar{m}-1)}{2} \  
\Big[ 2^{r_2+r_3} 2 |k_2 \, k_3| + 2^{\frac{r_2+r_3}{2}}
\eta_1 |k_2 \, k_3| + \eta_1 -  2^{\frac{r_2}{2}} |k_2| + 
2^{\frac{r_3}{2}} |k_3|  \Big]  \nonumber \\
   & + &  \Big[ \ \t_{oh}(2-) \ \Big] \ 
\frac{\bar{m} (\bar{m}+1)}{2} \  \Big[ 2^{r_2+r_3} 2 |k_2 \, k_3| - 
2^{\frac{r_2+r_3}{2}}
\eta_1 |k_2 \, k_3| - \eta_1 +  2^{\frac{r_2}{2}} |k_2| - 
2^{\frac{r_3}{2}} |k_3|  \Big]  \ ,
\eea
where $(0)$, $(\pm)$ and $(2 \pm)$ are shorthand notations for the 
arguments $(0,0,0)$, $(0;\pm z_2 \t;\pm z_3 \t)$ and 
$(0;\pm 2 z_2 \t;\pm 2 z_3 \t)$, respectively, and the characters with 
non-vanishing arguments actually denote restrictions to their massless parts.  

The resulting gauge groups are reported in Table \ref{cpz2z2m}, while  
the open unoriented spectra are displayed in Table \ref{ouz2z2m}.  
As expected from the previous discussion, chirality originates from two 
different sources. The first 
is the chiral asymmetry in the ``pure magnetic'' sector,
due to the misalignment introduced by the combined action of 
magnetic backgrounds and orbifold projections on the M\"obius amplitudes.
The second is the coupling between magnetized $D9$-branes and 
$D5$-branes longitudinal to the magnetized 
complex directions, familiar from the $T$-dual picture, where 
chiral fermions live in a natural way at
brane intersections \cite{intbranes}.
Whenever potentially anomalous $U(1)$'s are present, they call for a 
generalization of the 
Dine-Seiberg-Witten mechanism \cite{dinsw}, 
an option that, as in six dimensions  \cite{gss}, requires
generalized Green-Schwarz couplings in the Ramond-Ramond sectors 
\cite{ibanez}.

\subsection{Magnetized $Z_2 \times Z_2$ Shift-orientifolds}

In this Section we describe the magnetized versions of the 
$Z_2 \times Z_2$ shift-orientifolds introduced in \cite{shift}
and reviewed in Section 2.3.  As in the ``plain'' 
$Z_2 \times Z_2$ models of 
Section 4.1, 
chiral matter can be obtained if open strings stretched between magnetized 
$D9$-branes and $D5$-branes longitudinal to the $Z^2$ and/or $Z^3$ 
directions are present. An inspection of Table \ref{oldmodels} 
shows that the 
$p_3$, $w_2 p_3$ and $w_1 w_2 p_3$ models are potentially chiral, while the 
remaining models are not.  

The $p_3$ model requires a separate discussion, since 
in its undeformed version \cite{shift} 
it exhibits an 
$f$-twisted R-R tadpole condition, corresponding to the action of the 
$Z_2 \times Z_2$ element that fixes the $T^{67}$-torus, one of the
two along which we turn on  background magnetic fluxes 
(see Table \ref{clunp3} for the unoriented closed spectra and 
Table \ref{opunundef} for the unoriented open spectra with complex
Chan-Paton charges).  This tadpole condition can no longer be 
satisfied if the background magnetic field is present,
since some states are lifted in mass by a term depending solely on the field 
strength $H_2$ along $T^{67}$, rather than on the difference $H_2 - H_3$
as in the case 
for the tadpole conditions coming from the untwisted or from the $g$-twisted 
sectors.  The resulting models are thus anomalous as string vacua, 
because the magnetic deformations are, in the aforementioned 
sense, incompatible with the $p_3$ shift.  The natural geometric  
interpretation of this phenomenon is as follows: the $p_3$ shift is 
introducing a net number of fractional branes \cite{frabra} that, 
differently from what happens in the 
remaining models, are partly longitudinal and partly orthogonal to the 
magnetic fields carrying a non-vanishing twisted R-R charge, whose
excess can be canceled only turning off the background magnetic field.

In the following we shall analyze 
the chiral and non-chiral examples with selfdual configurations of the
magnetic field, {\it i.e.} at  supersymmetric points, 
leaving to  Section 5 the discussion
of models with ``brane supersymmetry breaking''.

\subsubsection{$w_2 p_3$ Models}

Let us first analyze in detail the $w_2 p_3$ class of orientifolds,
that captures all  interesting
features of the models discussed in this paper.  In the presence of the
NS-NS two-form $B_{ab}$, the unoriented truncation of the closed 
spectrum is obtained adding to 
the halved torus amplitude the Klein-bottle amplitude
\ba
{\cal K} &=& \frac{1}{8} \biggl\{ \bigl[ \  P_1\, P_2\, P_3\, 
+ 2^{-4}\, 
P_1\, W_2(B_2)\, W_3(B_3)\,  \nonumber \\ &+ & \!\!
 2^{-4}\,  W_1(B_1)\, P_2\,
W_3(B_3) \nonumber \\ &+ & \!\!  
2^{-4}\, W_1(B_1)\, (-1)^{n_2} W_2(B_2)\, (-1)^{m_3} P_3 \ \bigr] 
\, T_{oo} \nonumber \\ &+ & \!\! 
2 \times 16 \, \bigl[ \ 2^{-\frac{r_2}{2}-\frac{r_3}{2}}\, P_1\, T_{go} 
+  2^{-\frac{r_1}{2}-\frac{r_3}{2}}\, P_2^{\frac{1}{2}}\, T_{fo} 
\nonumber \\ &+ & \!\!  
2^{-\frac{r_1}{2}-\frac{r_2}{2}}\, 2^{-2}\, W_3^{\frac{1}{2}} 
(B_3)\, T_{ho} \ \bigr]\,  
\left( \frac{\eta}{\vartheta_4} \right)^2 \biggr\} \quad .
\label{kw2p3}
\ea
The resulting massless  unoriented closed spectra are reported
 in Table \ref{clunw2p3}.
The transverse-channel amplitude 
\ba
\tilde{\cal K}_0 &=& \frac{2^5}{8} \Biggl\{ \ \left( \sqrt{v_1v_2v_3} +
 2^{-\frac{r_2}{2}-\frac{r_3}{2}} \sqrt{\frac{v_1}{v_2 v_3}} + 
 2^{-\frac{r_1}{2}-\frac{r_3}{2}} \sqrt{\frac{v_2}{v_1 v_3}}
 \right)^2 \tau_{oo} \nonumber \\ 
&+&  \left( \sqrt{v_1v_2v_3} +
 2^{-\frac{r_2}{2}-\frac{r_3}{2}} \sqrt{\frac{v_1}{v_2 v_3}} - 
 2^{-\frac{r_1}{2}-\frac{r_3}{2}} \sqrt{\frac{v_2}{v_1 v_3}}
 \right)^2 \tau_{og} \nonumber \\ 
&+&  \left( \sqrt{v_1v_2v_3} -
 2^{-\frac{r_2}{2}-\frac{r_3}{2}} \sqrt{\frac{v_1}{v_2 v_3}} - 
 2^{-\frac{r_1}{2}-\frac{r_3}{2}} \sqrt{\frac{v_2}{v_1 v_3}}
 \right)^2 \tau_{oh} \nonumber \\ 
&+&  \left( \sqrt{v_1v_2v_3} -
 2^{-\frac{r_2}{2}-\frac{r_3}{2}} \sqrt{\frac{v_1}{v_2 v_3}} + 
 2^{-\frac{r_1}{2}-\frac{r_3}{2}} \sqrt{\frac{v_2}{v_1 v_3}}
 \right)^2 \tau_{of} \ \Biggr\} \quad 
\label{k0w2p3}
\ea
displays very neatly the 
presence of one conventional $O9$-plane and of the  $O5_1$ and $O5_2$ planes, 
while the $O5_3$-plane of the $Z_2 \times Z_2$-models in eq. (\ref{k0z2z2}) 
is no longer a fixed manifold of the combined orbifold and shifts, 
and is thus eliminated.

The annulus amplitude is more subtle.  In the presence of $B_{ab}$,  
$H_2$ along $T^{67}$ and $H_3$ along $T^{89}$, it is again 
the sum of three contributions:
\bea
\mc{A}_{(Q=0)} & = & \frac{1}{8} \biggl\{ \bigl[ 
\frac{N^2}{2} \ 2^{r - 6} \ 
P_1(B_1)  \ ( \ P_2(B_2) +  P_2^{1/2}(B_2) \ ) \ P_3(B_3) 
\nonumber \\ 
&+& \frac{2 m \bar{m}}{2}  \ 2^{r - 6} \ P_1(B_1) \ ( \ \tld{P}_2(B_2) + 
\tld{P}_2^{1/2}(B_2) \ ) \ \tld{P}_3(B_3) \nonumber \\
&+& \frac{D_1^2}{2} \ 2^{r_1 - 2} \ P_1(B_1) W_2 ( \ W_3 + W_3^{1/2} \ ) 
\nonumber \\ &+&   
\frac{D_2^2}{4} \ 2^{r_2 - 2} \ W_1 ( \ P_2(B_2) +  
P_2^{1/2}(B_2) \ ) \ ( \ W_3 + W_3^{1/2} \ ) \bigr] T_{oo}(0;0;0) \nonumber \\
&+& 2^{r_1 - 2} \bigl[ \frac{G^2}{2} +  \frac{G_1^2}{2} + 
\frac{2 m \bar{m}}{2} \bigr] T_{og}(0;0;0)  \ P_1(B_1) 
\Big( \frac{ 2 \eta}{\vth_2(0)} \Big)^2 \nonumber \\
&+&  2^{\frac{r_1}{2}+\frac{r_3}{2}} \ T_{go}(0;0;0) \ 
2 N D_1 \ 2^{r_1 - 2} \ P_1(B_1) 
\Big( \frac{ \eta}{\vth_4(0)} \Big)^2  \nonumber \\
&+&  2^{\frac{r_1}{2}+\frac{r_3}{2}} \ T_{gg}(0;0;0) \ 
2 G G_1 \ 2^{r_1 - 2} \ P_1(B_1) 
\Big( \frac{ \eta}{\vth_3(0)} \Big)^2  \nonumber \\
&+&  2^{\frac{r_1}{2}+\frac{r_3}{2}} \ T_{fo}(0;0;0) \ 
N D_2 \ 2^{r_2 - 2} \ ( \ P_2^{1/4}(B_2) +  P_2^{3/4}(B_2) \ ) \, 
\Big( \frac{ \eta}{\vth_4(0)} \Big)^2  \nonumber \\
&+&  2^{\frac{r_1}{2}+\frac{r_2}{2}} \ T_{ho}(0;0;0) \ 
2^{r_3 - 2} D_1 D_2 \ ( \ W_3^{1/4} + W_3^{3/4} \ ) \, 
\Big( \frac{ \eta}{\vth_4(0)} \Big)^2  \biggl\}
\eea
for the $Q=0$ sectors,
\bea
\mc{A}_{(Q=1)} & = & \frac{1}{8} \biggl\{ - 2^{r - 2} \, 2 \, m \, N \,
T_{oo}(0;z_2 \t;z_3 \t) P_1(B_1) \  
\frac{k_2 \eta}{\vth_1(z_2 \t)} \ \frac{k_3 \eta}{\vth_1(z_3 \t)} 
\nonumber \\
&-& 2^{r - 2} \, 2 \, \bar{m} \, N \, T_{oo}(0;- z_2 \t;- z_3 \t) P_1(B_1) \  
\frac{k_2 \eta}{\vth_1(- z_2 \t)} \ \frac{k_3 \eta}{\vth_1(- z_3 \t)} 
\nonumber \\
&+& \ T_{og}(0;z_2 \t;z_3 \t) \bigg[ 
\alpha_{m G} \, m \, G  \ 2^{r_1 - 2} \ P_1(B_1) \bigg] \, 
 \frac{ 2 \eta}{\vth_2(z_2 \t)} \  \frac{ 2 \eta}{\vth_2(z_3 \t)} \nonumber \\ 
&+& \ T_{og}(0;- z_2 \t; - z_3 \t) \bigg[ 
\bar{\alpha}_{m G} \, \bar{m} \, G  \ 2^{r_1 - 2} \ P_1(B_1) \bigg] \, 
 \frac{ 2 \eta}{\vth_2(- z_2 \t)} \  
\frac{ 2 \eta}{\vth_2(- z_3 \t)} \nonumber \\
&+&  2^{\frac{r_2}{2}+\frac{r_3}{2}} \ T_{go}(0;z_2 \t;z_3 \t) \bigg[ 
2 m D_1 \ 2^{r_1 - 2} \ P_1(B_1) \bigg] \, 
 \frac{ \eta}{\vth_4(z_2 \t)} \  \frac{ \eta}{\vth_4(z_3 \t)} \nonumber \\ 
&+&  2^{\frac{r_2}{2}+\frac{r_3}{2}} \ T_{go}(0;-z_2 \t;-z_3 \t) \bigg[ 
2 \bar{m} D_1 \ 2^{r_1 - 2} \ P_1(B_1) \bigg] \, 
 \frac{ \eta}{\vth_4(-z_2 \t)} \  \frac{ \eta}{\vth_4(-z_3 \t)} \nonumber \\ 
&+&  2^{\frac{r_2}{2}+\frac{r_3}{2}} \ T_{gg}(0;z_2 \t;z_3 \t) \bigg[ 
\alpha_{m G_1} \ 2 m G_1 \ 2^{r_1 - 2} \ P_1(B_1) \bigg] \, 
 \frac{ \eta}{\vth_3(z_2 \t)} \  \frac{ \eta}{\vth_3(z_3 \t)} \nonumber \\ 
&+&  2^{\frac{r_2}{2}+\frac{r_3}{2}} \ T_{gg}(0;- z_2 \t; - z_3 \t) \bigg[ 
\bar{\alpha}_{m G_1} \ 2 \bar{m} G_1 \ 2^{r_1 - 2} \ P_1(B_1) \bigg] \, 
 \frac{ \eta}{\vth_3(- z_2 \t)} \  \frac{ \eta}{\vth_3(- z_3 \t)} \nonumber \\
&+&  2^{\frac{r}{2}+\frac{r_2}{2}} \ T_{fo}(0;z_2 \t;z_3 \t) \bigg[ 
- 2 i m D_2 \bigg] \, \frac{ \eta}{\vth_4(0)} \ 
 \frac{ k_2 \eta}{\vth_1(z_2 \t)} \  \frac{ \eta}{\vth_4(z_3 \t)} \nonumber \\ 
&+&  2^{\frac{r}{2}+\frac{r_2}{2}} \ T_{fo}(0;-z_2 \t;-z_3 \t) \bigg[ 
2 i \bar{m} D_2 \bigg] \, \frac{ \eta}{\vth_4(0)} \ 
 \frac{ k_2 \eta}{\vth_1(-z_2 \t)} \ \frac{ \eta}{\vth_4(-z_3 \t)} \biggl\} , 
\eea
for the $Q=1$ sectors, and 
\bea
\mc{A}_{(Q=2)} & = & \frac{1}{8} \biggl\{- 2^{r - 2} m^2 
T_{oo}(0;2 z_2 \t;2 z_3 \t) P_1(B_1) \ 
\frac{2 k_2 \eta}{\vth_1(2 z_2 \t)} \ \frac{k_3 \eta}{2 \vth_1(z_3 \t)} 
\nonumber \\
&-& 2^{r - 2} \bar{m}^2  T_{oo}(0;- 2 z_2 \t;- 2 z_3 \t) P_1(B_1) \  
\frac{k_2 \eta}{\vth_1(- 2 z_2 \t)} \ 
\frac{k_3 \eta}{\vth_1(- 2 z_3 \t)} \nonumber \\
&+& 2^{r_1 - 2} \alpha_{m^2} \ \frac{m^2}{2}  
T_{og}(0; 2 z_2 \t; 2 z_3 \t) P_1(B_1) \ \frac{2 \eta}{\vth_2(2 z_2 \t)} \ 
\frac{2 \eta}{\vth_2(2 z_3 \t)} \nonumber \\
&+& 2^{r_1 - 2} \bar{\alpha}_{m^2} \ \frac{\bar{m}^2}{2}  
T_{og}(0; - 2 z_2 \t; - 2 z_3 \t) P_1(B_1) \ 
\frac{2 \eta}{\vth_2( - 2 z_2 \t)} \ 
\frac{2 \eta}{\vth_2( - 2 z_3 \t)} \biggl\} . 
\eea
for the $Q=2$ sectors, where the magnetized Chan-Paton charge 
multiplicity is denoted by $m$.  The coefficients 
$\alpha_{m G}$, $\bar{\alpha}_{m G}$, $\alpha_{m G_1}$, $\bar{\alpha}_{m G_1}$,
$\alpha_{m^2}$ and $\bar{\alpha}_{m^2}$ must be chosen in such a way that
the annulus amplitudes become real.
The M\"obius amplitude can be deduced 
in a similar way from the undeformed case, adding to the 
uncharged ($Q=0$) contributions the charged ($Q=2$) ones.  The result is 
\bea 
\mc{M}_{(Q=0)} & = & - \frac{1}{8} \biggl\{ \,
\bigl[\, 2^{\frac{r - 6}{2}} \, N \, P_1(B_1,\g_{\e_1}) \, 
P_2(B_2,\g_{\e_2}) \, P_3(B_3,\g_{\e_3}) \nonumber \\
   &+& 2^{\frac{r_1 - 6}{2}} \, D_1 \, 
P_1(B_1,\g_{\e_1}) W_2(B_2,\tld\g_{\e_2}) W_3(B_3,\tld\g_{\e_3}) \nonumber \\
   &+& 2^{\frac{r_2 - 6}{2}} \, D_2 \, 
W_1(B_1,\tld\g_{\e_1}) P_2(B_2,\g_{\e_2}) W_3(B_3,\tld\g_{\e_3}) 
\bigr] \, \hat{T}_{oo}(0;0;0)  \nonumber \\ 
&-& \! \bigl[ \,  2^{\frac{r_1 - 2}{2}} \, (N \!+\! D_1 ) 
P_1(B_1,\g_{\e_1}) \bigr] \, \hat{T}_{og}(0;0;0) 
\left( \frac{2 \hat{\eta}}{\hat{\theta_2}}\right)^2 \nonumber \\ 
&-& \! \bigl[ \, 2^{\frac{r_2 - 2}{2}} \, (N \!+\! D_2 ) 
P_2^{1/2}(B_2,\g_{\e_2}) \bigr] \, \hat{T}_{of}(0;0;0) 
\left( \frac{2 \hat{\eta}}{\hat{\theta_2}}\right)^2 \nonumber \\ 
&-& \! \bigl[ \, 2^{-1} \, (D_1 \!+\! D_2) W_3(B_3,\tld\g_{\e_3})
\bigr] \, \hat{T}_{oh}(0;0;0) 
\left( \frac{2 \hat{\eta}}{\hat{\theta_2}}\right)^2 \ \biggr\} \quad ,
\eea
and
\bea 
\mc{M}_{(Q=2)} &=& - \frac{1}{8} \biggl\{ \,
- 2^{\frac{r - 2}{2}} m \hat{T}_{oo}(0;2 z_2 \t;2 z_3 \t) \, 
P_1(B_1,\g_{\e_1}) \, \frac{2 k_2 \hat{\eta}}{\hat{\vth}_1(2 z_2 \t)} 
\frac{2 k_3 \hat{\eta}}{\hat{\vth}_1(2 z_3 \t)}  \nonumber \\
&-& 2^{\frac{r - 2}{2}} \bar{m} \hat{T}_{oo}(0;-2 z_2 \t;-2 z_3 \t) \, 
P_1(B_1,\g_{\e_1}) \, \frac{2 k_2 \hat{\eta}}{\hat{\vth}_1(-2 z_2 \t)} 
\frac{2 k_3 \hat{\eta}}{\hat{\vth}_1(-2 z_3 \t)}  \nonumber \\
&-& 2^{\frac{r_1 - 2}{2}} m \hat{T}_{og}(0;2 z_2 \t;2 z_3 \t) \, 
P_1(B_1,\g_{\e_1}) \, \frac{2 \hat{\eta}}{\hat{\vth}_2(2 z_2 \t)} 
\frac{2 \hat{\eta}}{\hat{\vth}_2(2 z_3 \t)}  \nonumber \\
&-& 2^{\frac{r_1 - 2}{2}} \bar{m} \hat{T}_{og}(0;-2 z_2 \t;-2 z_3 \t) \, 
P_1(B_1,\g_{\e_1}) \, \frac{2 \hat{\eta}}{\hat{\vth}_2(-2 z_2 \t)} 
\frac{2 \hat{\eta}}{\hat{\vth}_2(-2 z_3 \t)}  \nonumber \\
&-& 2^{\frac{r_2}{2}} i m \hat{T}_{of}(0;2 z_2 \t;2 z_3 \t) \, 
\frac{2 \hat{\eta}}{\hat{\vth}_2(0)}\, 
\frac{2 k_2 \hat{\eta}}{\hat{\vth}_1(2 z_2 \t)} 
\frac{2 \hat{\eta}}{\hat{\vth}_2(2 z_3 \t)}  \nonumber \\
&-& 2^{\frac{r_2}{2}} (- i) \bar{m} \hat{T}_{of}(0;-2 z_2 \t;-2 z_3 \t) \, 
\frac{2 \hat{\eta}}{\hat{\vth}_2(0)}\, 
\frac{2 k_2 \hat{\eta}}{\hat{\vth}_1(-2 z_2 \t)} 
\frac{2 \hat{\eta}}{\hat{\vth}_2(-2 z_3 \t)}  \biggr\} \quad .
\eea
In order to analyze in some detail the tadpole cancellation conditions,
it is worth  displaying the transverse (tree) channel amplitudes.  The 
annulus part comprises  untwisted and twisted terms
$$
\tld{\cal{A}} = \tld{\cal{A}}^U + \tld{\cal{A}}^T \qquad ,
$$
where 
\bea
\tld{\cal{A}}^U & = & \frac{2^{-5}}{8} \ \biggl\{ \ [ 2^{r-6} v_1 v_2 v_3 
\frac{N^2}{2} W_1(B_1) ( W_2(B_2)+ (-1)^{n_2} W_2(B_2)) 
W_3(B_3)\nonumber \\ 
&+& \ \frac{v_1}{v_2 v_3} \ 2^{r_1-2} \ 
\frac{D_1^2}{2} W_1(B_1) P_2 (P_3 + (-1)^{m_3} P_3) \nonumber \\ 
&+& \frac{v_2}{v_1 v_3}  \ 2^{r_2-2} \ \frac{D_2^2}{4} P_1 
( W_2(B_2) +( -1)^{n_2} W_2(B_2)) (P_3 + (-1)^{m_3} P_3) \nonumber \\
&+& 2^{r-6} v_1 v_2 v_3 \  \frac{2 m \bar{m}}{2}  
(1+q^2 H_2^2) (1+q^2 H_3^2) W_1(B_1) ( \tld{W_2}(B_2)  \nonumber \\
&+& (-1)^{n_2} 
\tld{W_2}(B_2)) \tld{W_3}(B_3)  \ ] \  
T_{oo}(0;0;0) \nonumber \\
 &+&  \ 2^{\frac{r_2}{2}+\frac{r_3}{2}} \  
 \Big[ 2^{r_1-2} \ 2 N D_1 v_1 W_1(B_1) \Big] \ T_{og}(0;0;0) \ \frac{2 \eta}{\vth_2(0)} \  
\frac{2 \eta}{\vth_2(0)} \nonumber \\
 &+&  \ 2^{\frac{r_1}{2}+\frac{r_3}{2}} \  
 \Big[ 2^{r_2-2} \ N D_2 v_2 ( (i)^{n_2} W_2(B_2) + (-i)^{n_2} W_2(B_2) ) 
\Big] \ T_{of}(0;0;0) \ \frac{2 \eta}{\vth_2(0)} \  \frac{2 \eta}{\vth_2(0)} \nonumber \\
 &+&  \ 2^{\frac{r_1}{2}+\frac{r_2}{2}} \  
 \Big[ D_1 D_2 \frac{1}{v_3} ( (i)^{m_3} P_3 + (-i)^{m_3} P_3 ) 
\Big] \ T_{oh}(0;0;0) \ \frac{2 \eta}{\vth_2(0)} \  \frac{2 \eta}{\vth_2(0)} \nonumber \\
& + &  2^{r - 2} \ 8 \ N \  v_1 W_1(B_1)
[ m T_{oo}(0;z_2;z_3)  +   \bar{m} 
T_{oo}(0;-z_2;-z_3) ] \  
\frac{k_2 \eta}{\vth_1(z_2)} \frac{k_3 \eta}{\vth_1(z_3)} \nonumber \\
 &+&  \ 2^{\frac{r}{2}+\frac{r_1}{2}-2} \ 
 2 D_1 v_1 W_1(B_1) \Big[ m T_{og}(0;z_2;z_3) + \bar{m} T_{og}(0;-z_2;-z_3) \Big] \ 
\frac{2 \eta}{\vth_2 (z_2)} \ \frac{2 \eta}{\vth_2 (z_3)} \nonumber \\           
&+&  \ 2^{\frac{r_2}{2}+\frac{r_3}{2}} \ 
 \Big[ 2^{r_1-2} \ 2 \bar{m} D_1 v_1 W_1(B_1) \Big] \ T_{og}(0;-z_2;-z_3) \
\frac{2 \eta}{\vth_2 (-z_2)} \ \frac{2 \eta}{\vth_2 (-z_3)} \nonumber \\ 
    &+&  \ 2^{\frac{r}{2}+\frac{r_2}{2}} \ 
 \ 2 m D_2 \ T_{of}(0;z_2;z_3) \  \frac{2 \eta}{\vth_2 (0)} \ 
\frac{2 k_2 \eta}{\vth_1 (z_2)} \ \frac{2 \eta}{\vth_2 (z_3)} \nonumber \\ 
              &-&  \ 2^{\frac{r}{2}+\frac{r_2}{2}} \ 
 \ 2 \bar{m} D_2 \ T_{of}(0;-z_2;-z_3) \ \frac{2 \eta}{\vth_2 (0)} \  
\frac{2 k_2 \eta}{\vth_1 (-z_2)} 
\ \frac{2 \eta}{\vth_2 (-z_3)} \\             
 & + & 2^{r - 2} \ 4 \ v_1 W_1(B_1) \ [  m^2 T_{oo}(0;2z_2;2z_3) + \bar{m}^2 
T_{oo}(0;-2z_2;-2z_3) ]  \ \frac{2 k_2 \eta}{\vth_1(2 z_2)} 
\frac{2 k_3 \eta}{\vth_1(2 z_3)} \ \biggr\} 
\quad , \nonumber
\eea
and
\bea
\tld{\cal{A}}^T & = & \frac{2^{-5}}{8} \ \biggl\{ \ 2^{r_1 - 2} \ 16 \ 
 v_1 W_1(B_1) \Big[ \frac{G^2}{2} + \frac{{G^2}_1}{2} \frac{2 m \bar{m}}{2}
 \Big] \ T_{go}(0;0;0) \ \frac{\eta}{\vth_4(0)} \ \frac{\eta}{\vth_4(0)} \nonumber \\
& - &  \ 2^{\frac{r}{2}+\frac{r_1}{2}-2} \
 v_1 W_1(B_1) \ 8 \ G G_1 \ T_{gg}(0;0;0) \ \frac{\eta}{\vth_3(0)} \  
\frac{\eta}{\vth_3(0)} \nonumber \\
& + & 2^{r_1 - 2} \ 16 \ G \ v_1 W_1(B_1) \Big[ \ \alpha_{mG} m T_{go}(0;z_2;z_3) +
 \bar{\alpha}_{mG} \bar{m} T_{go}(0;-z_2;-z_3) \Big] \ \frac{\eta}{\vth_4(z_2)} \  
\frac{\eta}{\vth_4(z_3)} \nonumber \\
& - & \  2^{\frac{r}{2}+\frac{r_1}{2}-2} \ 8 \ G_1 \ v_1 \ W_1(B_1) \ 
\Big[ \ \alpha_{mG_1} \ m \ T_{gg}(0;z_2;z_3) \nonumber\\ 
&+& \bar{\alpha}_{mG_1} \ \bar{m} \ T_{gg}(0;-z_2;-z_3) \ \Big] \ \frac{\eta}{\vth_3(z_2)} \  
\frac{\eta}{\vth_3(z_3)} \nonumber \\
 &+& 2^{r_1 - 2} \ 8 \ 
 v_1 \ W_1(B_1)\  \Big[ \ \alpha_{m^2} \ m^2 \ T_{go}(0;2z_2;2z_3) \nonumber \\
 &+&  \bar{\alpha}_{m^2} \ \bar{m}^2 \ T_{go}(0;-2z_2;-2z_3) \ \Big] \ 
\frac{\eta}{\vth_4(2 z_2)} \ \frac{\eta}{\vth_4(2 z_3)} \ \biggr\} 
\quad .
\eea
This is to be contrasted with the transverse M\"obius amplitude, that contains 
only  untwisted contributions  
\bea
\tld{\cal{M}} & = & - \frac{2}{8} \ \biggl\{  \ \Big[ \ 2^{\frac{r-6}{2}} \ 
N \ v_1 v_2 v_3 \ W^e_1(B_1,\g_{\e_1}) \ 
 W^e_2(B_2,\g_{\e_2}) \  W^e_3(B_3,\g_{\e_3})\nonumber \\ 
&+&  2^{\frac{r_1-6}{2}} \ \frac{v_1}{v_2 v_3} \ D_1 \ 
W^e_1(B_1,\g_{\e_1}) \ P^e_2(B_2,\tld\g_{\e_2}) \ P^e_3(B_3,\tld\g_{\e_3}) 
\nonumber \\
&+& 2^{\frac{r_2-6}{2}} \ \frac{v_2}{v_1 v_3} \ D_2 \ 
P^e_1(B_1,\tld\g_{\e_1}) \ W^e_2(B_2,\g_{\e_2}) \  
 P^e_3(B_3,\tld\g_{\e_3}) \ \Big] \ \hat{T}_{oo}(0;0;0)  
\nonumber \\
 &+&  2^{\frac{r_1}{2} -1} \ (N + D_1) \ v_1 \ W^e_1(B_1,\g_{\e_1}) 
\ \hat{T}_{og}(0;0;0) \ \frac{2 \hat{\eta}}
{\hat{\vth}_2(0)} \ \frac{2 \hat{\eta}}{\hat{\vth}_2(0)}\nonumber \\
          &+& 2^{\frac{r_2}{2} -1} \ (N + D_2) \ v_2 \ W^e_2(B_2,\g_{\e_2}) 
\ \hat{T}_{of}(0;0;0) \ \frac{2 \hat{\eta}}
{\hat{\vth}_2(0)} \ \frac{2 \hat{\eta}}{\hat{\vth}_2(0)}\nonumber \\
          &+&  \frac{2^{-1}}{v_3} \ (D_1 + D_2) \ \phi^{B_3} \ P^e_3(B_3,\tld\g_{\e_3}) 
\ \hat{T}_{oh}(0;0;0) \ \frac{2 \hat{\eta}}
{\hat{\vth}_2(0)} \ \frac{2 \hat{\eta}}{\hat{\vth}_2(0)}\nonumber \\
&+&  2^{\frac{r}{2} - 1} \ 4 \ v_1 W^e_1(B_1,\g_{\e_1}) \
[ \ m \hat{T}_{oo}(0;z_2;z_3) +  \bar{m} 
\hat{T}_{oo}(0;-z_2;-z_3) \ ] \ 
\frac{k_2 \hat{\eta}} {\hat{\vth}_1(z_2)} 
\frac{k_3 \hat{\eta}}{\hat{\vth}_1(z_3)} \nonumber \\
&+&  2^{\frac{r_1}{2}-1} \ v_1 W^e_1(B_1,\g_{\e_1}) \
[ \ m \hat{T}_{og}(0;z_2;z_3) +  \bar{m} 
\hat{T}_{og}(0;-z_2;-z_3) \ ]  \ 
\frac{2 \hat{\eta}}{\hat{\vth}_2(z_2)} \ 
\frac{2 \hat{\eta}}{\hat{\vth}_2(z_3)} \nonumber \\
&+&   2^{\frac{r_2}{2}} \ m \ \hat{T}_{of}(0;z_2;z_3) \ \frac{2 \hat{\eta}}{\hat{\vth}_2(0)}
\ \frac{2 k_2 \hat{\eta}}{\hat{\vth}_1(z_2)} \ \frac{2 \hat{\eta}}
{\hat{\vth}_2(z_3)} \nonumber \\
&-&  2^{\frac{r_2}{2}} \ \bar{m} \ \hat{T}_{of}(0;-z_2;-z_3) \ \frac{2 \hat{\eta}}{\hat{\vth}_2(0)}
\ \frac{2 k_2 \hat{\eta}}{\hat{\vth}_1(- z_2)} \ \frac{2 \hat{\eta}}
{\hat{\vth}_2(- z_3)} \biggr\} \quad  , 
\label{mtshiftz2}
\eea
where $\phi^{B_3}$ is a suitable phase that depends on the rank of
$B_{ab}$, not directly relevant for our discussion.
The untwisted tadpole cancellation conditions are related to the 
superposition of $\tilde{{\cal K}}$,  $\tilde{{\cal A}}$ and 
$\tilde{{\cal M}}$, and can be obtained as follows.  The residues corresponding
to the R-R part of the $\tau_{0 \alpha}$ character are 
\bea
&& \!\!\!\!\!\! \sqrt{v_1 v_2 v_3} \, \left\{ \, 2^{\frac{r}{2}} \, 
\left[ \, N + ( m + \bar{m} ) 
( 1 - 4 \pi^{2} {\alpha^{\prime}}^{2} q^{2} 
H_2 H_3 ) + ( m - \bar{m} ) \ 8 i \pi^{2} {\alpha^{\prime}}^{2} q^{2} 
( H_2 + H_3 ) \, \right] - 32 \, \right\} \nonumber \\
&+&  \lambda_{1 \alpha} \, \sqrt{\frac{v_1}{v_2 v_3}} \ 
\left[ \, 2^{\frac{r_1}{2}} D_1 \ - \ 32 
\, 2^{- \frac{r_2+r_3}{2}} \, \right] \ + \ \lambda_{2 \alpha} \, 
\sqrt{\frac{v_2}{v_1 v_3}} \  
\left[ \, 2^{\frac{r_2}{2}} D_2 \ - \ 32 \, 2^{- \frac{r_1+r_3}{2}} \, 
\right] \, = \, 0  ,
\label{untwisted}
\eea
together with the complex conjugates, 
where $\lambda_{1 \alpha}$ is $+1$ for $\alpha=o,g$ and is $-1$ 
for $\alpha=h,f$ while $\lambda_{2 \alpha}$ is $+1$ for $\alpha=o,f$ 
and is $-1$ for $\alpha=g,h$.  In order to obtain eqs. (\ref{untwisted}),
the conditions in (\ref{transversegamma}) for the signs 
$\gamma_\epsilon$ and $\tilde{\gamma}_\epsilon$ must be used, and, 
in order to 
ensure the vanishing of the imaginary part the numerical 
constraint, $m=\bar{m}$ must also be enforced.  Using the Dirac quantization 
condition in eq. (\ref{dirac}), it is interesting to notice that the
magnetized $D9$-branes contribute not only to the tadpole of
the R-R ten-form, but also to the tadpole of the R-R six-form.  
As in the six-dimensional examples, this signals the
phenomenon of brane transmutation.  In particular, disentangling the 
diverse contributions, one obtains
\be
\sqrt{v_1 v_2 v_3} \, \left[ \, 2^{\frac{r}{2}} \, 
( \, N +  m + \bar{m} ) \, \right] \ = \ \sqrt{v_1 v_2 v_3} \ 32 
\ee
for the $D9$-brane sector,  
\be
\sqrt{\frac{v_1}{v_2 v_3}} \ \left[ \, 2^{\frac{r_1}{2}} D_1 \ +  
2^{\frac{r}{2}} \, |k_2 k_3| \, ( m + \bar{m} ) \, \right]
\ = \ \sqrt{\frac{v_1}{v_2 v_3}} \,  \left[ \, 32 \, 
2^{- \frac{r_2+r_3}{2}} \, \right]
\ee
for the $D5_1$-brane sector and 
\be
\sqrt{\frac{v_2}{v_1 v_3}} \  
\left[ \, 2^{\frac{r_2}{2}} D_2 \right] \ = \ \sqrt{\frac{v_2}{v_1 v_3}} \, 
\left[ \, 32 \, 2^{- \frac{r_1+r_3}{2}} \, \right]
\ee
for the $D5_2$-brane sector.  

The twisted tadpole conditions determine the nature of the allowed 
Chan-Paton charges.  There are two options, that result in complex or real
Chan-Paton charges.  In the complex case, one must choose 
$$
\alpha_{m G} = - \bar{\alpha}_{m G} = \alpha_{m G_1} = 
- \bar{\alpha}_{m G_1} \ = \ i  \quad ; \qquad \alpha_{m^2} = \bar{\alpha}_{m^2} 
\ = \ -1 \quad ,
$$
and the tadpole cancellation condition can be written 
\be
(8 - 2^{r-r_1+1} ) \ [ \, G + i m - i \bar{m} \, ]^2 + 2 \ [ 
\, 2 \, G_1 + 2^{\frac{r-r_1}{2}} \, ( \, G + i m - i \bar{m} \, ) 
\, ]^2 = 0 \quad ,
\label{twisted}
\ee
while the real charges are determined by the choice
$$
\alpha_{m G} = - \bar{\alpha}_{m G} = \alpha_{m G_1} = 
- \bar{\alpha}_{m G_1} = \alpha_{m^2} = \bar{\alpha}_{m^2} 
\ = \ 1 ,
$$
and the corresponding tadpole cancellation condition can be written 
in the form
\be
(8 - 2^{r-r_1+1} ) \ [ \, G + m + \bar{m} \, ]^2 + 2 \ [
 \, 2 \, G_1 + 2^{\frac{r-r_1}{2}} \, ( \, G + m + \bar{m}\, ) 
\, ]^2 = 0 \quad .
\label{twistedreal}
\ee
The analysis of the open spectra is very similar to the one in Section 2, 
and therefore 
we shall not repeat it here.  With complex charges, after the choice of 
signs in eq. (\ref{directgamma}), one has to fix 
\be
\xi_2 \ \xi_3 =  \eta_2 \  \eta_3 \ = \ 1 \quad ,
\ee
while $\xi_1$ and $\eta_1$ are free signs.  In order to obtain 
amplitudes with a proper particle interpretation, the magnetic charges
must be rescaled by a factor of two, and a suitable parametrization of
the Chan-Paton multiplicities is
\ba
N &=& \ 2 \ ( \ n \ + \ \bar{n} \ ) \quad , \qquad 
G \ = \ 2 \ i \ ( \ n \ - \ \bar{n} \ ) \quad; \nonumber \\
D_1 &=&  2 \ ( \ d \ + \ \bar{d} \ ) \quad , \qquad 
G_1 \ = \ 2 \ i \ ( \ d \ - \ \bar{d} \ ) \quad; \nonumber \\
D_2 &=& 4 \ d_2 \quad .
\label{cpchaw2p3}
\ea
With this choice, the tadpole cancellation conditions are reported in Table 
\ref{cpw2p3}, together with  corresponding options for the Chan-Paton gauge
groups.  The resulting open spectra at the supersymmetric point 
are reported in Table \ref{opunw2p3},
and chirality emerges again both at brane intersections and due to the
chiral asymmetry present in the ``pure magnetic'' sector.  
It should be noticed that, as in \cite{aads}, the M\"obius-strip 
amplitudes must be suitably interpreted, since
naively they are not compatible with the corresponding annulus amplitudes.  As
is familiar from rational models, some missing parts must be identified with
differences of pairs of identical terms, 
one symmetrized and the other antisymmetrized by the
action of the open ``twist''\cite{zori1,review,croco}.  The real-charge solutions, present
only if the $B$-rank is non-vanishing, 
correspond to the choice 
\be
\xi_2 \ \xi_3 =  \eta_2 \  \eta_3 \ = \ - 1 \quad ,
\ee
with $\xi_1$ and $\eta_1$ again free signs.  In this case after 
rescaling the magnetic charge $m$ by a factor of two, a suitable 
parametrization for the Chan-Paton multiplicities is 
\ba
N &=& \ 2 \ ( \ n_1 + \ n_2) \quad , \qquad 
G \ = \ 2 \ ( \ n_1 - \ n_2 \ ) \quad; \nonumber \\
D_1 &=&  2 \ ( \ d_1 \ + \ d_2 \ ) \quad , \qquad 
G_1 \ = \ 2 \ ( \ d_1 \ - \ d_2 \ ) \quad; \nonumber \\
D_2 &=& 4 \ d_3 \quad .
\label{cpcharges}
\ea
Table \ref{cpw2p3r} displays the tadpole cancellation conditions 
and the allowed Chan-Paton gauge groups, while the resulting 
chiral open spectra are
exhibited in Table \ref{opunw2p3r}.

To conclude, let us mention that at the supersymmetric point,  {\it i.e.} 
for self-dual configurations of the background magnetic fields, the tadpoles
originating from the NS-NS sectors are also automatically canceled.
As in ref. \cite{aads}, they can be traced to corresponding derivatives of  
the Born-Infeld-type action for the untwisted sectors 
(the dilaton tadpole, for instance, is one of them).  
Moreover, the twisted NS-NS tadpoles are subtle: 
they are not perfect squares because of the behaviour of the 
magnetic field  under time reversal \cite{aads,review}.  
Still, they introduce additional couplings in the twisted NS-NS sectors
that are proportional to $H_2 - H_3$ and are thus canceled at the (self-dual)
supersymmetric point.

\subsubsection{$w_1 w_2 p_3$ Models}

Another interesting class of chiral orientifolds can be derived from 
deformations of the $w_1 w_2 p_3$ models.  
Since this is very similar to the $w_2 p_3$ case, we 
shall not perform a detailed description of all the amplitudes as 
in Section 4.2.1, but we shall just quote the results. 
The Klein bottle amplitude
\ba
{\cal K} &=& \frac{1}{8} \biggl\{ \bigl[ \  P_1\, P_2\, P_3\, + 2^{-4}\, 
P_1\, W_2(B_2)\, W_3(B_3)\, +\,  2^{-4}\,  W_1(B_1)\, P_2\,
W_3(B_3) \nonumber \\ &+ & \!\!  
2^{-4}\, (-1)^{n_1} \, W_1(B_1)\, (-1)^{n_2} W_2(B_2)\, 
(-1)^{m_3} P_3 \ \bigr] 
\, T_{oo} \nonumber \\ &+ & \!\! 
2 \times 16 \, \bigl[ \ 2^{-\frac{r_2}{2}-\frac{r_3}{2}}\, 
P_1^{\frac{1}{2}} \, T_{go} 
+  2^{-\frac{r_1}{2}-\frac{r_3}{2}}\, P_2^{\frac{1}{2}}\, T_{fo} 
\nonumber \\ &+ & \!\!  
2^{-\frac{r_1}{2}-\frac{r_2}{2}}\, 2^{-2}\, W_3^{\frac{1}{2}} 
(B_3)\, T_{ho} \ \bigr]\,  
\left( \frac{\eta}{\vartheta_4} \right)^2 \biggr\} \ ,
\label{kw1w2p3}
\ea
produces the unoriented closed spectra, whose massless part is reported in
Table \ref{clunw1w2p3}.  It should be noticed that the result does not depend 
on the rank of $B_{ab}$, in agreement with the 
considerations made in Section 2.1 
relating quantized values of $B_{ab}$ to shifts.  
An $S$ transformation of (\ref{kw1w2p3})
yields the transverse channel amplitude, that at the origin of the lattice
sums is identical to eq. (\ref{k0w2p3}).  As a result, the models
contain  $O9_+$,  $O5_{1+}$ and  $O5_{2+}$ planes.  
In order to neutralize the R-R charge, (magnetized) $D9$-branes, 
$D5_1$-branes and $D5_2$-branes are introduced.  Due to the triple shifts, 
the tadpole cancellation 
conditions derive only from the untwisted sectors, and their analysis is
very similar to the one in Section 4.2.1.  After a proper normalization, 
the Chan-Paton charge multiplicities are displayed in Table \ref{cpw1w2p3}, 
where
the allowed Chan-Paton gauge groups are also reported.  Apart from the 
$m$ charges, all others are real, as emerges from the open
unoriented chiral spectra shown in Table \ref{opunw1w2p3}.  

\subsubsection{Non-chiral Models}

In this Section we discuss the remaining models in Table \ref{oldmodels} 
that admit magnetic deformations, namely those containing $D5$-branes 
along $T^{45}$ that can
absorb the R-R charge flux of the magnetized $D9$-branes.  
It is easy to see that the $p_2 p_3$, $w_1 p_2$, $w_1 p_2 p_3$ and 
$w_1 p_2 w_3$ models do admit magnetic deformations, while the $p_1 p_2 p_3$,
$p_1 w_2 w_3$ and $w_1 w_2 w_3$ do not.
The four models are quite different, but inherit an effective world-sheet 
parity projection that allows one to express the Klein bottle amplitude in the 
following form:
\ba
{\cal K} &=& \frac{1}{8} \biggl\{ \bigl[ \  P_1\, P_2\, P_3\, + 2^{-4}\, 
P_1\, W_2(B_2)\, W_3(B_3)\, +\,  2^{-4}\,  W_1(B_1)\, (-1)^{\delta_2} \, P_2\,
(-1)^{\delta_3} \, W_3(B_3) \nonumber \\ &+ & \!\!  
2^{-4} \, (-1)^{\delta_1} W_1(B_1) \, (-1)^{\delta_2} \, W_2(B_2)\,
(-1)^{\delta_3} \, P_3 \ \bigr] 
\, T_{oo} \nonumber \\ &+ & \!\! 
2 \times 16 \ 2^{-\frac{r_2}{2}-\frac{r_3}{2}}\, {P_1}^{\lambda_1} \, T_{go} \ 
\left( \frac{\eta}{\vartheta_4} \right)^2 \biggr\} \ ,
\label{kp2p3}
\ea
where $\lambda_1$ is $0$ if $\delta_1 = p_1$ and $\frac{1}{2}$ if 
$\delta_1 = w_1$, while obviously 
the shifts affect the sums only if they are present 
in the corresponding $\sigma$ table and are of the same type as the 
lattice sums.
The transverse channel gives the $O$-plane content in the four cases, that is
expected to be the same for the four classes of models. 
Indeed, at the origin of the lattices
\ba
\tilde{\cal K}_0 &=& \frac{2^5}{8} \Biggl\{ \ \left( \sqrt{v_1v_2v_3} +
 2^{-\frac{r_2}{2}-\frac{r_3}{2}} \sqrt{\frac{v_1}{v_2 v_3}} \ \right)^2 
\ \left( \, \tau_{oo} + \tau_{og} \, \right) \nonumber \\ 
&+&  \left( \sqrt{v_1v_2v_3} +
 2^{-\frac{r_2}{2}-\frac{r_3}{2}} \sqrt{\frac{v_1}{v_2 v_3}} \ 
\right)^2 \  \left( \, \tau_{oh} + \tau_{of} \ \right) \, \Biggr\} \quad ,
\label{k0w2p3four}
\ea
so that in all  four classes of models only $O9_+$ and $O5_{1+}$ are present.
On the other hand, both the unoriented closed spectra and the open sectors are
quite distinct, as can be deduced from the diverse $D$-brane multiplet
configurations of the undeformed models in Table \ref{opunundef}.  
Of course, only (magnetized) $D9$-branes 
and $D5_1$-branes are needed, with a consequent lack of chirality.
The unoriented closed spectra of the four classes of models can 
be found in Tables \ref{clunp2p3}, \ref{clunw1p2} and \ref{clunw1p2p3}.  
Only the $p_2 p_3$ models show a dependence on the rank of $B_{ab}$, while
the two models with three shifts have identical massless closed spectra.

There are two different $p_2 p_3$ unoriented partition functions, 
that differ in the open-string sectors, 
depending on the sign freedom for the M\"obius projections.  With complex 
and properly normalized charges, untwisted
and twisted tadpole cancellation conditions are summarized in 
Table \ref{cpp2p3}, where the resulting  Chan-Paton gauge 
groups are also reported.  The open spectra can be read from 
Table \ref{opunp2p3}, where $R$ stands for the symmetric 
representation if $\eta_1=+1$, or for the antisymmetric representation
if $\eta_1=-1$.  
The second solution is linked to a real 
parametrization of the Chan-Paton charges that results into
tadpole cancellation conditions and gauge groups as in Table \ref{cpp2p3r}.
It should be noticed that in this case group factors, other than $U(m)$, 
must be all orthogonal or all symplectic.  The massless 
open spectra can be found in Table \ref{opunp2p3r}.

The $w_1 p_2$ models and the $w_1 p_2 p_3$ 
models are very similar and, independently of the presence of $H_i$, 
differ solely in their massive 
excitations.  In other words, the analysis of the massless
excitations is not sufficient to distinguish  these two classes of
models.  Their unoriented closed spectra are different, as emerges from
Tables \ref{clunw1p2} and  \ref{clunw1p2p3}, but they have identical open
spectra.  The tadpole cancellation conditions and the resulting 
Chan-Paton groups are reported in Table \ref{cpw1p2} 
for complex charges and in Table \ref{cpw1p2r}
for real charges.  The non-chiral and coincident open spectra are reported 
in Table \ref{opunw1p2} for the complex charge cases, and in Table 
\ref{opunw1p2r} for the real charge cases.

Finally, the $w_1 p_2 w_3$ models exhibit unoriented bulk spectra 
identical to the one of the $w_1 p_2 p_3$ models in Table 
\ref{clunw1p2p3}, but with tadpole
conditions and Chan-Paton groups as in Table \ref{cpw1p2w3}.
The resulting non-chiral massless open spectra are contained in Table 
\ref{opunw1p2w3}.

\section{Brane Supersymmetry Breaking}

In this Section we discuss one significant class of magnetized orientifolds
in which the field configurations are chosen so that 
magnetized $D9$-branes mimic anti-$D5$-branes rather than $D5$-branes, thus 
breaking supersymmetry in the open-string sector (``brane 
supersymmetry breaking'').  We analyze in some detail 
a variant of the $w_2p_3$ class of models 
extensively discussed at the supersymmetric point in Section 4.2.1. 
For simplicity, we shall confine ourselves
to the $B_{ab}=0$ case. The  oriented closed spectrum is always the one 
contained in Table \ref{clorz2z2sh}, but the Klein-bottle 
projection, described by 
\ba
{\cal K} &=& \frac{1}{8} \biggl\{ \bigl[ \  P_1\, P_2\, P_3\, + \, 
P_1\, W_2 \, W_3 \, + \,  W_1 \, P_2\,
W_3 + W_1 \, (-1)^{n_2} W_2 \, (-1)^{m_3} P_3 \ \bigr] 
\, T_{oo} \nonumber \\ &-& \!\! 
2 \times 16 \, \bigl[ \ T_{go} \, P_1 \,  
- \, T_{fo} \, {P_2}^{\frac{1}{2}} \, + \, W_3^{\frac{1}{2}} \, T_{ho} 
\ \bigr] \,  
\left( \frac{\eta}{\vartheta_4} \right)^2 \biggr\} \ ,
\label{kw2p3bsb}
\ea
is now different, due to the inversion of some signs, 
and produces the massless  unoriented closed spectra of 
Table \ref{clunw2p3bsb}.
The transverse channel amplitude at the lattice origin,
\ba
\tilde{\cal K}_0 &=& \frac{2^5}{8} \Biggl\{ \ \left( \sqrt{v_1v_2v_3} -
 \sqrt{\frac{v_1}{v_2 v_3}} - 
 \sqrt{\frac{v_2}{v_1 v_3}} \ \right)^2 \tau_{oo} \nonumber \\ 
&+&  \left( \sqrt{v_1v_2v_3} -
 \sqrt{\frac{v_1}{v_2 v_3}} +
 \sqrt{\frac{v_2}{v_1 v_3}}
 \ \right)^2 \tau_{og} \nonumber \\ 
&+&  \left( \sqrt{v_1v_2v_3} +
 \sqrt{\frac{v_1}{v_2 v_3}} +
 \sqrt{\frac{v_2}{v_1 v_3}}
 \ \right)^2 \tau_{oh} \nonumber \\ 
&+&  \left( \sqrt{v_1v_2v_3} +
 \sqrt{\frac{v_1}{v_2 v_3}} - 
 \sqrt{\frac{v_2}{v_1 v_3}}
 \ \right)^2 \tau_{of} \ \Biggr\} \quad ,
\label{ktw2p3bsb}
\ea
displays very neatly the 
presence of one $O9_+$-plane and of the two ``exotic'' $O5_{1-}$ and $O5_{2-}$ 
planes, that require the introduction of anti-$D5_1$-branes and 
anti-$D5_2$-branes, together with the (magnetized) $D9$-branes.  
In order to neutralize the 
global R-R charge, one has to sit at the antiself-dual background
field configuration, corresponding to $H_2=H_3$ in our conventions.
Only the R-R tadpole
cancellation conditions can be imposed, while the NS-NS tadpoles survive,
signaling, as customary, the need for a non-Minkowskian 
vacuum \cite{dilta}.  The results for the Chan-Paton gauge groups are 
displayed in Table \ref{cpw2p3bsb}, while the open
and unoriented massless spectra are displayed in Table 
\ref{opunw2p3bsb}.  As usual, in the models with ``brane supersymmetry 
breaking'' supersymmetry is
exact at tree level on the $D9$-branes  but it is only
non-linearly realized on the anti-$D5$-branes \cite{dudmo,prari}.  
This can be foreseen from Table 
\ref{opunw2p3bsb}, where modes originally in a given 
supermultiplet are assigned
to different gauge group representations.

\section{Conclusions}

In this paper we have analyzed in detail four dimensional orientifolds
originating from $Z_2 \times Z_2$ toroidal orbifolds and from freely acting 
$Z_2 \times Z_2$ shift-orbifolds of the type IIB superstring, in the
presence of uniform background magnetic fluxes
along four of the six internal directions and of a quantized NS-NS $B_{ab}$,
that has been shown to be equivalent to an asymmetric 
shift-orbifold projection.  These models are connected by
T-duality to models with branes intersecting at angles and   
contain magnetized
$D9$-branes charged also with respect to the R-R six-form, thus exhibiting 
several interesting novelties.  In particular, for suitable self-dual 
configurations of the
internal backgrounds, that in the T-dual picture correspond to 
suitable angles between the branes, it is possible to obtain non tachyonic four 
dimensional supersymmetric models with spectra containing in a natural way 
several families of matter fields whose numbers are
 related to the multiplicities
of the Landau levels.  Moreover, the instanton-like behaviour of
the ``magnetized'' $D9$-branes that mimic localized $D5$-branes produces
an interesting rank reduction of the Chan-Paton gauge groups.
As a bonus, if $D5$ branes longitudinal to the directions of the internal
magnetic fields are present, the models can acquire chiral spectra, due to the
unpairing of fermions at the intersections and to the chiral asymmetry 
in the ``pure magnetic'' sectors.  Geometrically, chirality is related to
configurations in which $D$-branes are not parallel to the corresponding
$O$-planes, differently from the models of ref. \cite{chiralas}, 
where the $D9$-branes are parallel to the $O9$-planes, but the
orbifold projection produces only left-handed fermions.  
Introducing antiself-dual
background fields, it is also possible to obtain non-tachyonic models
with ``brane supersymmetry breaking'', for which supersymmetry is
exact at tree level in the bulk and on the $D9$-branes, but is 
non-linearly realized and thus effectively broken at
the string scale on the anti-$D5$ branes and on the equivalent magnetized 
$D9$-branes.  The chiral four dimensional 
models can be used  to build
realistic extensions of the Standard Model in a brane-world like scenario,
introducing brane-antibrane pairs or Wilson lines. 
This is a very interesting and widely pursued effort, but the dynamical
stability of all these vacua is still an open question.  It would be 
also interesting to analyze in some detail 
mechanisms to reduce the number of moduli, 
for instance introducing  background fluxes, as recently discussed in refs.
\cite{fluxes}.

\vskip 24pt
\noindent{\bf Acknowledgements} 

It is a pleasure to acknowledge C. Angelantonj, P. Bantay, M Berg, M. Bianchi,
R. Blumenhagen, E. Dudas, L. Genovese, M. Haack, J. Mourad, Ya.S. Stanev and
A.M. Uranga for stimulating discussions and especially A. Sagnotti for 
the collaboration at the early stages of this work and for several
illuminating discussions.   G.P. is grateful to the Theoretical 
Physics Department of 
the E\"otv\"os Lor\'and University of Budapest for the kind hospitality 
while this work was being completed.  
This work was supported in part by I.N.F.N., by the
E.C. RTN programs HPRN-CT-2000-00122 and HPRN-CT-2000-00148, by the
INTAS contract 99-1-590, by the MIUR-COFIN contract 2001-025492 and
by the NATO contract PST.CLG.978785.
\clearpage

\appendix
\def\thesubsection{\Alph{section}.\arabic{subsection}}
\section{Lattice Sums in the Presence of a Quantized $B_{ab}$}

In this Appendix we collect the relevant lattice sums that enter the one-loop 
partition functions.
We follow mainly the notation of \cite{review}, and display  only the 
sums modified by the presence of an antisymmetric tensor $B_{ab}$. 
Since each  surface of vanishing Euler number has a different double cover, 
the sums also differ in their proper time dependence. 
We will denote with $\tau$
the loop channel modulus of each surface and with $\ell$ the modulus 
of the doubly covering tori.         
Let us begin by recalling that, in presence of a $B_{ab}$ background, 
the generalized $d$-dimensional momenta $p_{\rm L}$ and $p_{\rm R}$  
are  \cite{narain}:
\ba 
p_{{\rm L},a} = m_a + \frac{1}{\alpha'} ( g_{ab} - B_{ab} ) \, n^b
\, , \label{pleft} 
\\ 
p_{{\rm R},a} = m_a - \frac{1}{\alpha'} ( g_{ab} + B_{ab}
) \, n^b \, . \label{pright}
\ea 
The corresponding lattice sums on the torus take the form
\be 
\Lambda(B) \ = \ \sum_{m,n}
\frac{q^{\frac{\alpha'}{4} p_{\rm L}^T g^{-1} p_{\rm L}} \
\bar{q}^{\frac{\alpha'}{4} p_{\rm R}^T g^{-1} p_{\rm R}}}{|\eta(\tau)|^{2d}}
\, , 
\label{2dtorus}
\ee
as in ref. \cite{narain}. 
For the direct-channel Klein-bottle amplitudes, only the winding sums are 
modified and become
\be
W(B) =  \sum_{\epsilon =0,1} \sum_n 
{q^{{1\over 2\alpha '} n^{{\rm T}} g n} \, e^{{2 i \pi
\over \alpha '} n^{{\rm T}} B \epsilon} \over \eta ^2(2 i \tau)} \quad ,
\ee
while in the transverse channel the momentum sums are
\be
P(B) = \sum_{\epsilon=0,1} \sum_m { ( e^{-2\pi
\ell})^ {\alpha ' (m + {1\over \alpha '} B \epsilon )^{{\rm T}} g^{-1}
(m + {1\over \alpha '} B\epsilon)} \over \eta ^2(i \ell) } \quad.
\ee
In the annulus amplitudes the situation is reverted, 
and modified momentum sums 
\be
P(B) = \sum_{\epsilon =0,1} \sum_m
{q^{{\alpha '\over 2} (m + {1\over \alpha '} B\epsilon )^{{\rm T}}
g^{-1} (m + {1\over \alpha '} B\epsilon )} \over \eta ^2(i \tau/2)} \quad, 
\ee
appear in the direct channel, 
while modified winding sums 
\be
W(B) =  \sum_{\epsilon =0,1} \sum_n
{ (e^{-2\pi\ell})^{{1\over 4\alpha '} n^{{\rm T}} g n} 
e^{{2 i\pi \over \alpha '} n^{{\rm T}} B\epsilon} \over \eta ^2(i \ell)} 
\ee
appear in the transverse channel.
The direct M\"obius amplitudes involve 
\be
P(B,\gamma_{\epsilon}) = \sum_{\epsilon =0,1} \sum_m
{q^{{\alpha '\over 2} (m + {1\over \alpha '} B\epsilon )^{{\rm T}}
g^{-1} (m + {1\over \alpha '} B\epsilon )} \ \gamma_{\epsilon} \over 
{\hat{\eta}^2({i \tau \over 2} + {1 \over 2})}}  
\ee
and
\be
W(B, \tilde{\gamma}_{\epsilon}) =  \sum_{\epsilon =0,1} \sum_n 
{q^{{1\over 2\alpha '} n^{{\rm T}} g n} \, e^{{2 i \pi
\over \alpha '} n^{{\rm T}} B \epsilon}  \tilde{\gamma}_{\epsilon} 
\over \hat{\eta}^2({i \tau \over 2} + {1 \over 2})} \quad ,
\ee
while the transverse M\"obius amplitudes involve
\be
W(B,\gamma_{\epsilon}) =  \sum_{\epsilon =0,1} \sum_n
{ (e^{-2\pi\ell})^{{1\over 4\alpha '} n^{{\rm T}} g n} 
e^{{2 i\pi \over \alpha '} n^{{\rm T}} B\epsilon} \ \gamma_{\epsilon} 
\over \hat{\eta} ^2(i \ell)} 
\ee
and
\be
P(B,\tilde{\gamma}_{\epsilon}) = \sum_{\epsilon=0,1} \sum_m { ( e^{-2\pi
\ell})^ {\alpha ' (m + {1\over \alpha '} B \epsilon )^{{\rm T}} g^{-1}
(m + {1\over \alpha '} B\epsilon)} \ \tilde{\gamma}_{\epsilon} \over \hat{
\eta} ^2(i \ell) } \quad.
\ee
All sums displayed in this Appendix are two-dimensional, 
since for simplicity the six-dimensional internal torus
is chosen to be factorized 
as a product of two-dimensional tori, while the corresponding 
antisymmetric two-tensor is also 
chosen, for simplicity, 
in a block-diagonal form of two-by-two matrices.

\section{Characters for the $T^6/Z_2 \times Z_2$ Orbifolds}

In this Appendix we list the $Z_2 \times Z_2$ 
characters that enter the one-loop amplitudes. Using the conventions
of ref. \cite{review}, they may be
expressed as ordered products of the four SO(2) level-one 
characters, $O_2$, $V_2$, $S_2$ and $C_2$, as follows:

\ba
\tau_{oo}&=&V_2O_2O_2O_2+O_2V_2V_2V_2-S_2S_2S_2S_2-C_2C_2C_2C_2 \ , 
\nonumber \\
\tau_{og}&=&O_2V_2O_2O_2+V_2O_2V_2V_2-C_2C_2S_2S_2-S_2S_2C_2C_2 \ , 
\nonumber \\
\tau_{oh}&=&O_2O_2O_2V_2+V_2V_2V_2O_2-C_2S_2S_2C_2-S_2C_2C_2S_2 \ , 
\nonumber \\
\tau_{of}&=&O_2O_2V_2O_2+V_2V_2O_2V_2-C_2S_2C_2S_2-S_2C_2S_2C_2 \ , 
\nonumber \\
\tau_{go}&=&V_2O_2S_2C_2+O_2V_2C_2S_2-S_2S_2V_2O_2-C_2C_2O_2V_2 \ , 
\nonumber \\
\tau_{gg}&=&O_2V_2S_2C_2+V_2O_2C_2S_2-S_2S_2O_2V_2-C_2C_2V_2O_2 \ , 
\nonumber \\
\tau_{gh}&=&O_2O_2S_2S_2+V_2V_2C_2C_2-C_2S_2V_2V_2-S_2C_2O_2O_2 \ , 
\nonumber \\
\tau_{gf}&=&O_2O_2C_2C_2+V_2V_2S_2S_2-S_2C_2V_2V_2-C_2S_2O_2O_2 \ , 
\nonumber \\
\tau_{ho}&=&V_2S_2C_2O_2+O_2C_2S_2V_2-C_2O_2V_2C_2-S_2V_2O_2S_2 \ , 
\nonumber \\
\tau_{hg}&=&O_2C_2C_2O_2+V_2S_2S_2V_2-C_2O_2O_2S_2-S_2V_2V_2C_2 \ , 
\nonumber \\
\tau_{hh}&=&O_2S_2C_2V_2+V_2C_2S_2O_2-S_2O_2V_2S_2-C_2V_2O_2C_2 \ , 
\nonumber \\
\tau_{hf}&=&O_2S_2S_2O_2+V_2C_2C_2V_2-C_2V_2V_2S_2-S_2O_2O_2C_2 \ , 
\nonumber \\
\tau_{fo}&=&V_2S_2O_2C_2+O_2C_2V_2S_2-S_2V_2S_2O_2-C_2O_2C_2V_2 \ , 
\nonumber \\
\tau_{fg}&=&O_2C_2O_2C_2+V_2S_2V_2S_2-C_2O_2S_2O_2-S_2V_2C_2V_2 \ , 
\nonumber \\
\tau_{fh}&=&O_2S_2O_2S_2+V_2C_2V_2C_2-C_2V_2S_2V_2-S_2O_2C_2O_2 \ , 
\nonumber \\
\tau_{ff}&=&O_2S_2V_2C_2+V_2C_2O_2S_2-C_2V_2C_2O_2-S_2O_2S_2V_2 \ . 
\label{a3}
\ea

\section{Massless Spectra}

This Appendix collects the massless spectra of all the models in the paper.  
$N$ indicates the number of supersymmetries and $H$, $V$, 
$C$ and $C_{L,R}$ denote hypermultiplets, vector multiplets and 
chiral multiplets with a Majorana or a Weyl (left or right) 
spinor, respectively. 
$CY(h_{11},h_{12})$ is referred to 
the fact that the related orbifold is a singular limit of a 
Calabi-Yau compactification with Hodge numbers 
$(h_{11},h_{12})$, $\omega$ is the discrete 
torsion while $\omega_i$ are signs that satisfy 
$\omega_1 \, \omega_2 \, \omega_3 \, = \, \omega$ (Cfr. Section 2).  
$k_i$ are the integer multiplicities 
of the Landau Level degeneracies, 
$r_j$ is the rank of the two-by-two $j$-th block
of the $B_{ab}$ NS-NS antisymmetric tensor and $r=r_1+r_2+r_3$ is the 
total $B$-rank. The $\eta_i$ and  $\xi_i$, introduced in eq. 
(\ref{directgamma}) can be $\pm 1$, and
both choices are allowed if their values are not specified.
For what concerns the Chan-Paton gauge groups, $F$, $S$, $A$ and $Adj$ denote 
respectively the
Fundamental, Symmetric, Antisymmetric and Adjoint representations. 
When two Chan-Paton groups or two multiplets are within brackets, 
one of the two factors can be chosen independently.  
Finally, the notation related to the Chan-Paton charges or to the number of
branes reserves the $n$'s to the uncharged $D9$-branes, the $m$'s to the
magnetized $D9$-branes and the $d$'s to the $D5_i$-branes. 
\clearpage
\vskip 8pt
\subsection{Closed Spectra of the $Z_2 \times Z_2$ orbifolds}
\vskip 8pt
\begin{table}[ht]
\begin{center}
\begin{tabular}{||c||c|c|c|c|c|c||}\hline\hline
{}&{\rm untwisted}&{\rm untwisted}&
{untwisted}&{twisted}&{twisted}&{}\\ 
{\rm $\omega$}&{\rm SUGRA}&{H}&{V}&{H}&{V}&{}\\\hline\hline
{$+ 1$}&{$N=2$}&{$1+3$}&{3}&{$16+16+16$}&{$0$}&{CY $(3,51)$}\\\hline
{$- 1$}&{$N=2$}&{$1+3$}&{3}&{$0$}&{$16+16+16$}&{CY $(51,3)$}\\\hline
\end{tabular} 
\caption{ Oriented closed spectra of $Z_2 \times Z_2$ orbifolds.}
\label{clorz2z2}
\end{center}
\end{table}
\begin{table}[b]
\begin{center}
\begin{tabular}{||c||c||c|c|c|c||}\hline\hline
{}&{}&{\rm untwisted}&{untwisted}&{twisted}
&{twisted}\\ 
{\rm $(\omega_1,\omega_2,\omega_3)$}&{\rm $\omega$}&
{\rm SUGRA}&{C}&{C}&{V}\\\hline\hline
{$(+,+,+)$}&{+}&{$N=1$}&{$1+3+3$}&{$16+16+16$}&{$0$}\\\hline
{$(+,-,-)$}&{+}&{}&{}&{}&{}\\
{$(-,+,-)$}&{+}&{$N=1$}&{$1+3+3$}&{$16$}&{$16+16$}\\
{$(-,-,+)$}&{+}&{}&{}&{}&{}\\\hline\hline
{$(-,-,-)$}&{-}&{}&{}&{}&{}\\
{$(+,+,-)$}&{-}&{$N=1$}&{$1+3+3$}&{$16+16+16$}&{$0$}\\
{$(+,-,+)$}&{-}&{}&{}&{}&{}\\
{$(-,+,+)$}&{-}&{}&{}&{}&{}\\\hline\hline
\end{tabular}
\caption{Unoriented closed spectra of the $Z_2 \times Z_2$ orbifolds.}
\label{clunz2z2}
\end{center}
\end{table}
\begin{table}[ht]
\begin{center}
\begin{tabular}{||c|c|c||c|c|c|c||}\hline\hline
\multicolumn{3}{||c||}{\rm $B$ rank}&{\rm untwisted}&{untwisted}&{twisted}
&{twisted}\\ 
{\rm $r_1$}&{\rm $r_2$}&{\rm $r_3$}&{\rm SUGRA}&{C}&{C}&{V}\\\hline\hline
{$0$}&{$0$}&{$0$}&{$N=1$}&{$1+3+3$}&{$16+16+16$}&{$0$}\\\hline\hline
{$2$}&{$0$}&{$0$}&{$N=1$}&{$1+3+3$}&{$16+12+12$}&{$0+4+4$}\\\hline
{$0$}&{$2$}&{$0$}&{$N=1$}&{$1+3+3$}&{$12+16+12$}&{$4+0+4$}\\\hline
{$0$}&{$0$}&{$2$}&{$N=1$}&{$1+3+3$}&{$12+12+16$}&{$4+4+0$}\\\hline\hline
{$2$}&{$2$}&{$0$}&{$N=1$}&{$1+3+3$}&{$12+12+10$}&{$4+4+6$}\\\hline
{$0$}&{$2$}&{$2$}&{$N=1$}&{$1+3+3$}&{$10+12+12$}&{$6+4+4$}\\\hline
{$2$}&{$0$}&{$2$}&{$N=1$}&{$1+3+3$}&{$12+10+12$}&{$4+6+4$}\\\hline\hline
{$2$}&{$2$}&{$2$}&{$N=1$}&{$1+3+3$}&{$10+10+10$}&{$6+6+6$}\\\hline\hline
\end{tabular} 
\caption{Unoriented closed spectra of the $Z_2 \times Z_2$ orbifolds 
with $\omega_i=+1$.}
\label{clunomeg1}
\end{center}
\end{table}
\clearpage
\subsection
{Open spectra of the $Z_2 \times Z_2$ orientifolds with $\omega_i=+1$}
\begin{table}[ht]
\begin{center}
\begin{tabular}{c}
\hline\hline
{$\biggl(\begin{array}{c} USp(n) \\ SO(n) \end{array} \biggr) \ 
\otimes \ \biggl(\begin{array}{c} USp(d_1) \\ SO(d_1) \end{array} \biggr) \
 \otimes \ \biggl(\begin{array}{c} USp(d_2) \\ SO(d_2) \end{array} \biggr) \ 
\otimes \ \biggl(\begin{array}{c} USp(d_3) \\ SO(d_3) \end{array} 
\biggr)$}\\\hline\hline
{$n = d_1 = d_2 = d_3 = 16 \ 2^{-{r \over 2}}$}\\\hline\hline
\end{tabular}
\caption{Chan-Paton groups and tadpole conditions for the 
$[T^2 \times T^2 \times T^2]/Z_2 \times Z_2$
models.}
\label{cpz2z2}
\end{center}
\end{table}
\begin{table}
\begin{center}
\begin{tabular}{||c|c|c||}\hline\hline
{\rm Multiplets}&{Number}&{\rm Rep.}\\\hline\hline
{\rm $C$}&{$\biggl(\begin{array}{c}3\\0\end{array} \biggr)$ or 
$\biggl(\begin{array}{c}1\\2\end{array} \biggr)$ if $USp$ \ \ ; \ \ 
$\biggl(\begin{array}{c}2\\1\end{array} \biggr)$ or 
$\biggl(\begin{array}{c}0\\3\end{array} \biggr)$ if $SO$}&
{$\biggl(\begin{array}{c}  $(A,1,1,1)$ \\  $(S,1,1,1)$ \end{array} 
\biggr)$}\\ \hline
{\rm $C$}&{$\biggl(\begin{array}{c}3\\0\end{array} \biggr)$ or 
$\biggl(\begin{array}{c}1\\2\end{array} \biggr)$ if $USp$ \ \ ; \ \ 
$\biggl(\begin{array}{c}2\\1\end{array} \biggr)$ or 
$\biggl(\begin{array}{c}0\\3\end{array} \biggr)$ if $SO$}&
{$\biggl(\begin{array}{c}  $(1,A,1,1)$ \\  $(1,S,1,1)$ \end{array} 
\biggr)$}\\ \hline
{\rm $C$}&{$\biggl(\begin{array}{c}3\\0\end{array} \biggr)$ or 
$\biggl(\begin{array}{c}1\\2\end{array} \biggr)$ if $USp$ \ \ ; \ \ 
$\biggl(\begin{array}{c}2\\1\end{array} \biggr)$ or 
$\biggl(\begin{array}{c}0\\3\end{array} \biggr)$ if $SO$}&
{$\biggl(\begin{array}{c}  $(1,1,A,1)$ \\  $(1,1,S,1)$ \end{array} 
\biggr)$}\\ \hline
{\rm $C$}&{$\biggl(\begin{array}{c}3\\0\end{array} \biggr)$ or 
$\biggl(\begin{array}{c}1\\2\end{array} \biggr)$ if $USp$ \ \ ; \ \ 
$\biggl(\begin{array}{c}2\\1\end{array} \biggr)$ or 
$\biggl(\begin{array}{c}0\\3\end{array} \biggr)$ if $SO$}&
{$\biggl(\begin{array}{c}  $(1,1,1,A)$ \\  $(1,1,1,S)$ \end{array} 
\biggr)$}\\ \hline
{\rm $C$}&{$2^{\frac{r_2+r_3}{2}}$}&{$(F,F,1,1),(1,1,F,F) $}\\\hline
{\rm $C$}&{$2^{\frac{r_1+r_3}{2}}$}&{$(F,1,F,1),(1,F,1,F)$}\\\hline
{\rm $C$}&{$2^{\frac{r_1+r_1}{2}}$}&{$(F,1,1,F),(1,F,F,1)$}\\\hline
\end{tabular}
\caption{Open spectra of the $Z_2 \times Z_2$ orientifold with 
$\omega=1$.}
\label{ouz2z2}
\end{center}
\end{table}
\clearpage
\subsection{Spectra of the $[ T^2(H_2) \times T^2(H_3)] / Z_2$ 
orientifolds}
\begin{table}[ht]
\begin{center}
\begin{tabular}{||c|c|c||}\hline\hline
{\rm untwisted}&{\rm untwisted}&{twisted}\\ 
{\rm SUGRA}&{T}&{T}\\\hline\hline
{$N=(2,0)$}&{$1+4$}&{$16$}\\\hline\hline
\end{tabular} 
\vskip 5pt
\caption{Oriented closed spectra of the $[T^2 \times T^2]/Z_2$
 Orbifolds.}
\label{coz2d6}
\end{center}
\end{table}
\vskip 8pt
\begin{table}
\begin{center}
\begin{tabular}{||c||c|c|c|c|c||}\hline\hline
{B rank}&{\rm untwisted}&{\rm untwisted}&
{untwisted}&{twisted}&{twisted}\\ 
{\rm $r$}&{\rm SUGRA}&{H}&{T}&{H}&{T}\\\hline\hline
{$0$}&{$N=(1,0)$}&{$4$}&{1}&{$16$}&{$0$}\\\hline
{$2$}&{$N=(1,0)$}&{$4$}&{1}&{$14$}&{$2$}\\\hline
{$4$}&{$N=(1,0)$}&{$4$}&{1}&{$10$}&{$6$}\\\hline\hline
\end{tabular} 
\caption{Unoriented closed spectra of the 
$[T^2 \times T^2]/Z_2$ orientifolds.}
\label{cuz2d6}
\end{center}
\end{table}
\vskip 8pt
\begin{table}
\begin{center}
\begin{tabular}{c}
\hline\hline
{$U(n)\otimes U(d)\otimes U(m)$}\\\hline\hline
{$n + \bar{n} + m  + \bar{m}  =  32 \ 2^{-{r \over 2}}$}\\
{$d + \bar{d} + 2^{r \over 2} \ | k_2 k_3 | 
\ (m + \bar{m})   =  32 \  2^{-{r \over 2}}$}\\
{$ n = \bar{n} \qquad ; \qquad d  = \bar{d} \qquad ; \qquad 
m = \bar{m}$}\\\hline\hline
\end{tabular}
\caption{Chan-Paton groups and tadpole conditions for the 
$[T^2 \times T^2]/Z_2$ models (complex charges).}
\label{cpz2d6}
\end{center}
\end{table}
\vskip 8pt
\begin{table}
\begin{center}
\begin{tabular}{||c|c|c||}\hline\hline
{\rm Multiplets}&{Number}&{\rm Rep.}\\\hline\hline
{\rm H}&{$1$}&{$(A+\bar{A},1,1)$}\\\hline
{\rm H}&{$1$}&{$(1,A+\bar{A},1)$}\\\hline
{\rm H}&{$2^{r} \ |k_2 k_3| - 4$}&{$(F,1,F)$}\\\hline
{\rm H}&{$2^{r} \ |k_2 k_3| + 4$}&{$(\bar{F},1,F)$}\\\hline
{\rm H}&{$(2^{r} + 2^{r/2}) \, |k_2 k_3| -2$}&{$(1,1,A)$}\\\hline
{\rm H}&{$(2^{r} - 2^{r/2}) \, |k_2 k_3|$}&{$(1,1,S)$}\\\hline
{\rm H}&{$2^{r/2}$}&{$(F,\bar{F},1)$}\\\hline
{\rm H}&{$2^{r/2}$}&{$(1,\bar{F},F)$}\\\hline\hline
\end{tabular} 
\caption{Open spectra of the 
$[ T^2(H_2) \times T^2(H_3)] / Z_2$ orientifolds (complex charges).}
\label{ouz2d6}
\end{center}
\end{table}
\vskip 8pt
\begin{table}
\begin{center}
\begin{tabular}{c}
\hline\hline
{$USp(n_1)\otimes USp(n_2)\otimes USp(d_1)\otimes 
USp(d_2)\otimes U(m)$}\\\hline\hline
{$n_1 + n_2 + m  + \bar{m}  =  32 \ 2^{-{r \over 2}}$}\\
{$d_1 + d_2 + 2^{r \over 2} \ | k_2 k_3 | 
\ (m + \bar{m})   =  32 \  2^{- {r \over 2}}$}\\
{$ n_2 = n_1 + m + \bar{m} \qquad ; \qquad d_1  = d_2 \qquad ; \qquad 
m = \bar{m}$}\\\hline\hline
\end{tabular}
\caption{Chan-Paton groups and tadpole conditions for the 
$[T^2 \times T^2]/Z_2$ models (real charges).}
\label{cpz2d6r}
\end{center}
\end{table}
\vskip 8pt
\begin{table}
\begin{center}
\begin{tabular}{||c|c|c||}\hline\hline
{\rm Multiplets}&{Number}&{\rm Rep.}\\\hline\hline
{\rm H}&{$1$}&{$(F,F,1,1,1)$}\\\hline
{\rm H}&{$1$}&{$(1,1,F,F,1)$}\\\hline
{\rm H}&{$(2^{r} + 2^{r/2}) \, |k_2 k_3| - 2$}&{$(1,1,1,1,A)$}\\\hline
{\rm H}&{$(2^{r} - 2^{r/2}) \, |k_2 k_3|$}&{$(1,1,1,1,S)$}\\\hline
{\rm H}&{$2^{r} \ 2  \, |k_2 k_3| + 2$}&{$(F,1,1,1,F)$}\\\hline
{\rm H}&{$2^{r} \ 2  \, |k_2 k_3| - 2$}&{$(1,F,1,1,F)$}\\\hline
{\rm H}&{$2^{r/2 - 1}$}&{$(F,1,F,1,1)$}\\\hline
{\rm H}&{$2^{r/2 - 1}$}&{$(1,F,1,F,1)$}\\\hline
{\rm H}&{$2^{r/2}$}&{$(1,1,F,1,F)$}\\\hline\hline
\end{tabular} 
\caption{Open spectra of the 
$[ T^2(H_2) \times T^2(H_3)] / Z_2$ orientifolds (real charges).}
\label{ouz2d6r}
\end{center}
\end{table}
\clearpage
\subsection
{Open Spectra of the Magnetized $Z_2 \times Z_2$ Orientifolds 
with $\omega_i=+1$}
\begin{table}[ht]
\begin{center}
\begin{tabular}{c}
\hline\hline
{$\biggl(\begin{array}{c} USp(n) \\ SO(n) \end{array} \biggr) \ 
\otimes \ \biggl(\begin{array}{c} USp(d_1) \\ SO(d_1) \end{array} \biggr) \
 \otimes \ \biggl(\begin{array}{c} USp(d_2) \\ SO(d_2) \end{array} \biggr) \ 
\otimes \ \biggl(\begin{array}{c} USp(d_3) \\ SO(d_3) \end{array} \biggr) \ 
\otimes \ U(m)$}\\\hline\hline
{$n + m  + \bar{m}  =  16 \ 2^{-{r \over 2}}$}\\
{$d_1 + 2^{{r_2 \over 2} + {r_3 \over 2}} \ | k_2 k_3 | 
\ (m + \bar{m})   =  16 \  2^{-{r \over 2}}$}\\
{$ d_2 = 16 \ 2^{-{r \over 2}} \qquad ; \qquad d_3 = 
16 \ 2^{-{r \over 2}} \qquad ; \qquad 
m = \bar{m}$}\\\hline\hline
\end{tabular}
\caption{Chan-Paton groups and tadpole conditions for the magnetized 
$[T^2 \times T^2 \times T^2]/Z_2 \times Z_2$
models.}
\label{cpz2z2m}
\end{center}
\end{table}
\vskip 8pt
\begin{table}
\begin{center}
\begin{tabular}{||c|c|c||}\hline\hline
{\rm Multiplets}&{Number}&{\rm Rep.}\\\hline\hline
{\rm $C$}&{$\biggl(\begin{array}{c}3\\0\end{array} \biggr)$ or 
$\biggl(\begin{array}{c}1\\2\end{array} \biggr)$ if $USp$ \ \ ; \ \ 
$\biggl(\begin{array}{c}2\\1\end{array} \biggr)$ or 
$\biggl(\begin{array}{c}0\\3\end{array} \biggr)$ if $SO$}&
{$\biggl(\begin{array}{c}  $(A,1,1,1,1)$ \\  $(S,1,1,1,1)$ \end{array} 
\biggr)$}\\ \hline
{\rm $C$}&{$\biggl(\begin{array}{c}3\\0\end{array} \biggr)$ or 
$\biggl(\begin{array}{c}1\\2\end{array} \biggr)$ if $USp$ \ \ ; \ \ 
$\biggl(\begin{array}{c}2\\1\end{array} \biggr)$ or 
$\biggl(\begin{array}{c}0\\3\end{array} \biggr)$ if $SO$}&
{$\biggl(\begin{array}{c}  $(1,A,1,1,1)$ \\  $(1,S,1,1,1)$ \end{array} 
\biggr)$}\\ \hline
{\rm $C$}&{$\biggl(\begin{array}{c}3\\0\end{array} \biggr)$ or 
$\biggl(\begin{array}{c}1\\2\end{array} \biggr)$ if $USp$ \ \ ; \ \ 
$\biggl(\begin{array}{c}2\\1\end{array} \biggr)$ or 
$\biggl(\begin{array}{c}0\\3\end{array} \biggr)$ if $SO$}&
{$\biggl(\begin{array}{c}  $(1,1,A,1,1)$ \\  $(1,1,S,1,1)$ \end{array} 
\biggr)$}\\ \hline
{\rm $C$}&{$\biggl(\begin{array}{c}3\\0\end{array} \biggr)$ or 
$\biggl(\begin{array}{c}1\\2\end{array} \biggr)$ if $USp$ \ \ ; \ \ 
$\biggl(\begin{array}{c}2\\1\end{array} \biggr)$ or 
$\biggl(\begin{array}{c}0\\3\end{array} \biggr)$ if $SO$}&
{$\biggl(\begin{array}{c}  $(1,1,1,A,1)$ \\  $(1,1,1,S,1)$ \end{array} 
\biggr)$}\\ \hline
{\rm $C$}&{$3$}&{$(1,1,1,1,Adj)$}\\\hline
{\rm $C$}&{$2^{\frac{r_2+r_3}{2}}$}&{$(F,F,1,1,1),(1,1,F,F,1) $}\\\hline
{\rm $C$}&{$2^{\frac{r_1+r_3}{2}}$}&{$(F,1,F,1,1),(1,F,1,F,1)$}\\\hline
{\rm $C$}&{$2^{\frac{r_1+r_2}{2}}$}&{$(F,1,1,F,1),(1,F,F,1,1)$}\\\hline
{\rm $C$}&{$2^{\frac{r_2+r_3}{2}}$}&{$(1,F,1,1,F+ \bar{F})$}\\\hline
{\rm $C$}&{$2^{r_2+r_3} \ |k_2 \, k_3|$}&
{$(F,1,1,1,F+\bar{F})$}\\\hline\hline
{\rm $C_L$}&{$2^{r_2+r_3+1} |k_2 \, k_3| + 2^{\frac{r_2+r_3}{2}}
\eta_1 |k_2 \, k_3| + \eta_1 +  2^{\frac{r_2}{2}} \, |k_2| -  
 2^{\frac{r_3}{2}} \, |k_3|$}&{$(1,1,1,1,A)$}\\\hline
{\rm $C_L$}&{$2^{r_2+r_3+1} |k_2 \, k_3| - 2^{\frac{r_2+r_3}{2}}
\eta_1 |k_2 \, k_3| - \eta_1 -  2^{\frac{r_2}{2}} \, |k_2| + 
 2^{\frac{r_3}{2}} \, |k_3|$}&{$(1,1,1,1,S)$}\\\hline
{\rm $C_L$}&{$2^{r_2+r_3+1} |k_2 \, k_3| + 2^{\frac{r_2+r_3}{2}}
\eta_1 |k_2 \, k_3| + \eta_1 -  2^{\frac{r_2}{2}} \, |k_2| + 
 2^{\frac{r_3}{2}} \, |k_3|$}&{$(1,1,1,1,\bar{A})$}\\\hline
{\rm $C_L$}&{$2^{r_2+r_3+1} |k_2 \, k_3| - 2^{\frac{r_2+r_3}{2}}
\eta_1 |k_2 \, k_3| - \eta_1 +   2^{\frac{r_2}{2}} \, |k_2| -  
 2^{\frac{r_3}{2}} \, |k_3|$}&{$(1,1,1,1,\bar{S})$}\\
\hline\hline
{\rm $C_L$}&{$2^{\frac{r+r_2}{2}} \ |k_2|$}&{$(1,1,F,1,F)$}\\\hline
{\rm $C_R$}&{$2^{\frac{r+r_3}{2}} \ |k_3|$}&{$(1,1,1,F,F)$}\\\hline
\end{tabular}
\caption{Open spectra of the magnetized 
$Z_2 \times Z_2$ orientifolds with 
$\omega=1$.}
\label{ouz2z2m}
\end{center}
\end{table}
\clearpage
\subsection{Oriented Closed Spectra of the $Z_2 \times Z_2$ Shift-orientifolds}

\vskip 8pt
\begin{table}[ht]
\begin{center}
\begin{tabular}{||c||c|c|c|c|c|c||}\hline\hline
{}&{\rm untwisted}&{\rm untwisted}&
{untwisted}&{twisted}&{twisted}&{}\\ 
{\rm model}&{\rm SUGRA}&{H}&{V}&{H}&{V}&{}\\\hline\hline
{$p_3$}&{$N=2$}&{$1+3$}&{3}&{$16$}&{$16$}&{CY $(19,19)$}\\\hline\hline
{$p_2 p_3$}&{$N=2$}&{$1+3$}&{3}&{$8$}&{$8$}&{CY $(11,11)$}\\\hline
{$w_2 p_3$}&{$N=2$}&{$1+3$}&{3}&{$8$}&{$8$}&{CY $(11,11)$}\\\hline
{$w_1 p_2$}&{$N=2$}&{$1+3$}&{3}&{$8$}&{$8$}&{CY $(11,11)$}\\\hline\hline
{$p_1 p_2 p_3$}&{$N=2$}&{$1+3$}&{3}&{$0$}&{$0$}&{CY $(3,3)$}\\\hline
{$p_1 w_2 w_3$}&{$N=2$}&{$1+3$}&{3}&{$0$}&{$0$}&{CY $(3,3)$}\\\hline
{$w_1 p_2 p_3$}&{$N=2$}&{$1+3$}&{3}&{$0$}&{$0$}&{CY $(3,3)$}\\\hline
{$w_1 p_2 w_3$}&{$N=2$}&{$1+3$}&{3}&{$0$}&{$0$}&{CY $(3,3)$}\\\hline
{$w_1 w_2 p_3$}&{$N=2$}&{$1+3$}&{3}&{$0$}&{$0$}&{CY $(3,3)$}\\\hline
{$w_1 w_2 w_3$}&{$N=2$}&{$1+3$}&{3}&{$0$}&{$0$}&{CY $(3,3)$}\\\hline\hline
\end{tabular}
\caption{Oriented closed spectra of the $Z_2 \times Z_2$ shift-orbifolds.}
\label{clorz2z2sh}
\end{center}
\end{table}
\clearpage
\subsection{Unoriented Closed Spectra of the $p_3$ Models}
\vskip 8pt
\begin{table}[ht]
\begin{center}
\begin{tabular}{||c|c|c||c|c|c|c||}\hline\hline
\multicolumn{3}{||c||}{\rm $B$ rank}&{\rm untwisted}&{untwisted}&{twisted}
&{twisted}\\ 
{\rm $r_1$}&{\rm $r_2$}&{\rm $r_3$}&{\rm SUGRA}&{C}&{C}&{V}\\\hline\hline
{$0$}&{$0$}&{$0$}&{$N=1$}&{$1+3+3$}&{$16+16$}&{$0$}\\\hline\hline
{$2$}&{$0$}&{$0$}&{$N=1$}&{$1+3+3$}&{$14+14$}&{$2+2$}\\\hline
{$0$}&{$2$}&{$0$}&{$N=1$}&{$1+3+3$}&{$14+14$}&{$2+2$}\\\hline
{$0$}&{$0$}&{$2$}&{$N=1$}&{$1+3+3$}&{$12+12$}&{$4+4$}\\\hline\hline
{$2$}&{$2$}&{$0$}&{$N=1$}&{$1+3+3$}&{$12+12$}&{$4+4$}\\\hline
{$0$}&{$2$}&{$2$}&{$N=1$}&{$1+3+3$}&{$11+11$}&{$5+5$}\\\hline
{$2$}&{$0$}&{$2$}&{$N=1$}&{$1+3+3$}&{$11+11$}&{$5+5$}\\\hline\hline
{$2$}&{$2$}&{$2$}&{$N=1$}&{$1+3+3$}&{$10+10$}&{$6+6$}\\\hline\hline
\end{tabular} 
\caption{Unoriented closed spectra of the $p_3$ models.}
\label{clunp3}
\end{center}
\end{table}
\clearpage
\subsection{Orientifolds of the $w_2 p_3$ Models}
\vskip 8pt
\begin{table}[ht]
\begin{center}
\begin{tabular}{||c||c|c|c|c||}\hline\hline
{\rm $B$ rank}&{\rm untwisted}&{untwisted}&{twisted}&{twisted}\\ 
{\rm $r_2+r_3$}&{\rm SUGRA}&{C}&{C}&{V}\\\hline\hline
{$0$}&{$N=1$}&{$1+3+3$}&{$8+8$}&{$0$}\\\hline\hline
{$2$}&{$N=1$}&{$1+3+3$}&{$6+6$}&{$2+2$}\\\hline
{$4$}&{$N=1$}&{$1+3+3$}&{$5+5$}&{$3+3$}\\\hline
\end{tabular}
\caption{Unoriented closed spectra of the $w_2 p_3$ models.}
\label{clunw2p3}
\end{center}
\end{table}
\begin{table}
\begin{center}
\begin{tabular}{c}
\hline\hline
{$U(n) \ \otimes \ U(d_1) \ \otimes \ \biggl(\begin{array}{c} USp(d_2) 
\\ SO(d_2) \end{array} \biggr) \ \otimes \ U(m)$}\\\hline\hline
{$n + \bar{n} + m  + \bar{m}  =  
16 \ 2^{-{r \over 2}}$}\\
{$d_1 + \bar{d_1} + 2^{{r_2 \over 2} + {r_3 \over 2}} \ | k_2 k_3 | \ (m + \bar{m})   =  16 \  2^{-{r \over 2}}$}\\
{$ d_2  =  8 \ 2^{-{r \over 2}} \qquad ; \qquad 
n = \bar{n} \qquad ; \qquad d_1 = \bar{d}_1  \qquad ; \qquad m = \bar{m}$}\\\hline\hline
\end{tabular}
\caption{Chan-Paton groups and tadpole conditions for the 
$w_2 p_3$ models (complex charges).}
\label{cpw2p3}
\end{center}
\end{table}
\vskip 8pt
\begin{table}
\begin{center}
\begin{tabular}{||c|c|c||}\hline
{\rm Mult.}&{Number}&{\rm Rep.}\\\hline\hline
{\rm $C$}&{$1$}&{$(Adj,1,1,1),(1,Adj,1,1)$}\\
{}&{}&{$(1,1,1,Adj)$}\\\hline
{\rm $C$}&{$\biggl(\begin{array}{c}2\\0\end{array} \biggr)$ if $USp$ \ \ \ \
; \ \ \ \  
$\biggl(\begin{array}{c}0\\2\end{array} \biggr)$ if $SO$} &
{$\biggl(\begin{array}{c} (1,S+ \bar{S},1,1),(S+\bar{S},1,1,1)\\
(A+\bar{A},1,1,1), (1,A+\bar{A},1,1)\end{array} \biggr)$}\\\hline
{\rm $C$}&{$3$}&{$(1,1,A,1)$ or $(1,1,S,1)$}\\\hline
{\rm $C$}&{$2^{r_2+r_3} \ |k_2 \, k_3| + 2$}&{$(F,1,1,F)$,
$(\bar{F},1,1,\bar{F})$}\\\hline
{\rm $C$}&{$2^{r_2+r_3} \ |k_2 \, k_3| - 2$}&{$(\bar{F},1,1,F)$,
$(F,1,1,\bar{F})$}\\\hline
{\rm $C$}&{$2 \ 2^{\frac{r_2+r_3}{2}}$}&{$(F,F,1,1)$,$(\bar{F},
\bar{F},1,1)$}\\\hline
{\rm $C$}&{$2 \ 2^{\frac{r_2+r_3}{2}}$}&{$(1,F,1,F)$,
$(1,\bar{F},1\bar{F})$}\\\hline\hline
{\rm $C_L$}&{$2^{r_2+r_3} \ 2 \ |k_2 \, k_3| + 1 + 2^{\frac{r_2+r_3}{2}}
\eta_1 |k_2 \, k_3| + \eta_1 + 2^{\frac{r_2}{2}} |k_2|$}
&{$(1,1,1,A)$}\\\hline
{\rm $C_L$}&{$2^{r_2+r_3} \ 2 \ |k_2 \, k_3| + 1 - 2^{\frac{r_2+r_3}{2}}
\eta_1 |k_2 \, k_3| - \eta_1 - 2^{\frac{r_2}{2}} |k_2|$}
&{$(1,1,1,S)$}\\\hline
{\rm $C_L$}&{$2^{r_2+r_3} \ 2 \ |k_2 \, k_3| + 1 + 2^{\frac{r_2+r_3}{2}}
\eta_1 |k_2 \, k_3| + \eta_1 - 2^{\frac{r_2}{2}} |k_2|$}
&{$(1,1,1,\bar{A})$}\\\hline
{\rm $C_L$}&{$2^{r_2+r_3} \ 2 \ |k_2 \, k_3| + 1 - 2^{\frac{r_2+r_3}{2}}
\eta_1 |k_2 \, k_3| - \eta_1 + 2^{\frac{r_2}{2}} |k_2|$}
&{$(1,1,1,\bar{S})$}\\\hline\hline
{\rm $C_L$}&{$2^{\frac{r_1+r_3}{2}+r_2} \ 2 \ |k_2|$}
&{$(1,1,F,F)$}\\\hline
\end{tabular}
\caption{Open spectra of the $w_2 p_3$ models (complex charges).}
\label{opunw2p3}
\end{center}
\end{table}

\vskip 8pt
\begin{table}
\begin{center}
\begin{tabular}{c}
\hline\hline
{$ \biggl(\begin{array}{c} USp(n_1) \ \otimes \ USp(n_2) \ \otimes \ 
USp(d_1) \ \otimes USp(d_2)
\\ SO(n_1) \ \otimes SO(n_2) \ \otimes SO(d_1) \ \otimes SO(d_2) 
\end{array} \biggr) \otimes \ \biggl(\begin{array}{c} USp(d_3) 
\\ SO(d_3) \end{array} \biggr) \ \otimes \ U(m)$}\\\hline\hline
{$n_1 + n_2 + m  + \bar{m}  =  
16 \ 2^{-{r \over 2}}$}\\
{$d_1 + d_2 + 2^{{r_2 \over 2} + {r_3 \over 2}} \ | k_2 k_3 | \ (m + \bar{m})   =  16 \  2^{-{r \over 2}}$}\\
{$ d_3  =  8 \ 2^{-{r \over 2}} \qquad ; \qquad 
n_2 = n_1 + m + \bar{m}  \qquad ; \qquad d_1 = d_1  \qquad ; \qquad m = \bar{m}$}\\\hline\hline
\end{tabular}
\caption{Chan-Paton groups and tadpole conditions for the $w_2 p_3$ 
models (real charges).}
\label{cpw2p3r}
\end{center}
\end{table}

\vskip 8pt
\begin{table}
\begin{center}
\begin{tabular}{||c|c|c||}\hline
{\rm Mult.}&{Number}&{\rm Rep.}\\\hline\hline
{\rm $C$}&{$1$}&{$(1,1,1,1,1,Adj)$}\\\hline
{\rm $C$}&{$2$}&{$(F,F,1,1,1,1),(1,1,F,F,1,1)$}\\\hline
{\rm $C$}&{$\biggl(\begin{array}{c}1\\0\end{array} \biggr)$ if $USp$ \ \ \ \
; \ \ \ \  
$\biggl(\begin{array}{c}0\\1\end{array} \biggr)$ if $SO$} &
{$\biggl(\begin{array}{c} (S,1,1,1,1,1),(1,S,1,1,1,1)\\
(A,1,1,1,1,1), (1,A,1,1,1,1)\end{array} \biggr)$}\\\hline
{\rm $C$}&{$\biggl(\begin{array}{c}1\\0\end{array} \biggr)$ if $USp$ \ \ \ \
; \ \ \ \  
$\biggl(\begin{array}{c}0\\1\end{array} \biggr)$ if $SO$} &
{$\biggl(\begin{array}{c} (1,1,S,1,1,1),(1,1,1,S,1,1)\\
(1,1,A,1,1,1), (1,1,1,A,1,1)\end{array} \biggr)$}\\\hline
{\rm $C$}&{$\biggl(\begin{array}{c}3\\0\end{array} \biggr)$ if $USp$ \ \ \ \
; \ \ \ \  
$\biggl(\begin{array}{c}0\\3\end{array} \biggr)$ if $SO$} &
{$\biggl(\begin{array}{c} (1,1,1,1,1,S)\\
(1,1,1,1,1,A)\end{array} \biggr)$}\\\hline
{\rm $C$}&{$2^{r_2+r_3} \ |k_2 \, k_3| + 2$}&{$(1,F,1,1,1,F+\bar{F})$}\\\hline
{\rm $C$}&{$2^{r_2+r_3} \ |k_2 \, k_3| - 2$}&{$(F,1,1,1,1,F+\bar{F})$}\\\hline
{\rm $C$}&{$2 \ 2^{\frac{r_2+r_3}{2}}$}&{$(F,1,1,F,1,1),(1,F,F,1,1,1)$}\\\hline
{\rm $C$}&{$2 \ 2^{\frac{r_2+r_3}{2}}$}&{$(1,1,1,F,1,F+\bar{F})$}\\\hline\hline
{\rm $C_L$}&{$2^{r_2+r_3} \ 2 \ |k_2 \, k_3| - 1 + 2^{\frac{r_2+r_3}{2}}
\eta_1 |k_2 \, k_3| + \eta_1 + 2^{\frac{r_2}{2}} |k_2|$}
&{$(1,1,1,1,1,A)$}\\\hline
{\rm $C_L$}&{$2^{r_2+r_3} \ 2 \ |k_2 \, k_3| - 1 - 2^{\frac{r_2+r_3}{2}}
\eta_1 |k_2 \, k_3| - \eta_1 - 2^{\frac{r_2}{2}} |k_2|$}
&{$(1,1,1,1,1,S)$}\\\hline
{\rm $C_L$}&{$2^{r_2+r_3} \ 2 \ |k_2 \, k_3| - 1 + 2^{\frac{r_2+r_3}{2}}
\eta_1 |k_2 \, k_3| + \eta_1 - 2^{\frac{r_2}{2}} |k_2|$}
&{$(1,1,1,1,1,\bar{A})$}\\\hline
{\rm $C_L$}&{$2^{r_2+r_3} \ 2 \ |k_2 \, k_3| - 1 - 2^{\frac{r_2+r_3}{2}}
\eta_1 |k_2 \, k_3| - \eta_1 + 2^{\frac{r_2}{2}} |k_2|$}
&{$(1,1,1,1,1,\bar{S})$}\\\hline\hline
{\rm $C_L$}&{$2^{\frac{r_1+r_3}{2}+r_2} \ 2 \ |k_2|$}
&{$(1,1,1,1,F,F)$}\\\hline
\end{tabular}
\caption{Open spectra of the $w_2 p_3$ models (real charges).}
\label{opunw2p3r}
\end{center}
\end{table}
\clearpage
\subsection{Orientifolds of the $w_1 w_2 p_3$ Models}

\vskip 8pt
\begin{table}[ht]
\begin{center}
\begin{tabular}{||c|c|c|c||}\hline\hline
{\rm untwisted}&{untwisted}&{twisted}&{twisted}\\ 
{\rm SUGRA}&{C}&{C}&{V}\\\hline\hline
{$N=1$}&{$1+3+3$}&{$0$}&{$0$}\\\hline\hline
\end{tabular} 
\caption{Unoriented closed spectra of the $w_1 w_2 p_3$ models.}
\label{clunw1w2p3}
\end{center}
\end{table}
\vskip 8pt
\begin{table}
\begin{center}
\begin{tabular}{ccccccc}
\hline\hline
{$\biggl(\begin{array}{c} USp(n) \\ SO(n) \end{array} \biggr)$}&
{$\otimes$}&
{$\biggl(\begin{array}{c} USp(d_1) \\ SO(d_1) \end{array} \biggr)$}&
{$\otimes$}&
{$\biggl(\begin{array}{c} USp(d_2) \\ SO(d_2) \end{array} \biggr)$}&
{$\otimes$}&
{$U(m)$}\\\hline\hline
\multicolumn{7}{c}{$n + m  + \bar{m}  =  8 \ 2^{-{r \over 2}}$}\\
\multicolumn{7}{c}{$d_1 + 2^{-{r_2 \over 2} - {r_3 \over 2}} \ | k_2 k_3 | 
\ (m + \bar{m})   =  8 \  2^{-{r \over 2}}$}\\
\multicolumn{7}{c}{$ d_2  =  8 \ 2^{-{r \over 2}} \qquad ; \qquad 
m = \bar{m}$}\\\hline\hline
\end{tabular}
\caption{Chan-Paton groups and tadpole conditions for the $w_1 w_2 p_3$ 
models.}
\label{cpw1w2p3}
\end{center}
\end{table}
\vskip 8pt
\begin{table}
\begin{center}
\begin{tabular}{||c|c|c||}\hline\hline
{\rm Mult.}&{Number}&{\rm Rep.}\\\hline\hline
{\rm $C$}&{$3$}&{$(Adj,1,1,1),(1,Adj,1,1)$}\\
{}&{}&{$(1,1,Adj,1)$, $(1,1,1,Adj)$}\\\hline
{\rm $C$}&{$2^{r_2+r_3} \ 2 \ |k_2 \, k_3|$}&{$(F,1,1,F+\bar{F})$}\\
\hline\hline
{\rm $C_L$}&{$2^{r_2+r_3} \ 4 \ |k_2 \, k_3| + 2^{\frac{r_2+r_3}{2}}
\eta_1 2 |k_2 \, k_3| + 2^{\frac{r_2}{2}} 2 |k_2|$}
&{$(1,1,1,A)$}\\\hline
{\rm $C_L$}&{$2^{r_2+r_3} \ 4 \ |k_2 \, k_3| - 2^{\frac{r_2+r_3}{2}}
\eta_1 2 |k_2 \, k_3| - 2^{\frac{r_2}{2}} 2 |k_2|$}
&{$(1,1,1,S)$}\\\hline
{\rm $C_L$}&{$2^{r_2+r_3} \ 4 \ |k_2 \, k_3| + 2^{\frac{r_2+r_3}{2}}
\eta_1 2 |k_2 \, k_3| - 2^{\frac{r_2}{2}} 2 |k_2|$}
&{$(1,1,1,\bar{A})$}\\\hline
{\rm $C_L$}&{$2^{r_2+r_3} \ 4 \ |k_2 \, k_3| - 2^{\frac{r_2+r_3}{2}}
\eta_1 2 |k_2 \, k_3| + 2^{\frac{r_2}{2}} 2 |k_2|$}
&{$(1,1,1,\bar{S})$}\\\hline\hline
{\rm $C_L$}&{$2^{\frac{r_1+r_3}{2}+r_2} \ 4 \ |k_2|$}
&{$(1,1,F,F)$}\\\hline
\end{tabular}
\caption{Open spectra of the $w_1w_2 p_3$ models.}
\label{opunw1w2p3}
\end{center}
\end{table}
\clearpage
\subsection{Non-chiral Orientifolds}

\vskip 8pt
\begin{table}[ht]
\begin{center}
\begin{tabular}{||c||c|c|c|c||}\hline\hline
{\rm $B$ rank}&{\rm untwisted}&{untwisted}&{twisted}&{twisted}\\ 
{\rm $r_2+r_3$}&{\rm SUGRA}&{C}&{C}&{V}\\\hline\hline
{$0$}&{$N=1$}&{$1+3+3$}&{$8+8$}&{$0$}\\\hline\hline
{$2$}&{$N=1$}&{$1+3+3$}&{$6+6$}&{$2+2$}\\\hline
{$4$}&{$N=1$}&{$1+3+3$}&{$5+5$}&{$3+3$}\\\hline
\end{tabular} 
\caption{Unoriented closed spectra of the $p_2 p_3$ models.}
\label{clunp2p3}
\end{center}
\end{table}

\vskip 8pt
\begin{table}
\begin{center}
\begin{tabular}{||c|c|c|c||}\hline\hline
{\rm untwisted}&{untwisted}&{twisted}&{twisted}\\ 
{\rm SUGRA}&{C}&{C}&{V}\\\hline\hline
{$N=1$}&{$1+3+3$}&{$8$}&{$8$}\\\hline\hline
\end{tabular} 
\caption{Unoriented closed spectra of the $w_1 p_2$ models.}
\label{clunw1p2}
\end{center}
\end{table}

\vskip 8pt 

\begin{table} 
\begin{center}
\begin{tabular}{||c|c|c|c||}\hline\hline
{\rm untwisted}&{untwisted}&{twisted}&{twisted}\\ 
{\rm SUGRA}&{C}&{C}&{V}\\\hline\hline
{$N=1$}&{$1+3+3$}&{$0$}&{$0$}\\\hline\hline
\end{tabular} 
\caption{Unoriented closed spectra of the $w_1 p_2 p_3$ and 
$w_1 p_2 w_3$ models.}
\label{clunw1p2p3}
\end{center}
\end{table}
\vskip 8pt
\begin{table}
\begin{center}
\begin{tabular}{c}
\hline\hline
{$U(n_1) \ \otimes \ U(n_2) \ \otimes \ U(d) \ \otimes \ 
U(m)$}\\\hline\hline
{$n_1 + \bar{n}_1 + n_2 + \bar{n}_2 + m  + \bar{m}  =  
32 \ 2^{-{r \over 2}}$}\\
{$d + \bar{d} + 2^{{r_2 \over 2} + {r_3 \over 2}} \ 
| k_2 k_3 | \ (m + \bar{m})   =  32 \  2^{-{r \over 2}}$}\\
{$n_1 = \bar{n}_1 \qquad ; \qquad 
n_2 = \bar{n}_2 \qquad ; \qquad d = \bar{d}  \qquad ; 
\qquad m = \bar{m}$}\\\hline\hline
\end{tabular}
\caption{Chan-Paton groups and tadpole conditions for the $p_2 p_3$ 
models (complex charges).}
\label{cpp2p3}
\end{center}
\end{table}

\begin{table}
\begin{center}
\begin{tabular}{||c|c|c||}\hline\hline
{\rm Multiplets}&{\rm Number}&{\rm Rep.}\\ 
\hline\hline
{\rm C}&{$1$}&{$(F,\bar{F},1,1)$,$(\bar{F},F,1,1)$}\\
{}&{}&{$(1,1,Adj,1)$}\\\hline
{\rm C}&{$1$}&{$(F,F,1,1)$,$(\bar{F},\bar{F},1,1)$}\\
{}&{}&{$(1,1,R+\bar{R},1)$}\\\hline
{\rm C}&{$1$}&{$(R+\bar{R},1,1,1), (1,R+\bar{R},1,1)$}\\ 
{}&{}&{$(1,1,R+\bar{R},1)$}\\\hline
{\rm C}&{$\frac{|k_2 \, k_3|}{4} \ 2^{r_2+r_3} + 1$}&{$(F,1,1,F), (1,F,1,F)$}\\
{}&{}&{$(\bar{F},1,1,\bar{F}), (1,\bar{F},1,\bar{F})$}\\\hline
{\rm C}&{$\frac{|k_2 \, k_3|}{4} \  2^{r_2+r_3} - 1$}&{$(\bar{F},1,1,F), 
(1,\bar{F},1,F)$}\\
{}&{}&{$(F,1,1,\bar{F}), (1,F,1,\bar{F})$}\\\hline
{\rm C}&{$2^{\frac{r_2+r_3}{2}}$}&{$(1,1,F,F), (1,1,\bar{F},\bar{F})$}\\\hline
{\rm C}&{$\frac{|k_2 k_3|}{2} \ (2^{r_2+r_3} + \eta_1 \ 
2^{\frac{r_2+r_3}{2}}) + 1$}&{$(1,1,1,A+\bar{A})$}\\\hline
{\rm C}&{$\frac{|k_2 k_3|}{2} \ (2^{r_2+r_3} - \eta_1 \ 
2^{\frac{r_2+r_3}{2}})$}&{$(1,1,1,S + \bar{S})$}\\\hline\hline
\end{tabular} 
\caption{Open spectra of the magnetized $p_2p_3$ models (complex charges).}
\label{opunp2p3}
\end{center}
\end{table}
\vskip 8pt
\begin{table}
\begin{center}
\begin{tabular}{c}
\hline\hline
{$ \biggl(\begin{array}{c} USp(n_1) \ \otimes \ USp(n_2) \ \otimes \ 
USp(n_3) \ \otimes \ USp(n_4) \ \otimes \
USp(d_1) \ \otimes \ USp(d_2)
\\ SO(n_1) \ \otimes \ SO(n_2) \ \otimes \ SO(n_3) \ \otimes \ SO(n_4) \
\otimes \ SO(d_1) \ \otimes \ SO(d_2) 
\end{array} \biggr) \ \otimes \ U(m)$}\\\hline\hline
{$n_1 + n_2 + n_3 + n_4 + m  + \bar{m}  =  
32 \ 2^{-{r \over 2}}$}\\
{$d_1 + d_2 + 2^{{r_2 \over 2} + {r_3 \over 2}} \ | k_2 k_3 | \ 
(m + \bar{m})   =  32 \  2^{-{r \over 2}}$}\\
{$ n_3 + n_4 = n_1 + n_2 +  m  + \bar{m} 
\qquad ; \qquad d_1 = d_2  \qquad ; \qquad m = \bar{m}$}\\\hline\hline
\end{tabular}
\caption{Chan-Paton groups and tadpole conditions for the $p_2 p_3$ 
models (real charges).}
\label{cpp2p3r}
\end{center}
\end{table}

\begin{table}
\begin{center}
\begin{tabular}{||c|c|c||}\hline\hline
{\rm Multiplets}&{\rm Number}&{\rm Rep.}\\ 
\hline\hline
{\rm C}&{$1$}&{$(F,F,1,1,1,1,1)$,$(1,1,F,F,1,1,1)$}\\
{}&{}&{$(1,1,1,1,1,Adj,1)$,$(1,1,1,1,Adj,1,1)$}\\\hline
{\rm C}&{$1$}&{$(F,1,F,1,1,1,1)$,$(1,F,1,F,1,1,1)$}\\
{}&{}&{$(1,1,1,1,F,F,1)$}\\\hline
{\rm C}&{$1$}&{$(F,1,1,F,1,1,1)$,$(1,F,F,1,1,1,1)$}\\
{}&{}&{$(1,1,1,1,F,F,1)$}\\\hline
{\rm C}&{$2^{\frac{r_2+r_3}{2}}$}&{$(F,1,1,1,1,F,1)$, $(1,F,1,1,1,F,1)$}\\
{}&{}&{$(1,1,F,1,F,1,1)$, $(1,1,1,F,F,1,1)$}\\\hline
{\rm C}&{$2^{\frac{r_2+r_3}{2}}$}&{$(1,1,1,1,1,F,F+\bar{F})$}\\\hline
{\rm C}&{$\frac{|k_2 \, k_3|}{4} \ 2^{r_2+r_3} + 1$}&{
$(F,1,1,1,1,1,F+\bar{F})$, $(1,F,1,1,1,1,F+\bar{F})$}\\\hline
{\rm C}&{$\frac{|k_2 \, k_3|}{4} \ 2^{r_2+r_3} - 1$}&{
$(1,1,F,1,1,1,F+\bar{F})$, $(1,1,1,F,1,1,F+\bar{F})$}\\\hline
{\rm C}&{$\frac{|k_2 k_3|}{2} \ (2^{r_2+r_3} + \eta_1 \ 
2^{\frac{r_2+r_3}{2}}) + 1$}&{$(1,1,1,1,1,1,A+\bar{A})$}\\\hline
{\rm C}&{$\frac{|k_2 k_3|}{2} \ (2^{r_2+r_3} - \eta_1 \ 
2^{\frac{r_2+r_3}{2}})$}&{$(1,1,1,1,1,1,S + \bar{S})$}\\\hline\hline
\end{tabular} 
\caption{Open spectra of the magnetized $p_2p_3$ models (real charges).}
\label{opunp2p3r}
\end{center}
\end{table}
\vskip 8pt
\begin{table}
\begin{center}
\begin{tabular}{c}
\hline\hline
{$U(n) \otimes \biggl(\begin{array}{c} USp(d) \\
SO(d) \end{array} \biggr) \otimes U(m)$}\\\hline\hline
{$n + \bar{n} + m  + \bar{m}  =  
16 \ 2^{-{r \over 2}}$}\\
{$2 d_1 + 2^{{r_2 \over 2} + {r_3 \over 2}} \ 
| k_2 k_3 | \ (m + \bar{m})   =  16 \  2^{-{r \over 2}}$}\\
{$n = \bar{n} \qquad ; 
\qquad m = \bar{m}$}\\\hline\hline
\end{tabular}
\caption{Chan-Paton groups and tadpole conditions for the $w_1 p_2$ and 
$w_1 p_2 p_3$ models (complex charges).}
\label{cpw1p2}
\end{center}
\end{table}
\begin{table}
\begin{center}
\begin{tabular}{||c|c|c||}\hline\hline
{\rm Multiplets}&{\rm Number}&{\rm Rep.}\\ 
\hline\hline
{\rm C}&{$1$}&{$(Adj,1,1)$, $(1,Adj,1)$}\\
{}&{}&{$(1,1,Adj)$}\\\hline
{\rm C}&{$2$}&{$(1,Adj,1)$}\\
{}&{}&{$(A+\bar{A},1,1)$ or $(S+\bar{S},1,1)$}\\\hline
{\rm C}&{$\frac{|k_2 \, k_3|}{2} \ 2^{r_2+r_3}$}
&{$(F+\bar{F},1,F+\bar{F})$}\\\hline
{\rm C}&{$2 \ |k_2 k_3| \ (2^{r_2+r_3} + \eta_1 \ 
2^{\frac{r_2+r_3}{2}})$}&{$(1,1,A+\bar{A})$}\\\hline
{\rm C}&{$2 \ |k_2 k_3| \ (2^{r_2+r_3} - \eta_1 \ 
2^{\frac{r_2+r_3}{2}})$}&{$(1,1,S + \bar{S})$}\\\hline\hline
\end{tabular} 
\caption{Open spectra of the magnetized $w_1 p_2$ and $w_1p_2p_3$ 
models (complex charges).}
\label{opunw1p2}
\end{center}
\end{table}
\vskip 8pt
\begin{table}
\begin{center}
\begin{tabular}{c}
\hline\hline
{$\biggl(\begin{array}{c} USp(n_1) \ \otimes \ USp(n_2)\\ 
SO(n_1) \ \otimes \ SO(n_2)\end{array} \biggr) \ \otimes \ 
\biggl(\begin{array}{c} USp(d_1) \\ SO(d_1) \end{array} \biggr)
\ \otimes \ U(m)$}\\\hline\hline
{$n_1 + n_2 + m  + \bar{m}  =  
16 \ 2^{-{r \over 2}}$}\\
{$2 d_1 + 2^{{r_2 \over 2} + {r_3 \over 2}} \ 
| k_2 k_3 | \ (m + \bar{m})   =  16 \  2^{-{r \over 2}}$}\\
{$m = \bar{m}$}\\\hline\hline
\end{tabular}
\caption{Chan-Paton groups and tadpole conditions for the $w_1 p_2$ and 
the $w_1 p_2 p_3$ models (real charges).}
\label{cpw1p2r}
\end{center}
\end{table}
\vskip 18pt
\begin{table}
\begin{center}
\begin{tabular}{||c|c|c||}\hline\hline
{\rm Multiplets}&{\rm Number}&{\rm Rep.}\\ 
\hline\hline
{\rm C}&{$1$}&{$(Adj,1,1,1)$, $(1,Adj,1,1)$}\\
{}&{}&{$(1,1,Adj,1)$, $(1,1,1,Adj)$}\\\hline
{\rm C}&{$2$}&{$(1,1,Adj,1)$, $(F,F,1,1)$}\\\hline
{\rm C}&{$\frac{|k_2 \, k_3|}{2} \ 2^{r_2+r_3}$}
&{$(F,1,1,F+\bar{F})$, $(1,F,1,F+\bar{F})$}\\\hline
{\rm C}&{$2 \ |k_2 k_3| \ (2^{r_2+r_3} + \eta_1 \ 
2^{\frac{r_2+r_3}{2}})$}&{$(1,1,1,A+\bar{A})$}\\\hline
{\rm C}&{$2 \ |k_2 k_3| \ (2^{r_2+r_3} - \eta_1 \ 
2^{\frac{r_2+r_3}{2}})$}&{$(1,1,1,S + \bar{S})$}\\\hline\hline
\end{tabular} 
\caption{Open spectra of the magnetized $w_1 p_2$ and $w_1p_2p_3$
models (real charges).}
\label{opunw1p2r}
\end{center}
\end{table}
\vskip 8pt
\begin{table}
\begin{center}
\begin{tabular}{c}
\hline\hline
{$\biggl(\begin{array}{c} USp(n) \\ SO(n) \end{array} \biggr) \ 
\otimes \ \biggl(\begin{array}{c} USp(d) \\ SO(d) \end{array} \biggr) \ 
\otimes \ U(m)$}\\\hline\hline
{$n + m  + \bar{m}  =  
8 \ 2^{-{r \over 2}}$}\\
{$d + 2^{{r_2 \over 2} + {r_3 \over 2}} \ 
| k_2 k_3 | \ (m + \bar{m})   =  8 \  2^{-{r \over 2}}$}\\
{$m = \bar{m}$}\\\hline\hline
\end{tabular}
\caption{Chan-Paton groups of the $w_1 p_2 w_3$ models.}
\label{cpw1p2w3}
\end{center}
\end{table}
\vskip 18pt
\begin{table}
\begin{center}
\begin{tabular}{||c|c|c||}\hline\hline
{\rm Multiplets}&{\rm Number}&{\rm Rep.}\\ 
\hline\hline
{\rm C}&{$3$}&{$(Adj,1,1)$, $(1,Adj,1)$}\\
{}&{}&{$(1,1,Adj)$}\\\hline
{\rm C}&{$ 2 \ |k_2 \, k_3| \ 2^{r_2+r_3}$}
&{$(F,1,F+\bar{F})$}\\\hline
{\rm C}&{$2 \ |k_2 k_3| \ (2 \ 2^{r_2+r_3} + \eta_1 \ 
2^{\frac{r_2+r_3}{2}})$}&{$(1,1,A+\bar{A})$}\\\hline
{\rm C}&{$2 \ |k_2 k_3| \ (2 \ 2^{r_2+r_3} - \eta_1 \ 
2^{\frac{r_2+r_3}{2}})$}&{$(1,1,S + \bar{S})$}\\\hline\hline
\end{tabular} 
\caption{Open spectra of the magnetized $w_1p_2w_3$
models.}
\label{opunw1p2w3}
\end{center}
\end{table}

\clearpage
\subsection{$w_2p_3$ Models with ``Brane Supersymmetry Breaking''}
\vskip 8pt
\begin{table}[ht]
\begin{center}
\begin{tabular}{||c||c|c|c|c||}\hline\hline
{}&{\rm untwisted}&{untwisted}&{twisted}&{twisted}\\ 
{\rm model}&{\rm SUGRA}&{C}&{C}&{V}\\\hline\hline
{$w_2p_3$}&{$N=1$}&{$1+3+3$}&{$8$}&{$8$}\\\hline\hline
\end{tabular}
\caption{Unoriented closed spectra of the $w_2 p_3$ models with
 ``brane supersymmetry breaking''.}
\label{clunw2p3bsb}
\end{center}
\end{table}
\vskip 8pt
\begin{table}
\begin{center}
\begin{tabular}{c}
\hline\hline
{$SO(n_1) \otimes SO(n_2) \otimes Usp(d_1) \otimes 
Usp(d_2) \otimes Usp(d_3) \otimes U(m)$}\\\hline\hline
{$n_1 + n_2 + m  + \bar{m}  =  
16$}\\
{$d_1 + d_2 +  | k_2 k_3 | \ (m + \bar{m})   =  16$}\\
{$ d_3  =  8 \qquad ; \qquad 
n_2 = n_1 + m + \bar{m}  \qquad ; \qquad m = \bar{m}$}\\\hline\hline
\end{tabular}
\caption{Chan-Paton groups of the $w_2 p_3$ models 
with ``brane supersymmetry breaking''.}
\label{cpw2p3bsb}
\end{center}
\end{table}
\vskip 8pt
\begin{table}
\begin{center}
\begin{tabular}{||c|c|c||}\hline\hline
{\rm States}&{Number}&{\rm Rep.}\\\hline\hline
{\rm Scalars}&{$2$}&{$(Adj,1,1,1,1,1),(1,Adj,1,1,1,1),(1,1,Adj,1,1,1)$}\\
{}&{}&{$(1,1,1,Adj,1,1),(1,1,1,1,Adj,1),(1,1,1,1,1,Adj)$}\\\hline
{\rm $C$}&{$2$}&{$(Adj,1,1,1,1,1),(1,Adj,1,1,1,1),(1,1,A,1,1,1)$}\\
{}&{}&{$(1,1,1,A,1,1),(1,1,1,1,A,1),(1,1,1,1,1,Adj)$}\\\hline
{\rm Scalars}&{$4$}&{$(F,F,1,1,1,1),(1,1,F,F,1,1),(1,1,1,1,Adj,1)$}\\\hline
{\rm $C$}&{$2$}&{$(F,F,1,1,1,1),(1,1,F,F,1,1),(1,1,1,1,Adj,1)$}\\\hline
{\rm Scalars}&{$4$}&{$(F,1,F,1,1,1),(1,F,1,F,1,1)$}\\\hline
{\rm $C$}&{$2$}&{$(1,F,F,1,1,1),(F,1,1,F,1,1)$}\\\hline
{\rm $C$}&{$|k_2 \, k_3|/2 + 2$}&{$(F,1,1,1,1,F+\bar{F})$}\\\hline
{\rm $C$}&{$|k_2 \, k_3|/2 - 2$}&{$(1,F,1,1,1,F+\bar{F})$}\\\hline
{\rm Scalars}&{$|k_2 \, k_3| - 4 $}&{$(F,1,1,1,1,F+\bar{F})$}\\\hline
{\rm Scalars}&{$|k_2 \, k_3| + 4$}&{$(1,F,1,1,1,F+\bar{F})$}\\\hline
{\rm $C$}&{$2$}&{$(1,1,1,F,1,F+\bar{F})$}\\\hline
{\rm Scalars}&{$4$}&{$(1,1,F,1,1,F+\bar{F})$}\\\hline
{\rm Scalars}&{$3 \ |k_2 \, k_3| - 2 - |k_2|$}&
{$(1,1,1,1,1,A+\bar{A})$}\\\hline
{\rm Scalars}&{$|k_2 \, k_3| + |k_2|$}&{$(1,1,1,1,1,S+\bar{S})$}\\\hline
{\rm Scalars}&{$3 \ |k_2 \, k_3| - 2 + |k_2|$}&
{$(1,1,1,1,1,A+\bar{A})$}\\\hline
{\rm Scalars}&{$|k_2 \, k_3| - |k_2|$}&{$(1,1,1,1,1,S+\bar{S})$}\\\hline\hline
{\rm $C_L$}&{$3 \ |k_2 \, k_3| + 2 + |k_2|$}&{$(1,1,1,1,1,A)$}\\\hline
{\rm $C_L$}&{$|k_2 \, k_3| - |k_2|$}&{$(1,1,1,1,1,S)$}\\\hline
{\rm $C_R$}&{$3 \ |k_2 \, k_3| + 2 - |k_2|$}&{$(1,1,1,1,1,A)$}\\\hline
{\rm $C_R$}&{$|k_2 \, k_3| + |k_2|$}&{$(1,1,1,1,1,S)$}\\\hline\hline
{\rm $C_L$}&{$|k_2|$}&{$(1,1,1,1,F,F)$}\\\hline\hline
{\rm Scalars}&{$|k_2|$}&{$(1,1,1,1,F,F+\bar{F})$}\\\hline
\end{tabular}
\caption{Open spectra of the $w_2 p_3$ models with 
``brane supersymmetry breaking''.}
\label{opunw2p3bsb}
\end{center}
\end{table}
\clearpage
\subsection{Open Spectra of the Undeformed 
$Z_2 \times Z_2$ Shift-orientifolds}
\vspace{2cm}
\begin{table}
\begin{center}
\begin{tabular}{||c|c|c|c|c||}\hline\hline
{model}&{CP group}&{constraints}
&{susy}&{chiral multiplets}\\ 
\hline\hline
{$p_{3}$}&{$[U(n_1)\times U(n_2)]_9 \times$}&{$n_1+n_2=16$}&{N=1}&
{$(A+\bar{A},1,1,1)_{99}$ $(1,A+\bar{A},1,1)_{99}$} \\
{}&{$U(d_1)_{5_1} \times U(d_2)_{5_2}$}&{$d_1=d_2=8$}&{}&{$(F+\bar{F},
F+\bar{F},1,1)_{99}$ }\\
{}&{}&{}&{}&{$(1,1,A,1)_{55}$ $(1,1,1,A)_{55}$} \\
{}&{}&{}&{}&{$(F,1,F,1)_{59}$ $(F,1,1,F)_{59}$}\\
{}&{}&{}&{}&{$(1,\bar{F},1,F)_{59}$ $(1,F,F,1)_{59}$}\\ \hline
{$p_{23}$}&{$[U(n_1)\times U(n_2)]_9 \times$}&{$n_1+n_2=16$}&{N=1}&
{$(A+\bar{A},1,1)_{99}$ $(1,A+\bar{A},1)_{99}$} \\
{}&{$U(d)_{5_1}$}&{$d=8$}&{}&{$(F+\bar{F},F+\bar{F},1)_{99}$ $(1,1,A+\bar{A})_{55}$ } \\
{}&{}&{}&{}&{$(F,1,F)_{59}$ $(\bar{F},1,\bar{F})_{59}$}\\
{}&{}&{}&{}&{$(1,F,F)_{59}$ $(1,\bar{F},\bar{F})_{59}$}\\ \hline 
{$p_{123}$}&{$SO(n_o) \times SO(n_g) \times$}&{$\sum_i n_i =32$}&{N=1}&
{$(F,F,1,1)_{99}$ $(F,1,F,1)_{99}$} \\
{}&{$SO(n_h) \times SO(n_f)$}&{}&{}&
{$(F,1,1,F)_{99}$ $(1,F,F,1)_{99}$} \\
{}&{}&{}&{}&{$(1,F,1,F)_{99}$ $(1,1,F,F)_{99}$} \\ 
\hline 
\hline\hline 
{\rm model}&{\rm CP group}&{constraints}&{susy}&{hypermultiplets}\\ 
\hline\hline
{$w_2p_3$}&{$U(n)_9 \times$}
&{$n=8$}&{N=2}&
{$2$ $(A,1,1)_{99}$ \  $2$ $(1,A,1,)_{5_1 5_1}$} \\
{}&{$U(d_1)_{5_1} \times SO(d_2)_{5_2}$}&{$d_1=d_2=8$}&{}&{$2$ $(F,F,1)_{9 5_1}$}\\ \hline
{$w_1p_2$}&{$U(n)_9 \times SO(d_1)_{5_1}$}
&{$n=d_1=8$}&{N=2}&
{$2$ $(A,1)_{99}$} \\  \hline
{$w_1p_2p_3$}&{$U(n)_9 \times SO(d_1)_{5_1}$}
&{$n=d_1=8$}&{N=2}&
{$2$ $(A,1)_{99}$} \\  \hline
{$w_1w_2p_3$}&{$SO(n)_9 \times $}
&{$n=8$}&{N=4}&
{ -- } \\
{}&{$SO(d_1)_{5_1} \times  SO(d_2)_{5_2}$}&{$d_1=d_2=8$}&{}&
{ -- } \\
\hline 
{$w_1p_2w_3$}&{$SO(n)_9 \times SO(d_1)_{5_1}$}
&{$n=d_1=8$}&{N=4}&
{ -- } \\  \hline 
{$p_1w_2w_3$}&{$U(n)_9$}
&{$n=8$}&{N=4}&
{ -- } \\  \hline
{$w_1w_2w_3$}&{$SO(n)_9$}
&{$n=8$}&{N=4}&
{ -- } \\  \hline\hline
\end{tabular} 
\caption{ Open spectra of the undeformed $Z_2 \times Z_2$ 
shift-orientifolds.}
\label{opunundef}
\end{center}
\end{table}



\clearpage
\begin{thebibliography}{99}

\bibitem{cargese} A.~Sagnotti, in: Cargese '87, 
Non-Perturbative Quantum Field Theory,\\
eds. G. Mack et al. (Pergamon Press, Oxford, 1988) p. 521,
arXiv:hep-th/0208020.

\bibitem{zori1} M.~Bianchi and A.~Sagnotti,
Phys.\ Lett.\ B {\bf 247} (1990) 517,
Nucl.\ Phys.\ B {\bf 361} (1991) 519.

\bibitem{ps} G.~Pradisi and A.~Sagnotti,
Phys.\ Lett.\ B {\bf 216} (1989) 59.

\bibitem{review}
C.~Angelantonj and A.~Sagnotti,
Phys.\ Rept.\  {\bf 371} (2002) 1
[arXiv:hep-th/0204089].

\bibitem{dudas} E.~Dudas,
Class.\ Quant.\ Grav.\  {\bf 17} (2000) R41
[arXiv:hep-ph/0006190].

\bibitem{mtheory} C.~M.~Hull and P.~K.~Townsend,
Nucl.\ Phys.\ B {\bf 438} (1995) 109
[arXiv:hep-th/9410167];
 P.~K.~Townsend,
Phys.\ Lett.\ B {\bf 350} (1995) 184
[arXiv:hep-th/9501068];
 E.~Witten,
Nucl.\ Phys.\ {\bf B443} (1995) 85
[hep-th/9503124];
for a review, see e.g.:   
M.~J.~Duff,
Int.\ J.\ Mod.\ Phys.\ A {\bf 11} (1996) 5623
[arXiv:hep-th/9608117];
M.~Li,
arXiv:hep-th/9811019.

\bibitem{aahdd}
I.~Antoniadis, N.~Arkani-Hamed, S.~Dimopoulos and G.~R.~Dvali,
Phys.\ Lett.\ B {\bf 436} (1998) 257
[arXiv:hep-ph/9804398];
for reviews, see e.g. \cite{dudas}, \cite{anto}.

\bibitem{anto} I.~Antoniadis,
hep-th/0102202.

\bibitem{die1} For a review, see K.~R.~Dienes,
Phys.\ Rept.\ {\bf 287} (1997) 447
[hep-th/9602045].

\bibitem{laq} J.~D.~Lykken,
Phys.\ Rev.\ D {\bf 54} (1996) 3693
[hep-th/9603133]; K.~R.~Dienes, E.~Dudas and T.~Gherghetta,
Nucl.\ Phys.\ B {\bf 537} (1999) 47
[arXiv:hep-ph/9806292].

\bibitem{bw}
N.~Arkani-Hamed, S.~Dimopoulos and G.~Dvali,
Phys.\ Lett.\ B {\bf 429} (1998) 263
[hep-ph/9803315];
Z.~Kakushadze and S.~H.~Tye,
Nucl.\ Phys.\ B {\bf 548} (1999) 180
[arXiv:hep-th/9809147];
L.~Randall and R.~Sundrum,
Phys.\ Rev.\ Lett.\  {\bf 83} (1999) 3370
[arXiv:hep-ph/9905221],
Phys.\ Rev.\ Lett.\  {\bf 83} (1999) 4690
[arXiv:hep-th/9906064];
for a review, see R.~Dick,
Class.\ Quant.\ Grav.\  {\bf 18} (2001) R1
[arXiv:hep-th/0105320].

\bibitem{exper} J.~C.~Long and J.~C.~Price,
arXiv:hep-ph/0303057.

\bibitem{type0} L.~J.~Dixon and J.~A.~Harvey,
Nucl.\ Phys.\ B {\bf 274} (1986) 93;
N.~Seiberg and E.~Witten,
Nucl.\ Phys.\ B {\bf 276} (1986) 272.

\bibitem{zori2} A.~Sagnotti,
arXiv:hep-th/9509080, 
Nucl.\ Phys.\ Proc.\ Suppl.\  {\bf 56B} (1997) 332
[arXiv:hep-th/9702093];
G.~Pradisi,
Nuovo Cim.\ B {\bf 112} (1997) 467
[arXiv:hep-th/9603104].

\bibitem{zori3} C.~Angelantonj,
Phys.\ Lett.\ B {\bf 444} (1998) 309
[arXiv:hep-th/9810214];
C.~Angelantonj and A.~Armoni,
Nucl.\ Phys.\ B {\bf 578} (2000) 239
[arXiv:hep-th/9912257];
R.~Blumenhagen, A.~Font and D.~Lust,
Nucl.\ Phys.\ B {\bf 558} (1999) 159
[arXiv:hep-th/9904069];
R.~Blumenhagen and A.~Kumar,
Phys.\ Lett.\ B {\bf 464} (1999) 46
[arXiv:hep-th/9906234];
K.~Forger,
Phys.\ Lett.\ B {\bf 469} (1999) 113
[arXiv:hep-th/9909010].

\bibitem{branes0} E.~Dudas, J.~Mourad and A.~Sagnotti,
Nucl.\ Phys.\ B {\bf 620} (2002) 109
[arXiv:hep-th/0107081].

\bibitem{scsc} J.~Scherk and J.~H.~Schwarz,
Nucl.\ Phys.\ B {\bf 153} (1979) 61.

\bibitem{sshet} R.~Rohm,
Nucl.\ Phys.\ B {\bf 237} (1984) 553;
S.~Ferrara, C.~Kounnas and M.~Porrati,
Phys.\ Lett.\ B {\bf 181} (1986) 263,
Nucl.\ Phys.\ B {\bf 304} (1988) 500.
Phys.\ Lett.\ B {\bf 206} (1988) 25;
C.~Kounnas and M.~Porrati,
Nucl.\ Phys.\ B {\bf 310} (1988) 355.
I.~Antoniadis, C.~Bachas, D.~C.~Lewellen and T.~N.~Tomaras,
Phys.\ Lett.\ B {\bf 207} (1988) 441;
S.~Ferrara, C.~Kounnas, M.~Porrati and F.~Zwirner,
Nucl.\ Phys.\ B {\bf 318} (1989) 75;
C.~Kounnas and B.~Rostand,
Nucl.\ Phys.\ B {\bf 341} (1990) 641;
I.~Antoniadis,
Phys.\ Lett.\ B {\bf 246} (1990) 377.
I.~Antoniadis and C.~Kounnas,
Phys.\ Lett.\ B {\bf 261} (1991) 369.
E.~Kiritsis and C.~Kounnas,
Nucl.\ Phys.\ B {\bf 503} (1997) 117
[arXiv:hep-th/9703059].

\bibitem{ssopen}
I.~Antoniadis, E.~Dudas and A.~Sagnotti,
Nucl.\ Phys.\ B {\bf 544} (1999) 469
[hep-th/9807011];
I.~Antoniadis, G.~D'Appollonio, E.~Dudas and A.~Sagnotti,
Nucl.\ Phys.\ B {\bf 553} (1999) 133
[hep-th/9812118];
R.~Blumenhagen and L.~Gorlich,
Nucl.\ Phys.\ B {\bf 551} (1999) 601
[hep-th/9812158];
C.~Angelantonj, I.~Antoniadis and K.~Forger,
Nucl.\ Phys.\ B {\bf 555} (1999) 116
[hep-th/9904092];
A.~L.~Cotrone,
Mod.\ Phys.\ Lett.\ A {\bf 14} (1999) 2487
[arXiv:hep-th/9909116].

\bibitem{shift} I.~Antoniadis, G.~D'Appollonio, E.~Dudas and A.~Sagnotti,
Nucl.\ Phys.\ B {\bf 565} (2000) 123
[arXiv:hep-th/9907184].

\bibitem{bantib}
G.~Aldazabal and A.~M.~Uranga,
JHEP {\bf 9910} (1999) 024
[hep-th/9908072]; G.~Aldazabal, L.~E.~Ibanez, F.~Quevedo and A.~M.~Uranga,
JHEP {\bf 0008} (2000) 002 [hep-th/0005067].

\bibitem{aadds}
C.~Angelantonj, I.~Antoniadis, G.~D'Appollonio, E.~Dudas and A.~Sagnotti,
Nucl.\ Phys.\ B {\bf 572} (2000) 36
[hep-th/9911081].

\bibitem{abg}
C.~Angelantonj, R.~Blumenhagen and M.~R.~Gaberdiel,
Nucl.\ Phys.\ B {\bf 589} (2000) 545
[hep-th/0006033].

\bibitem{bsb} I.~Antoniadis, E.~Dudas and A.~Sagnotti,
Phys.\ Lett.\ B {\bf 464} (1999) 38
[arXiv:hep-th/9908023].

\bibitem{bmp} M.~Bianchi, J.~F.~Morales and G.~Pradisi,
Nucl.\ Phys.\ B {\bf 573} (2000) 314
[arXiv:hep-th/9910228].

\bibitem{sugi} S.~Sugimoto,
Prog.\ Theor.\ Phys.\  {\bf 102} (1999) 685
[arXiv:hep-th/9905159].

\bibitem{magwit} E.~Witten,
Phys.\ Lett.\ B {\bf 149} (1984) 351.

\bibitem{bachas} C.~Bachas,
{\it ``A Way to break supersymmetry,''} 
arXiv:hep-th/9503030.

\bibitem{penta} M.~Bianchi and Y.~S.~Stanev,
Nucl.\ Phys.\ B {\bf 523} (1998) 193
[arXiv:hep-th/9711069].

\bibitem{ft}
E.~S.~Fradkin and A.~A.~Tseytlin,
Phys.\ Lett.\ B {\bf 158} (1985) 316;
A.~Abouelsaood, C.~G.~Callan, C.~R.~Nappi and S.~A.~Yost,
Nucl.\ Phys.\ B {\bf 280} (1987) 599;

\bibitem{aads} C.~Angelantonj, I.~Antoniadis, E.~Dudas and A.~Sagnotti,
Phys.\ Lett.\ B {\bf 489} (2000) 223
[arXiv:hep-th/0007090].

\bibitem{magnbsb} C.~Angelantonj and A.~Sagnotti,
arXiv:hep-th/0010279.

\bibitem{ted} R.~Blumenhagen, L.~Goerlich, B.~Kors and D.~Lust,
JHEP {\bf 0010} (2000) 006
[arXiv:hep-th/0007024],
Fortsch.\ Phys.\  {\bf 49} (2001) 591
[arXiv:hep-th/0010198];
R.~Blumenhagen, B.~Kors and D.~Lust,
JHEP {\bf 0102} (2001) 030
[arXiv:hep-th/0012156].

\bibitem{noi} M.~Larosa,
arXiv:hep-th/0111187,
arXiv:hep-th/0212109; 
G.~Pradisi,
arXiv:hep-th/0210088.

\bibitem{nielsole} N.~K.~Nielsen and P.~Olesen,
Nucl.\ Phys.\ B {\bf 144} (1978) 376;
J.~Ambjorn, N.~K.~Nielsen and P.~Olesen,
Nucl.\ Phys.\ B {\bf 152} (1979) 75;
H.~B.~Nielsen and M.~Ninomiya,
Nucl.\ Phys.\ B {\bf 169} (1980) 309.

\bibitem{intbranes} M.~Berkooz, M.~R.~Douglas and R.~G.~Leigh,
Nucl.\ Phys.\ B {\bf 480} (1996) 265
[arXiv:hep-th/9606139];
V.~Balasubramanian and R.~G.~Leigh,
Phys.\ Rev.\ D {\bf 55} (1997) 6415
[arXiv:hep-th/9611165].

\bibitem{phen1}  R.~Blumenhagen, B.~Kors, D.~Lust and T.~Ott,
Nucl.\ Phys.\ B {\bf 616} (2001) 3
[arXiv:hep-th/0107138];
R.~Blumenhagen, V.~Braun, B.~Kors and D.~Lust,
JHEP {\bf 0207} (2002) 026
[arXiv:hep-th/0206038];
R.~Blumenhagen, L.~Gorlich and T.~Ott,
JHEP {\bf 0301} (2003) 021
[arXiv:hep-th/0211059];
D.~Lust and S.~Stieberger,
arXiv:hep-th/0302221.

\bibitem{phen2} G.~Aldazabal, S.~Franco, L.~E.~Ibanez, 
R.~Rabadan and A.~M.~Uranga,
J.\ Math.\ Phys.\  {\bf 42} (2001) 3103
[arXiv:hep-th/0011073],
JHEP {\bf 0102} (2001) 047
[arXiv:hep-ph/0011132];
L.~E.~Ibanez, F.~Marchesano and R.~Rabadan,
JHEP {\bf 0111} (2001) 002
[arXiv:hep-th/0105155];
D.~Cremades, L.~E.~Ibanez and F.~Marchesano,
JHEP {\bf 0207} (2002) 009
[arXiv:hep-th/0201205],
JHEP {\bf 0207} (2002) 022
[arXiv:hep-th/0203160],
arXiv:hep-th/0205074;
A.~M.~Uranga,
arXiv:hep-th/0208014.

\bibitem{csu} M.~Cvetic, G.~Shiu and A.~M.~Uranga,
Nucl.\ Phys.\ B {\bf 615} (2001) 3
[arXiv:hep-th/0107166].
 
\bibitem{phen3} M.~Cvetic, G.~Shiu and A.~M.~Uranga,
Phys.\ Rev.\ Lett.\  {\bf 87} (2001) 201801
[arXiv:hep-th/0107143];
M.~Cvetic, P.~Langacker and G.~Shiu,
Phys.\ Rev.\ D {\bf 66} (2002) 066004
[arXiv:hep-ph/0205252],
Nucl.\ Phys.\ B {\bf 642} (2002) 139
[arXiv:hep-th/0206115];
M.~Cvetic, I.~Papadimitriou and G.~Shiu,
arXiv:hep-th/0212177;
M.~Cvetic and I.~Papadimitriou,
arXiv:hep-th/0303083,
arXiv:hep-th/0303197.

\bibitem{phen4} C.~Kokorelis,
JHEP {\bf 0208} (2002) 036
[arXiv:hep-th/0206108],
arXiv:hep-th/0207234,
arXiv:hep-th/0209202,
arXiv:hep-th/0210200,
arXiv:hep-th/0212281.

\bibitem{phen5} S.~Forste, G.~Honecker and R.~Schreyer,
Nucl.\ Phys.\ B {\bf 593} (2001) 127
[arXiv:hep-th/0008250],
JHEP {\bf 0106} (2001) 004
[arXiv:hep-th/0105208].
 
\bibitem{phen6} D.~Bailin, G.~V.~Kraniotis and A.~Love,
arXiv:hep-th/0108127,
Phys.\ Lett.\ B {\bf 530} (2002) 202
[arXiv:hep-th/0108131],
arXiv:hep-th/0208103,
JHEP {\bf 0302} (2003) 052
[arXiv:hep-th/0212112].

\bibitem{fluxes} R.~Blumenhagen, D.~Lust and T.~R.~Taylor,
arXiv:hep-th/0303016;
J.~F.~Cascales and A.~M.~Uranga,
arXiv:hep-th/0303024.

\bibitem{toroidal} M.~Bianchi, G.~Pradisi and A.~Sagnotti,
Nucl.\ Phys.\ B {\bf 376} (1992) 365;
M.~Bianchi,
Nucl.\ Phys.\ B {\bf 528} (1998) 73
[arXiv:hep-th/9711201];
E.~Witten,
JHEP {\bf 9802} (1998) 006
[arXiv:hep-th/9712028].

\bibitem{carlo} Z.~Kakushadze, G.~Shiu and S.~H.~Tye,
Phys.\ Rev.\ D {\bf 58} (1998) 086001
[arXiv:hep-th/9803141];
C.~Angelantonj,
Nucl.\ Phys.\ B {\bf 566} (2000) 126
[arXiv:hep-th/9908064].

\bibitem{disto} C.~Vafa,
Nucl.\ Phys.\ B {\bf 273} (1986) 592;
A.~Font, L.~E.~Ibanez and F.~Quevedo,
Phys.\ Lett.\ B {\bf 217} (1989) 272;
A.~Font, L.~E.~Ibanez, F.~Quevedo and A.~Sierra,
Nucl.\ Phys.\ B {\bf 337} (1990) 119;
C.~Vafa and E.~Witten,
J.\ Geom.\ Phys.\  {\bf 15} (1995) 189
[arXiv:hep-th/9409188].

\bibitem{erice} M.Bianchi, Ph.D. Thesis, preprint ROM2F-92/13; 
A.~Sagnotti,
{\it{``Anomaly cancellations and open string theories,''}}
in Erice 1992, Proceedings, 
{\it{``From superstrings to supergravity,''}}
arXiv:hep-th/9302099.

\bibitem{bl} M.~Berkooz and R.~G.~Leigh,
Nucl.\ Phys.\ B {\bf 483} (1997) 187
[arXiv:hep-th/9605049].

\bibitem{kaku} Z.~Kakushadze,
Int.\ J.\ Mod.\ Phys.\ A {\bf 15} (2000) 3113
[arXiv:hep-th/0001212].

\bibitem{sethi} J.~de Boer, R.~Dijkgraaf, K.~Hori, A.~Keurentjes, J.~Morgan, 
D.~R.~Morrison and S.~Sethi,
Adv.\ Theor.\ Math.\ Phys.\  {\bf 4} (2002) 995
[arXiv:hep-th/0103170].

\bibitem{ab} C.~Angelantonj and R.~Blumenhagen,
Phys.\ Lett.\ B {\bf 473} (2000) 86
[arXiv:hep-th/9911190].

\bibitem{bgk} R.~Blumenhagen, L.~Gorlich and B.~Kors,
Nucl.\ Phys.\ B {\bf 569} (2000) 209
[arXiv:hep-th/9908130];
R.~Blumenhagen, L.~Gorlich and B.~Kors,
JHEP {\bf 0001} (2000) 040
[arXiv:hep-th/9912204].

\bibitem{z3} G.~Pradisi,
Nucl.\ Phys.\ B {\bf 575} (2000) 134
[arXiv:hep-th/9912218].

\bibitem{croco} D.~Fioravanti, G.~Pradisi and A.~Sagnotti,
Phys.\ Lett.\ B {\bf 321} (1994) 349
[arXiv:hep-th/9311183];
G.~Pradisi, A.~Sagnotti and Y.~S.~Stanev,
Phys.\ Lett.\ B {\bf 354} (1995) 279
[arXiv:hep-th/9503207],
Phys.\ Lett.\ B {\bf 356} (1995) 230
[arXiv:hep-th/9506014];
Phys.\ Lett.\ B {\bf 381} (1996) 97
[arXiv:hep-th/9603097];
related reviews are: G.~Pradisi in ref. \cite{zori2};
A.~Sagnotti and Y.~S.~Stanev,
Fortsch.\ Phys.\  {\bf 44} (1996) 585
[Nucl.\ Phys.\ Proc.\ Suppl.\  {\bf 55B} (1997) 200]
[arXiv:hep-th/9605042];
Y.~S.~Stanev,
arXiv:hep-th/0112222.

\bibitem{usedici} E.~G.~Gimon and J.~Polchinski,
Phys.\ Rev.\ D {\bf 54} (1996) 1667
[arXiv:hep-th/9601038].

\bibitem{bdl} M.~Li,
Nucl.\ Phys.\ B {\bf 460} (1996) 351
[arXiv:hep-th/9510161];
M.~R.~Douglas,
arXiv:hep-th/9512077;
C.~Schmidhuber,
Nucl.\ Phys.\ B {\bf 467} (1996) 146
[arXiv:hep-th/9601003];
M.~B.~Green, J.~A.~Harvey and G.~W.~Moore,
Class.\ Quant.\ Grav.\  {\bf 14} (1997) 47
[arXiv:hep-th/9605033];
J.~Mourad,
Nucl.\ Phys.\ B {\bf 512} (1998) 199
[arXiv:hep-th/9709012];
Y.~K.~Cheung and Z.~Yin,
Nucl.\ Phys.\ B {\bf 517} (1998) 69
[arXiv:hep-th/9710206];
R.~Minasian and G.~W.~Moore,
JHEP {\bf 9711} (1997) 002
[arXiv:hep-th/9710230].

\bibitem{smallins}
E.~Witten,
Nucl.\ Phys.\ B {\bf 460} (1996) 541
[arXiv:hep-th/9511030].

\bibitem{dudmo} E.~Dudas and J.~Mourad,
Phys.\ Lett.\ B {\bf 514} (2001) 173
[arXiv:hep-th/0012071].

\bibitem{prari} G.~Pradisi and F.~Riccioni,
Nucl.\ Phys.\ B {\bf 615} (2001) 33
[arXiv:hep-th/0107090].

\bibitem{chiralas} C.~Angelantonj, M.~Bianchi, G.~Pradisi, 
A.~Sagnotti and Y.~S.~Stanev,
Phys.\ Lett.\ B {\bf 385} (1996) 96
[arXiv:hep-th/9606169].

\bibitem{dinsw} M.~Dine, N.~Seiberg and E.~Witten,
Nucl.\ Phys.\ B {\bf 289} (1987) 589.

\bibitem{gss} A.~Sagnotti,
Phys.\ Lett.\ B {\bf 294} (1992) 196
[arXiv:hep-th/9210127].

\bibitem{ibanez} L.~E.~Ibanez, R.~Rabadan and A.~M.~Uranga,
Nucl.\ Phys.\ B {\bf 542} (1999) 112
[arXiv:hep-th/9808139].
G.~Aldazabal, D.~Badagnani, L.~E.~Ibanez and A.~M.~Uranga,
JHEP {\bf 9906} (1999) 031
[arXiv:hep-th/9904071].

\bibitem{narain} K.~S.~Narain,
Phys.\ Lett.\ B169 (1986) 41;
K.~S.~Narain, M.~H.~Sarmadi and E.~Witten,
Nucl.\ Phys.\ B279 (1987) 369.

\bibitem{dilta}E.~Dudas and J.~Mourad,
Phys.\ Lett.\ B {\bf 486} (2000) 172
[arXiv:hep-th/0004165];
R.~Blumenhagen and A.~Font,
Nucl.\ Phys.\ B {\bf 599} (2001) 241
[arXiv:hep-th/0011269].
 
\bibitem{frabra} M.~R.~Douglas,
JHEP {\bf 9707} (1997) 004
[arXiv:hep-th/9612126];
M.~R.~Douglas, B.~R.~Greene and D.~R.~Morrison,
Nucl.\ Phys.\ B {\bf 506} (1997) 84
[arXiv:hep-th/9704151];
D.~E.~Diaconescu, M.~R.~Douglas and J.~Gomis,
JHEP {\bf 9802} (1998) 013
[arXiv:hep-th/9712230].


\end{thebibliography}
\end{document}